%% file: ms.tex
\pdfoutput=1
\documentclass[a4paper, 11pt, abstracton, titlepage, oneside]{scrartcl}
\makeatletter
\input{resources/packages.tex}
\input{resources/configuration.tex}
\begin{document}
\emergencystretch 3em
% first put your acronyms and abbrevs.
\input{./contents/acronyms}
\input{./resources/notation}
\input{./resources/titlepage.tex}
% and your chapters
\input{./contents/Introduction.tex}
\input{./contents/ProblemSetting.tex}
\input{./contents/Methodology.tex}
\input{./contents/DesignOfExperiments.tex}
\input{./contents/Results.tex}
\input{./contents/Conclusion.tex}

%\onehalfspacing

%
%% finally the bib file
\singlespacing{
%%\footnotesize
\bibliographystyle{model5-names}%\bibliographystyle{model5-names}%\biboptions{authoryear}
%\bibliography{./bibliography/main}} % if more than one, comma separated
\bibliography{ms}}
\newpage
%% and the appendices
\onehalfspacing
\begin{appendices}
	\normalsize
	\input{./contents/Appendices.tex}
\end{appendices}
\end{document}

%% file: resources/packages.tex
\usepackage[T1]{fontenc}
\usepackage[utf8]{inputenc}
\usepackage[onehalfspacing]{setspace}
\usepackage[headsepline]{scrlayer-scrpage}
\clearscrheadfoot % reset standard format
\usepackage{layout}
\usepackage{geometry}
\usepackage{adjustbox}
\usepackage{authblk}
\usepackage[titletoc,title]{appendix}

\usepackage[hyphens]{url}
\usepackage{color}

\usepackage{amsmath,amssymb,amsfonts,amsthm, mathtools, bm}
\usepackage{thmtools, thm-restate}
\usepackage{nicefrac}
\usepackage{array}

\usepackage[normal]{threeparttable}
\usepackage{threeparttablex}
\usepackage{tabularx}
\usepackage{longtable}
\usepackage{booktabs}
\usepackage{colortbl}
\usepackage{multirow}
\usepackage{makecell}

\usepackage[acronym]{glossaries}

\usepackage{floatrow}
\usepackage{graphicx}
\usepackage{placeins}

\usepackage{tikz}
\usetikzlibrary{arrows.meta}
\usetikzlibrary{math}
\usetikzlibrary{shapes}
\usepackage{pgfplots}
\usepgfplotslibrary{groupplots}
\usepgfplotslibrary{dateplot}
\usepackage{tikzscale}

\usepackage{enumerate}
\usepackage{multicol}
\usepackage[author=Patrick]{pdfcomment}
\usepackage{xparse}
\usepackage{twoopt}
\usepackage{etoolbox}

\usepackage{eurosym}
\usepackage{ellipsis}
\usepackage{natbib}
\usepackage{paralist}
\usepackage{ulem}

\usepackage{graphicx}
\usepackage{subfigure, epsfig}
\usepackage{booktabs}
\usepackage{pgfplots}
\usepackage{algorithm}
\usepackage[noend]{algpseudocode}
\usepackage{placeins}
\usepackage{listings}
\usepackage{bbm}
\usepackage{array}
\usepackage{makecell}
\usepackage{multirow}
\usepackage{enumitem}
\usepackage{mathtools, bm}

%% file: resources/configuration.tex
\AtBeginDocument{
	\newgeometry{
		textheight=9in,
		textwidth=6.5in,
		top=1in,
		headheight=14pt,
		headsep=25pt,
		footskip=30pt
	}
}
\newsavebox{\abstractbox}
\renewenvironment{abstract}
{\begin{lrbox}{0}\begin{minipage}{\textwidth}
			\begin{center}\normalfont\sectfont\abstractname\end{center}\quotation}
		{\endquotation\end{minipage}\end{lrbox}%
	\global\setbox\abstractbox=\box0 }

\makeatletter
\expandafter\patchcmd\csname\string\maketitle\endcsname
{\vskip\z@\@plus3fill}
{\vskip\z@\@plus2fill\box\abstractbox\vskip\z@\@plus1fill}
{}{}
\makeatother

\addtokomafont{captionlabel}{\footnotesize\bfseries} % style for caption labels
\addtokomafont{caption}{\footnotesize\bfseries} % and the caption
\bibpunct[, ]{(}{)}{,}{a}{}{,}%
\newtheorem{proposition}{Proposition}

\counterwithin*{equation}{section}

\newenvironment{myfont}{\fontfamily{ccr}\selectfont}{\par}
\DeclareTextFontCommand{\textmyfont}{\myfont}

\AtBeginDocument{%
	\abovedisplayskip=7.5pt plus 0pt minus 0pt
	\abovedisplayshortskip=7.5pt plus 0pt minus 0pt
	\belowdisplayskip=7.5pt plus 0pt minus 0pt
	\belowdisplayshortskip=7.5pt plus 0pt minus 0pt
}
\newcolumntype{L}[1]{>{\raggedright\let\newline\\\arraybackslash\hspace{0pt}}p{#1}}
\newcolumntype{C}[1]{>{\centering\let\newline\\\arraybackslash\hspace{0pt}}p{#1}}
\newcolumntype{R}[1]{>{\raggedleft\let\newline\\\arraybackslash\hspace{0pt}}p{#1}}
\renewcommand{\emph}[1]{\textit{#1}}

\usepackage{natbib}
\bibpunct[, ]{(}{)}{,}{a}{}{,}%

\newcommand{\location}[1]{region}
\newcommand{\locationCAP}[1]{Region}
\newcommand{\timeStep}{time step}

\usepackage{tikz}

\usepackage{tikz}
\usepackage{etoolbox}
\newcommand{\circled}[2][]{%
	\tikz[baseline=(char.base)]{%
		\node[shape = circle, draw, inner sep = 1pt]
		(char) {\phantom{\ifblank{#1}{#2}{#1}}};%
		\node at (char.center) {\makebox[0pt][c]{#2}};}}
\robustify{\circled}

\newtheorem{result}{Result}
\newtheorem{remark}{Remark}

%% file: contents/acronyms.tex
% a
\newacronym{ADP}{ADP}{approximate dynamic programming}

\newacronym{LMD}{LMD}{last-mile delivery}
\newacronym{LP}{LP}{linear program}
% b
% c
\newacronym{CD}{CD}{crowdsourced driver}
\newacronym{CTMDP}{CTMDP}{Continous Time \textsc{Markov} Decision Process}
\newacronym{CAGR}{CAGR}{compound annual growth rate}
% d
\newacronym{DP}{DP}{dynamic program}
\newacronym{DDP}{DDP}{dynamic delivery problem}
\newacronym{DPDP}{DPDP}{dynamic pick-up and delivery problem}
\newacronym{DTMDP}{DTMDP}{Discrete-Time \textsc{Markov} Decision Process}
% e
\newacronym{eu}{EU}{European Union}
\newacronym{ecv}{ECV}{electric commercial vehicle}
\newacronym{FD}{FD}{fixed driver}
\newacronym{FTE}{FTE}{full-time equivalent}
\newacronym{FA}{FA}{fluid approximation}
%\newacronym{exact}{exact}{exact}
% f
% g
\newacronym{ghg}{GHG}{greenhouse gas}
\newacronym{GW}{GW}{gigworker}
% h
% i
% j
% k
% l
\newacronym{LSP}{LSP}{logistics service provider}
% m
\newacronym{MDP}{MDP}{Markov decision process}
\newacronym{MDRP}{MDRP}{meal delivery routing problem}
\newacronym{MY}{MY}{myopic policy}
% n
% o
\newacronym{OD}{OD}{occasional driver}
\newacronym{vfa}{VFA}{value function approximation}
\newacronym{bdp}{BDP}{backward dynamic programming}
% p
\newacronym{PL-VFA}{PL-VFA}{piecewise linear value function approximation}
% q
% r
% s
\newacronym{SMDP}{SMDP}{Semi-\textsc{Markov} Decision Process}
% t
% u
% v
\newacronym{VRP}{VRP}{vehicle routing problem}
\newacronym{VRPOD}{VRPOD}{vehicle routing problem with occasional driver}
% w
\newacronym{WP}{WP}{workforce planning}

%% file: resources/titlepage.tex
\title{\large Strategic Workforce Planning in Crowdsourced Delivery with Hybrid Driver Fleets}

% and the authors
\author[1]{\normalsize Julius Luy}
\author[1]{\normalsize Gerhard Hiermann}
\author[2]{\normalsize Maximilian Schiffer}
\affil{\small 
	TUM School of Management, Technical University of Munich, 80333 Munich, Germany
	
	\scriptsize julius.luy@tum.de; gerhard.hiermann@tum.de
	
	\small
	\textsuperscript{2}TUM School of Management \& Munich Data Science Institute,
	
	Technical University of Munich, 80333 Munich, Germany
	
	\scriptsize schiffer@tum.de}

% if you like - a date
\date{}

% in case you have a headline - otherwise outcomment
\lehead{\pagemark}
%\rehead{\normalfont\scriptsize\textbf{Schiffer et. al.:} \textit{Perspectives for electric commercial vehicles}}
%\lohead{\normalfont\scriptsize\textbf{Schiffer et. al.:} \textit{Perspectives for electric commercial vehicles}}
\rohead{\pagemark}

% finally your abstract
\begin{abstract}
\begin{singlespace}
\input{./contents/acronyms.tex}
{\small\noindent \input{./contents/abstract.tex}}
\end{singlespace}
\end{abstract}

% don't forget to make the tile
\maketitle

%% file: contents/abstract.tex
Nowadays, \glspl{LSP} increasingly consider using a crowdsourced workforce on the last mile to
	fulfill customers' expectations regarding same-day or on-demand delivery at reduced costs. The crowdsourced workforce's availability is, however, uncertain. Therefore, \glspl{LSP} often hire additional fixed employees to perform deliveries when the availability of crowdsourced drivers is low. In this context, the reliability versus flexibility trade-off which \glspl{LSP} face over a longer period, e.g., a year, remains unstudied. Against this background, we jointly study a workforce planning problem that considers \glspl{FD} and the temporal development of the \gls{CD} fleet over a long-term time horizon. We consider two types of \glspl{CD}, \glspl{GW} and \glspl{OD}. While \glspl{GW} are not sensitive
	to the request's destination and typically exhibit high availability, \glspl{OD} only serve requests whose
	origin and destination coincide with their own private route's origin and destination. Moreover, to account for time horizon-specific dynamics, we consider stochastic turnover for both \glspl{FD} and \glspl{CD} as well as stochastic \gls{CD} fleet growth. We formulate the resulting workforce planning problem as a \gls{MDP} whose reward function reflects total costs, i.e., wages and operational costs arising from serving demand with \glspl{FD} and \glspl{CD}, and solve it via \gls{ADP}. Applying our approach to an environment based on real-world demand data from GrubHub, we find that in fleets consisting of \glspl{FD} and \glspl{CD}, \gls{ADP}-based hiring policies can outperform myopic hiring policies by up to $19\,\%$ in total costs. In the studied setting, we observed that \glspl{GW} reduce the \gls{LSP} 's total costs more than \glspl{OD}. When we account for \glspl{CD}' increased resignation probability when not being matched with enough requests, the amount of required \glspl{FD} increases.

%% file: contents/Introduction.tex
\section{Introduction}\label{sec:introduction}
\glsresetall
In recent years, on-demand home delivery services experienced significant growth, especially in urban areas. Global e-commerce sales grew by $38\,\%$ in a year-over-year comparison in 2021 \citep{Forbes2021}, and consumers increasingly use meal and grocery delivery apps, of which meal delivery has globally tripled in revenue since 2017 to $150\,\$$ billion in 2022 \citep{AhujaChandraEtAl2022}. This rapid growth in demand for delivery services leads to increased customer expectations regarding same-hour or same-day delivery. Herein, \glspl{LSP} face a dilemma as they must provide sufficient delivery capacity to rapidly serve increasing demand while maintaining low costs to remain competitive. 

To address this challenge, \glspl{LSP} increasingly consider crowdsourcing: they outsource delivery requests to \glspl{CD}, i.e., independent contractors who flexibly decide when and where to work and are paid per request and not per hour. \glspl{CD} can decide to leave the \gls{LSP}'s fleet at any time. The \gls{LSP} profits from the \glspl{CD} as an easily scalable workforce at the price of uncertainty in the \glspl{CD}' availability, both from a strategic long-term and an operational short-term perspective. While many \glspl{LSP} base their business models exclusively on \glspl{CD}, e.g., Doordash or Grubhub, some companies operate hybrid driver fleets, i.e., fleets consisting of \glspl{CD} and \glspl{FD}, which are permanent employees, to reduce the uncertainty arising from the use of \glspl{CD}. For example, Bringg's partnership with WorkWhile aims at providing additional delivery capacity through a pool of \glspl{CD} to \glspl{LSP} with existing \gls{FD} fleets~{\citep{Freightwaves2022}}. 

To account for heterogeneous \gls{CD} behavior, we consider two predominant types of \glspl{CD}: the first type of \glspl{CD} are \glspl{GW}, whose request acceptance behavior is not sensitive to the request's destination and who typically exhibit high availability. Example companies relying on this type of workforce are Postmates, Instacart, DoorDash (in the US), or Rappi (in South America). \glspl{GW} install an app on their phone and receive a notification when a new delivery request arises. They can then either accept the request or wait for another request. When serving the request, they receive compensation, typically proportional to the distance of the request's route. Second, we consider \glspl{OD}, which only serve requests whose origin and destination coincide with their private route's origin and destination. For example, the company Roadie relies on this type of driver. As such a concept leverages pre-existing routes, it potentially reduces delivery traffic, emissions, and costs.

One major challenge of a mixed fleet of \glspl{FD} and \glspl{CD} is to ensure minimum service levels, which the \gls{LSP} achieves by hiring the right number of \glspl{FD} based on the expected demand and the uncertain \glspl{CD} supply unfolding throughout the planning horizon. While initially hired \glspl{FD} might become obsolete if the number of \glspl{CD} grows, not hiring enough \glspl{FD} early negatively impacts the \gls{LSP}'s service level in early time periods. Studying this cost versus service level trade-off is the main focus of this paper. 

Since the level of required \glspl{FD} to match the demand depends on operational aspects, e.g., request route patterns, we develop a framework that integrates decision-making on two planning levels: the strategic level, where the \gls{LSP} makes hiring decisions, and the operational level, where the \gls{LSP} decides on how to route its \glspl{FD} and which request to outsource to \glspl{CD}. In the remainder of this section, we relate our work to the existing literature (Section \ref{sec:literature}), state our contribution (Section \ref{sec:contribution}), and outline the paper's structure~(Section \ref{sec:organization}).

\FloatBarrier
\subsection{Related Literature}\label{sec:literature}
Three streams of literature relate to our work: \glspl{VRP} with \glspl{CD}, strategic workforce planning problems with conventional employees, and studies combining workforce planning and crowdsourcing for general on-demand service platforms and last-mile delivery companies. We detail these streams in the following.

A large body of literature emerged in the field of \glspl{VRP} with \glspl{CD}. First studies consider an \gls{LSP} that optimizes route plans for deliveries from a single depot for \glspl{FD} and expected \glspl{CD} in a static day-ahead manner \citep{ArchettiSavelsberghEtAl2016, GdowskaVianaEtAl2018, TorresGendreauEtAl2020}. Other papers study delivery with \glspl{CD} in a multi-depot \citep{SampaioSavelsberghEtAl2017} or many-to-many network \citep{RavivTenzer2018,VoigtKuhn2022}. Further works consider a \gls{DDP} in a crowdsourced context \citep{ArslanZuidwijk2016, DayarianSavelsbergh2020, Mak2020}, which is a special class of a dynamic pick-up and delivery problem \citep{BerbegliaCordeauEtAl2010}, wherein drivers do not change their route or pick-up another request once they began serving the current one. More recently, the \gls{MDRP} led to an increased focus on the \gls{DDP} with \glspl{CD}. In the \gls{MDRP}, requests arise dynamically at random \location{}s and must be delivered instantly to their destinations. In this context, some works consider random \gls{CD} supply \citep{ReyesEreraEtAl2018}, while others assume that \glspl{CD}' availability is known to the \gls{LSP} \citep{YildizSavelsbergh2019, UlmerBarettEtAl2021}. Our problem corresponds to a \gls{DDP} in a many-to-many network using \glspl{CD}, and we refer to the literature review of crowdsourced delivery in \cite{AlnaggarGzaraEtAl2019} and \cite{SavelsberghUlmer2022} for a comprehensive overview. So far, studies on the dynamic delivery setting consider relatively small instance sizes to benchmark their order matching and courier routing policies, e.g., 24 drivers \citep{UlmerBarettEtAl2021}. Moreover, to the best of our knowledge, all works considering the dynamic delivery problem with \glspl{CD} envision \glspl{CD} to behave like \glspl{GW} and neglect the potential of synchronizing demand with \glspl{OD}.

%\textbf{Conventional workforce management:} 
In the strategic workforce planning problem with conventional employees, the objective minimizes costs from hiring, compensating, promoting, and operating a workforce over a certain time horizon. Several studies model employee hiring, training and learning, and turnover dynamics as a sequential decision-making problem formalized as a \gls{MDP}. \cite{GansZhou2002} consider the employee hiring problem of a service organization that wants to serve uncertain demand. They model hiring decisions, up-skilling transitions, employees' turnover rates and formulate a total cost minimization objective, including an operational cost element. Similar studies include firing decisions \citep{AhnRighterEtAl2005}, propose heuristics to solve large instances \citep{SongHuang2008}, consider worker heterogeneity \citep{ArlottoEtAl2014}, account for inter-departmental worker mobility \citep{DimitriouEtAl2013}, model decisions on multiple organizational levels \citep{GuerryAndFeyter2012}, or focus on a specific application case, e.g., healthcare \citep{HuLavieriEtAl2016}. Further works use multi-stage stochastic programming combined with linearizations, Bender's decomposition, or conic optimization \citep[cf.][]{ZhuAndSherali2009, FeyterEtAl2017,JailletEtAl2022}.
Similar to these studies, we aim at finding total cost minimizing \glspl{FD} hiring policies over a long-term planning horizon. None of these works, however, considers the presence of a partially uncertain workforce whose size cannot be controlled. Incorporating such an uncertain workforce in our long-term \gls{FD} hiring problem is the focus of our work.

Some studies investigate workforce management in a crowdsourced context. One stream of works analyzes general on-demand platforms controlling the supply of crowdsourced workers indirectly by adjusting the compensation offered for a service. \cite{GurvichLariviereEtAl2019} study such a platform and consider self-scheduling agents that decide to work based on expected compensation and their availabilities. Similar works focus on surge pricing to balance demand and supply \citep{CachonDanielsEtAl2017}, on the influence of agents' independence and customers' delay sensitivity \citep{Taylor2018}, and platform commission schemes \citep{ZhouLinEtAl2019}. Similarly to these works, we consider self-scheduling agents as part of our workforce. However, our problem formulation differs significantly from existing works, as we consider them jointly with conventional employees (the \glspl{FD}) and control our workforce solely through the hiring process of \glspl{FD}. Finally, studies combining workforce planning and crowdsourced delivery are closest to our work. \cite{DaiLiuEtAl2017} study a problem with in-house drivers (equivalent to permanent employees), part- and full-time \glspl{CD}, and derive optimal in-house driver and \gls{CD} staffing levels at different depots and times of one day based on a deterministic demand scenario. Similarly, \cite{BehrendtEtAl2022}, \cite{ChengEtAl2023} and \cite{GoyalEtAl2023} consider hybrid crowdsourced fleets with joint \gls{FD} fleet-sizing and operational decision making, respectively focusing on warehouse allocation decisions, robust workforce management, and order pricing. All of these studies are restricted to a time horizon of one day, similar to \cite{UlmerSavelsbergh2020} and \cite{BehrendtEtAl2022_b}, who focus on pure \gls{CD} fleets and consider two types of \glspl{CD}: scheduled \glspl{CD} that announce their availability prior to the operational time horizon and unscheduled couriers, that arrive ad-hoc while the \gls{LSP} already operates. They aim to find the optimal set of schedules for one day to minimize fixed costs associated with scheduled \glspl{CD} and operational costs. While the former employ a classical value function approximation approach, the latter use neural networks to find the optimal set of shifts. Finally, \cite{LeiJasinEtal2020} also consider a one-day planning horizon and an entirely crowdsourced delivery platform and study mechanisms to reduce demand-supply imbalance by outsourcing excess requests to drivers' willingness to prolong their scheduled shifts. While these works study joint \gls{FD} acquisition and operational planning, they consider short-term planning horizons. Hence, these works do not account for long-term dynamics, e.g., workforce turnover or stochastic \gls{CD} fleet growth. Moreover, the \glspl{LSP}' contractual commitment when hiring \glspl{FD} reduces itself to one day in the studies above. However, in many legislative systems, contracts for fixed employees must have a minimum duration of a year, even when considering temporary contracts. Changing demand levels or increasing \gls{CD} supply might make these fixed employees obsolete before their minimum contract duration terminates. Our work will address this untouched issue by considering long-term time horizons.

In conclusion, our work closes three gaps in the literature, combining crowdsourced delivery and workforce management. First, to the best of our knowledge, no work considers joint \gls{FD} fleet sizing and operational decision-making on a long-term time horizon, thereby neglecting dynamics such as workforce turnover or stochastic \gls{CD} fleet growth. Second, all works considering the dynamic delivery problem with \glspl{CD} envision \glspl{CD} to behave like \glspl{GW}, hence disregarding the potential to synchronize demand with \glspl{OD}. Third, studies on the dynamic delivery setting with \glspl{CD} consider relatively small instance sizes. Yet, the instant delivery market, especially in urban areas, is expected to grow significantly, thus calling for studies accounting for large demand scenarios and large delivery fleets. 

\subsection{Contribution}\label{sec:contribution}
To close the research gaps outlined above, we develop a novel framework to study the long-term workforce planning problem in the context of hybrid crowdsourced delivery fleets. To account for the interplay between workforce planning and operations, we integrate hiring decisions for a long-term time horizon with operational decisions regarding \gls{FD} relocation and outsourcing of demand to \glspl{CD}. Moreover, we consider two \gls{CD} types, \glspl{GW} and \glspl{OD},  which exhibit distinct request acceptance behaviors. While the former is less sensitive to a request's origin and destination and typically exhibits higher availability, the latter only accepts requests whose origin and destination coincide with their private route's origin and destination.

Specifically, our contribution is threefold. First, we formalize the strategic level planning problem as a novel stochastic workforce planning problem, wherein the \gls{LSP} needs to decide on how many \glspl{FD} to hire or fire while taking into account uncertain \gls{CD} supply. We model the strategic level as a finite-horizon \gls{MDP}. Here, the objective is to minimize total costs arising from \gls{FD} wages and operational costs. To obtain the latter term for large fleets within reasonable computation times, we approximate the operational problem with a fluid model. Second, we prove the value function's convexity along the \gls{FD} dimension and use this property to develop a look-ahead policy based on \gls{PL-VFA}, which approximately solves our strategic problem. Third, we conduct numerical studies based on real-world data provided by \cite{Grubhub2018}, wherein we benchmark our look-ahead with a myopic policy and evaluate sensitivities of strategic and operational levels' parameters, e.g., joining and resignation rates of \glspl{CD}. Our main findings are as follows: 
\begin{enumerate}
	\item A hiring policy obtained from \gls{PL-VFA} can yield up to $19\,\%$ lower total costs than a myopic hiring policy. It does so by hiring less \glspl{FD} than required in early time steps and by relying on future \gls{CD} supply.
	\item \glspl{FD} remain an important cost driver of total costs in the hybrid fleet, constituting up to $50\,\%$ of total costs. Thus, developing better hiring \gls{FD} policies can significantly impact total costs. \glspl{GW} are the main cost driver among \glspl{CD}, whereas \glspl{OD} bear a significant potential when their spatial and temporal patterns are synchronized with request patterns. 
	\item When we take into account that \glspl{CD} leave the \gls{LSP}'s platform with a higher likelihood if they are matched to a lower number of requests, we observe a lower effective \gls{CD} supply, which leads to a higher amount of required \glspl{FD}.
\end{enumerate}

\subsection{Structure}\label{sec:organization}
We structure the remainder of this work as follows. In Section \ref{sec:problem_setting}, we introduce our problem setting, formalizing the strategic level as an \gls{MDP} and introduce a closed queueing network to model the operational level. In Section \ref{sec:methodology}, we describe the \gls{PL-VFA} for finding the optimal number of \glspl{FD} and derive a fluid approximation for our operational planning problem. We detail the design of experiments for our numerical study in Section \ref{sec:design_of_experiments} and discuss results in Section \ref{sec:results}. We conclude this paper with a short synthesis in Section \ref{sec:conclusion}.

%% file: contents/ProblemSetting.tex
\section{Problem Setting}\label{sec:problem_setting}
We focus on an \gls{LSP} providing on-demand delivery services in an urban area. Requests arise dynamically in different \location{}s within the urban area and must be dispatched instantaneously. To serve their demand, the \gls{LSP} operates a mixed fleet of drivers consisting of \glspl{FD}, \glspl{GW}, and \glspl{OD}. While \glspl{FD} are hired as fixed employees, \glspl{CD} register on the \gls{LSP}'s crowdsourced delivery platform, e.g., via an app, through which they receive notifications about and can accept potential delivery requests. They can de-register from the platform at any point in time. The \gls{LSP} seeks to minimize total costs arising from serving requests and paying wages to or laying off their \glspl{FD}  over the strategic level's time horizon $\mathcal{T}$. The horizon $\mathcal{T}$ can span multiple months to a couple of years, wherein the time steps represent, for example, weeks. 

\begin{figure}[htbp]
	\centering
	\includegraphics[scale=0.7]{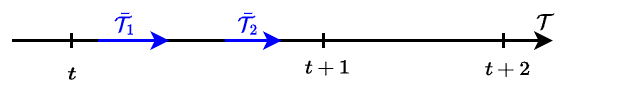}
	\caption{Time horizon example with $K=2$.}\label{fig:time_horizons}
\end{figure}

We first describe the evolution of the driver pool available to the \gls{LSP} before detailing the total cost's composition. Let $n_t^{\mathrm{\gls{FD}}}$, $n_t^{\mathrm{\gls{GW}}}$, $n_t^{\mathrm{\gls{OD}}}$ describe the number of \glspl{FD}, \glspl{GW}, and \glspl{OD} available to the \gls{LSP} in time step $t\in\mathcal{T}$.  Let $a_t$ denote the net number of newly hired ($a_t>0$) or laid-off \glspl{FD} ($a_t<0$) in time step $t$. At the beginning of each \timeStep, the \gls{LSP} can decide to hire or lay off $a_t$ \glspl{FD}. At the end of each time step $t$, some \glspl{FD} and \glspl{CD} resign, while some new \glspl{CD} decide to join the \gls{LSP}'s platform. Let $\tilde{x}^{\alpha}$ denote the random variable describing the number of drivers resigning at the end of $t$ with $\alpha\in\left\{\mathrm{\gls{FD}},\mathrm{\gls{GW}},\mathrm{\gls{OD}}\right\}$. We let $\tilde{x}^{\alpha}$ follow a probability distribution $\mathcal{X}^{\alpha}$, which we will detail in Section \ref{sec:design_of_experiments}. Analogously, we model the number of newly joining \glspl{CD}, $\tilde{y}^{\alpha}$, to follow a probability distribution $\mathcal{Y}^{\alpha}$. We assume the distributions~${\mathcal{X}^{\alpha}}$ and $\mathcal{Y}^{\alpha}$ to be independent. This is plausible since they represent different types of workforce exhibiting distinct behaviors and motivations to work for the \gls{LSP}.
The evolution of the number of drivers available to the \gls{LSP}, from one time step $t$ to $t+1$, then reads
\begin{align}
n_{t+1}^{\mathrm{\gls{FD}}} = n_{t}^{\mathrm{FD}}+a_t-\tilde{x}^{\mathrm{\gls{FD}}}; \qquad 
n_{t+1}^{\mathrm{\gls{GW}}} = n_{t}^{\mathrm{GW}}+\tilde{y}^{\mathrm{\gls{GW}}}-\tilde{x}^{\mathrm{\gls{GW}}}; \qquad 
n_{t+1}^{\mathrm{\gls{OD}}} = n_{t}^{\mathrm{OD}}+\tilde{y}^{\mathrm{\gls{OD}}}-\tilde{x}^{\mathrm{\gls{OD}}}. \label{eq:driver_transitions} % \label{eq:fd_transition}, ,\label{eq:gw_transition}
\end{align}
Let the constant $C^{\mathrm{fix}}$ denote the wage per \gls{FD} and time step $t$. Moreover, let $C^{\mathrm{sev}}$ denote the severance payment per laid-off \gls{FD} and time step. Finally, let the function $C_t^{\mathrm{ops}}(n_t^{\mathrm{FD}},n_t^{\mathrm{GW}},n_t^{\mathrm{OD}},\mathcal{R}^t)$ denote costs from serving requests. Here, $\mathcal{R}^{t}$ represents the demand to be served in time step $t$. We assume a deterministic demand curve over the strategic time horizon $\mathcal{T}$. 
%\begin{assumption}
%	\label{ass:deterministic_demand}
%\end{assumption}
%\begin{assumption}
We assume there are no shortages in \gls{FD} supply. Thus, the \gls{LSP} can hire enough \glspl{FD} to cover the entire demand in each time step $t$. % \label{ass:fd_supply}
%\end{assumption}
We make operational decisions on a more granular time scale than strategic ones. To this end, we define a second time horizon $\bar{\mathcal{T}}$ on the operational level, wherein $\bar{t}\in\mathcal{\bar{T}}$ denotes the respective time step, which describes, e.g., hours or minutes. We visualize the dependency between strategic and operational level time horizons in Figure \ref{fig:time_horizons}. We can define multiple operational level time horizons between two strategic level time steps $t$ and $t+1$. Let $K$ denote the number of operational level time horizons, then $\mathcal{\bar{T}} = \bigcup_{k=0}^K \mathcal{\bar{T}}_k$. Note that the $\mathcal{\bar{T}}_k$ do not need to cover the entire period between $t$ and $t+1$. This is the case when, for example, $t$ and $t+1$ describe months and operational level's time horizons describe only the afternoon time windows of every working day.

Let $\mathcal{M}$ denote a set of \location{}s. Requests arise dynamically in some \location{} $i\in\mathcal{M}$ and need to be delivered instantaneously to region $j\in\mathcal{M}$. We model request arrivals in \location{} $i$ with a \textsc{Poisson} process with arrival rates $\lambda^{\mathrm{R}}_{it}\in\mathcal{R}^t$. We describe requests' destinations by a request origin-destination matrix $P^{R}_{ijt}$. Requests not being served in time step $\bar{t}\in\mathcal{\bar{T}}$ disappear from the system and result in a penalty. Likewise, we let \gls{CD} arrivals follow a \textsc{Poisson} process. We link arrival rates at \location{}s $i$, denoted by $\lambda^{\mathrm{\gls{GW}}}_{it}$ and $\lambda^{\mathrm{\gls{OD}}}_{it}$, to the number of \glspl{CD} currently active on the \gls{LSP}'s platform, to $\zeta^{\alpha}$, and to demand and area specific mobility patterns $I_i^{\mathrm{\gls{GW}}}$ and $I_i^{\mathrm{\gls{OD}}}$ for \glspl{GW} and \glspl{OD} respectively. The constants $\zeta^{\alpha}$ quantify the \glspl{CD}' temporal patterns, i.e., the share of driver arrivals of type $\alpha$ occurring within $\bar{\mathcal{T}}$.  
The arrival rates, therefore, result to 
\begin{equation}
\lambda^{\mathrm{\gls{GW}}}_{it} = n_t^{\mathrm{\gls{GW}}}\,\zeta^{\mathrm{\gls{GW}}}\,I^{\mathrm{\gls{GW}}}_i, \qquad \lambda^{\mathrm{\gls{OD}}}_{it} = n_t^{\mathrm{\gls{OD}}}\,\zeta^{\mathrm{\gls{OD}}}\,I^{\mathrm{\gls{OD}}}_i, \label{eq:lambda_rates}
\end{equation}
where $\zeta^{\alpha}$ and $I_i^{\alpha}$ depend on the problem instances. 

We assume that requests and \glspl{CD} who are not matched to a driver or request within $\bar{t}$ leave the system. This assumption aligns with the on-demand delivery context we study, wherein requests need to be delivered instantaneously, for example, because they consist of perishable goods. The penalty cost $c_{ij}^{\emptyset}$ accounts for both opportunity costs and actual costs of paying an expensive third-party courier to perform the delivery. The assumption regarding \glspl{CD} is sensible as \glspl{CD} find better outside options if not being matched because they often register at different delivery platforms simultaneously \citep{Wired2018}. 
Let $R_{ij}(\bar{t})$ denote the number of \glspl{FD} relocated from $i$ to $j$ in time step $\bar{t}$ and let $A_{ij}^{\beta}(\bar{t})$ be the number of requests matched to delivery option~$\beta\in\left\{\mathrm{\gls{FD}},\mathrm{\gls{GW}},\mathrm{\gls{OD}},\emptyset\right\}$, with $R^{\emptyset}_{ij}(\bar{t})$ being the number of requests not matched to any driver. We consider that a request arriving in \location{} $i$ is directly matched to the cheapest option available. Finally, $c_{ij}^{\beta}$ denotes the costs of serving a request with delivery option $\beta$. 
We assume that $c_{ij}^{\mathrm{\gls{FD}}}<c_{ij}^{\mathrm{\gls{GW}}}$ and $c_{ij}^{\mathrm{\gls{FD}}}<c_{ij}^{\mathrm{\gls{OD}}}$. This is plausible since \gls{FD}'s variable costs only include mileage costs, whereas \gls{CD}'s variable costs need to cover mileage costs and a profit margin which motivates them to serve the request.

The operational costs then read
\begin{equation}
C_t^{\mathrm{ops}}(n_t^{\mathrm{FD}}+a_t,n_t^{\mathrm{GW}},n_t^{\mathrm{OD}},\mathcal{R}^t) = \sum_{k}\left(\underset{R_{ij}^k(\bar{t})}{\mathrm{min}}\sum_{\bar{t}\in\mathcal{\bar{T}}_k}\left[\sum_{ij}\left(\sum_{\beta}\left[c_{ij}^{\beta}\,A_{ij}^{\beta,k}(\bar{t})\right]+c_{ij}^{\mathrm{\gls{FD}}}\,R_{ij}^k(\bar{t})\right)\right]\right). \label{eq:ops_costs}
\end{equation}
In the remainder of this paper, we consider a simplified case, wherein all $\bar{\mathcal{T}}_k$ are equal and the summation over $k$ in Equation \eqref{eq:ops_costs} is, hence, equivalent to multiplication with the number of operational level time horizons $K$
\begin{equation}
C_t^{\mathrm{ops}}(n_t^{\mathrm{FD}}+a_t,n_t^{\mathrm{GW}},n_t^{\mathrm{OD}},\mathcal{R}^t) = K\underset{R_{ij}^k(\bar{t})}{\mathrm{min}}\sum_{\bar{t}\in\mathcal{\bar{T}}_k}\left[\sum_{ij}\left(\sum_{\beta}\left[c_{ij}^{\beta}\,A_{ij}^{\beta,k}(\bar{t})\right]+c_{ij}^{\mathrm{\gls{FD}}}\,R_{ij}^k(\bar{t})\right)\right]. \label{eq:ops_costs_mult}
\end{equation}
Considering all $\mathcal{\bar{T}}_k$ being equal, we can drop index $k$ in the following. We refer the interested reader to Appendix \ref{app:multiple_ops_time_horizons} for details on how to adapt our methodology to unequal $\mathcal{\bar{T}}_k$.
Let us for now assume that we obtain some approximation for $C_t^{\mathrm{ops}}(n_t^{\mathrm{FD}}+a_t,n_t^{\mathrm{GW}},n_t^{\mathrm{OD}},\mathcal{R}^t)$. Then, total costs $C_t^{\mathrm{tot}}$ in time step $t$ then result to
\begin{equation}
C_t^{\mathrm{tot}}(n_t^{\mathrm{FD}}+a_t,n_t^{\mathrm{GW}},n_t^{\mathrm{OD}},\mathcal{R}^t)  =C_t^{\mathrm{ops}}(n_t^{\mathrm{FD}}+a_t,n_t^{\mathrm{GW}},n_t^{\mathrm{OD}},\mathcal{R}^t) + K\,\left(C^{\mathrm{fix}}\cdot (n_t^{\mathrm{FD}}+a_t) + C^{\mathrm{sev}}\,|\mathrm{min}(0,a_t)|\right). \label{eq:total_costs}
\end{equation}
The \gls{LSP} aims at minimizing its expected total costs over the time horizon $\mathcal{T}$. Due to the stochastic nature of the problem, this objective formally results to
\begin{equation}
\mathrm{min}\mathop{\mathbb{E}}\left[\sum_{t=0}^{|\mathcal{T}|}C_t^{\mathrm{tot}}\right]=\mathrm{min}\mathop{\mathbb{E}}\left[\sum_{t=0}^{|\mathcal{T}|} \left(C_{t}^{\mathrm{ops}}(n_t^{\mathrm{\gls{FD}}}+a_t,\cdot) +K\,\left(C^{\mathrm{fix}}\,(n_t^{\mathrm{FD}}+a_t)+C^{\mathrm{sev}}\,|\mathrm{min}(0,a_t)|\right)\right)\right]. \label{eq:lsp_objective}
\end{equation}

Some comments on our modeling choices and assumptions are in order.  
First, we do not account for uncertainty in the parameters governing the evolution of future demand levels, i.e., the set of \textsc{Poisson} rates $\mathcal{R}^t$ in each time step $t$ of the strategic level's problem, as we want to exclusively analyze the effect of uncertain \gls{CD} supply. Therefore, we assume that robust forecasts concerning these parameters can be carried out on the strategic level's planning horizon. This is plausible because \glspl{LSP} indirectly control future demand levels and distributions through the contracts they set up with demand sources, e.g., restaurants or supermarkets, before the strategic level's planning horizon starts.

Second, we assume that there are no \gls{FD} supply shortages, as we restrict our problem to urban areas that typically have an abundant workforce supply, especially in the gig economy sector.

Finally, we consider a finite time horizon on the strategic level since \glspl{LSP}' strategic workforce planning process relies on finite horizons, for which they can leverage a robust forecast. This is in line with works in the strategic workforce planning literature \citep[cf.][]{GansZhou2002}.
\FloatBarrier

%% file: contents/Methodology.tex
\section{Methodology}\label{sec:methodology}
This section formalizes the problem setting presented in Section \ref{sec:problem_setting}. We model the strategic level's problem as an \gls{MDP} (Section \ref{sec:strategic_level}) and the operational level's problem as a closed queueing network (Section~\ref{sec:ops_problem}). Finally, we present an approximate dynamic programming approach to solve large instance sizes in Section~\ref{sec:dynamic_programming}.
\subsection{Strategic Level Workforce Planning}\label{sec:strategic_level}
\FloatBarrier
In this section, we formalize the \gls{LSP}'s workforce planning problem, outlined in Section \ref{sec:problem_setting}, as an \gls{MDP}. In the following, we will successively describe the state, feasible actions, the state transition, the policy, and the objective function. 

\textbf{Pre-decision state:} We denote pre-decision states that represent the fleet composition in time step $t$ by~$s_t =  \left(n^\mathrm{\gls{FD}}_t,n^\mathrm{\gls{GW}}_t,n^\mathrm{\gls{OD}}_t\right) \in \mathbb{N}_0^3$ with $s\in \mathcal{S} = \left\{0,\ldots,N^{\mathrm{\gls{FD}}}\right\}\times\left\{0,\ldots,N^{\mathrm{\gls{GW}}}\right\}\times\left\{0,\ldots,N^{\mathrm{\gls{OD}}}\right\}\times \left\{0,\ldots,T\right\}$, where~ $\mathcal{S}$ denotes the state space. Pre-decision states describe the fleet composition at the beginning of time step $t$ before making any decisions. 
Variables $N^{\mathrm{\gls{FD}}}$, $N^{\mathrm{\gls{GW}}}$, and $N^{\mathrm{\gls{OD}}}$ represent the maximum number of drivers attainable for each driver type, e.g., the maximum number of individuals with the intention to work for the \gls{LSP} within urban area $\mathcal{M}$. Similarly, we denote the state space in time step $t$ by~${\mathcal{S}_t = \left\{0,\ldots,N^{\mathrm{\gls{FD}}}\right\}\times\left\{0,\ldots,N^{\mathrm{\gls{GW}}}\right\}\times\left\{0,\ldots,N^{\mathrm{\gls{OD}}}\right\}}$.

\textbf{Feasible actions and post-decision state:}
The \gls{LSP} decides on the number of \glspl{FD} to hire or fire. The action space $\mathcal{A}_t = \left\{-n_t^{\mathrm{\gls{FD}}},\ldots,N^{\mathrm{\gls{FD}}}\right\} \in \mathbb{Z}$ describes the possible hiring decisions that the \gls{LSP} can make. The maximum number of \glspl{FD} that can be fired depends on the current time step, as the \gls{LSP} cannot fire more \glspl{FD} than currently employed. When the \gls{LSP} decides on an action $a_t\in\mathcal{A}_t$, we reach the post-decision state~${s_t^a = (n_t^{\mathrm{\gls{FD}}}+a_t,n_t^{\mathrm{\gls{GW}}},n_t^{\mathrm{\gls{OD}}})}$ and evaluate the operational problem to approximate the operational costs~$C_t^{\mathrm{ops}}(s_t^a,\mathcal{R}^t)$.

\textbf{State transition:} We transition to the next time step by following a resignation process for all drivers and a joining process for \glspl{CD}, described by Equations \eqref{eq:driver_transitions}. Let $P(s_{t+1}|s_t^a)\in \left[0,1\right]^{|\mathcal{S}|\times|\mathcal{S}|}$ denote the transition probability matrix, describing the probability of transitioning to state $s_{t+1}$ when being in state $s_t^a$. Accordingly, as the distributions $\mathcal{X}^{\alpha}$ and $\mathcal{Y}^{\alpha}$ are assumed to be independent, each entry of $P$ reads
\begin{equation}
P(s_{t+1}|s_t^a) =  \mathbb{P}(\tilde{x}^{\mathrm{\gls{FD}}}_t|s_t^a) \cdot \mathbb{P}(\tilde{x}^{\mathrm{\gls{GW}}}_t|s_t^a) \cdot \mathbb{P}(\tilde{x}^{\mathrm{\gls{OD}}}_t|s_t^a)\cdot \mathbb{P}(\tilde{y}^{\mathrm{\gls{GW}}}_t|s_t^a)\cdot \mathbb{P}(\tilde{y}^{\mathrm{\gls{OD}}}_t|s_t^a). \label{eq:transition_probabilities}
\end{equation}

\textbf{Policy:} We denote a deterministic state-dependent hiring and firing policy by $\pi:\mathcal{S} \to \mathcal{A}$. It assigns an action $a_t\in\mathcal{A}(s_t)$ to each pre-decision state $s_t\in\mathcal{S}_t$. Moreover, we denote the set of all possible policies by~$\Pi$.

\textbf{Objective:} Starting in an initial state $s_0$, the \gls{LSP}'s objective is to minimize expected future total costs over the time horizon $\mathcal{T}$, formally
\begin{equation}
V_0(s_0) = \underset{\pi\in\Pi}{\mathrm{min}}\mathop{\mathbb{E}}\left[\sum_{t=0}^{T} \gamma^{t}\cdot \left( C_{t}^{\mathrm{ops}}(n_t^{\mathrm{\gls{FD}}}+a_t,\cdot) +K\,\left(C^{\mathrm{fix}}\,(n_t^{\mathrm{FD}}+a_t)+C^{\mathrm{sev}}\,|\mathrm{min}(0,a_t)|\right)\right)|s_0\right], \label{eq:objective_strategic_level}
\end{equation}
wherein $\gamma$ denotes the discount factor. The following section introduces the model for determining operational costs $C_t^{\mathrm{ops}}$. 
\subsection{Formalization of the Operational Level's Problem} \label{sec:ops_problem}
To approximate operational costs, we formalize the operational level's problem as a closed queueing network. We base our formulation on the model from \cite{ZhangPavone2014}. Firstly, we consider $|\mathcal{M}|$ single server queues, which we denote by $E_{ii}(\bar{t})$, having service rate $\lambda^{\mathrm{R}}_{it}$ and describing \glspl{FD} idling in \location{}~$i$. Secondly, additional $|\mathcal{M}|^2-|\mathcal{M}|$ infinite server queues, denoted by~$E_{ij}(\bar{t})$, have service rate~$\mu_{ij}$ and describe \glspl{FD} relocating from $i$ to $j$. Finally, further $|\mathcal{M}|^2$ infinite server queues, denoted by $F_{ij}(\bar{t})$, have service rate~$\mu_{ij}$ and describe \glspl{FD} serving a request from $i$ to $j$. Here, $1/\mu_{ij}$ denotes the travel time matrix. 

Finding a policy $Q$ that minimizes the sum of operational costs over time horizon $\bar{\mathcal{T}}$ becomes computationally intractable. Thus, we adapt an approach proposed by \cite{BravermanDaiEtAl2019}, wherein we study a fluid approximation of the closed queueing network. Moreover, we consider steady-state conditions, i.e.,~$\bar{t}\to \infty$. This l.  holds approximately in on-demand delivery services in urban areas if the operational planning horizon is small enough. 

The fluid approximation reformulates the closed-queueing system as a network flow problem, whose counterparts to the closed-queueing system's queue lengths $E_{ij}$ and $F_{ij}$ are network flows from \location{} $i$ to $j$. We denote these network flows by $e_{ij}$ and $f_{ij}$. They correspond to single and infinite server queues in steady state conditions respectively and read
\begin{align}
e_{ij} = \frac{E_{ij}(\bar{t}\to \infty)}{n_t^{\mathrm{\gls{FD}}}}; \qquad f_{ij} = \frac{F_{ij}(\bar{t}\to \infty)}{n_t^{\mathrm{\gls{FD}}}}.
\end{align}
Moreover, we denote by $a_i^{\mathrm{\gls{FD}}}$ the fraction of requests in \location{} $i$ matched to \glspl{FD} in \location{} $i$ in steady state conditions. Analogously, let us denote by $a_{ij}^{\mathrm{\gls{GW}}}$ and $a_{ij}^{\mathrm{\gls{OD}}}$, the corresponding fractions of requests matched to \glspl{GW} and \glspl{OD} respectively on routes $i,j$. Consider the following \gls{LP}, whose full derivation we detail in Appendix \ref{app:fluid_model}. It represents the network flow problem whose objective value is the operational cost over time horizon $\bar{\mathcal{T}}$
\begin{subequations}
	\begin{align}
	\underset{e,f,a^\beta}{\mathrm{min}} \sum_{i}\sum_{j} \left[\lambda_{it}^{\mathrm{R}}\,\left(\sum_{\beta} \left(c_{ij}^{\beta}\cdot a_{ij}^{\beta}\right)+ c_{ij}^{\mathrm{\gls{FD}}}\cdot P_{ij}^{\mathrm{R}} \cdot a_{i}^{\mathrm{\gls{FD}}}\right)+ (1-\delta_{ij})\,c_{ij}^\mathrm{FD}\,n^{\mathrm{FD}}_t\,e_{ij}\right], \text{ }\notag\\ \forall i,j \in \mathcal{M}, \beta \in \left\{\mathrm{\gls{GW}},\mathrm{\gls{OD}},\emptyset\right\}  \label{eq:objective_operational_level_fluid}
	\end{align}
	\begin{align}
	(\lambda_{it}^{\mathrm{R}}/n_t^\mathrm{FD})\cdot a_i^{\mathrm{FD}}\cdot P_{ij}^{\mathrm{R}} & = \mu_{ij}\cdot f_{ij}\text{ } & \forall i,j \in \mathcal{M},  & \text{ } \label{eq:little1}\\
	\lambda_{it}^{\mathrm{R}}\cdot a_{ij}^{\mathrm{GW}} & \le \lambda_{it}^{\mathrm{GW}}\,P_{ij}^{\mathrm{\gls{GW}}}\text{ } & \forall i,j \in \mathcal{M},&\text{ }  \label{eq:little2} \\ %s_{ij}^{\mathrm{GW}}
	\lambda_{it}^{\mathrm{R}}\cdot a_{ij}^{\mathrm{OD}} & \le \lambda_{it}^{\mathrm{OD}}\,P_{ij}^{\mathrm{OD}} \text{ } & \forall i,j \in \mathcal{M},&\text{ }\label{eq:little3}\\ % + s_{ij}^{\mathrm{OD}} 
	\mu_{ij}\,e_{ij} \le \sum_{k}\mu_{ki} f_{ki}, \text{ }& \text{  } i \neq j & \forall i,j \in \mathcal{M},  &\text{ } \label{eq:flow_cons1}\\
	\sum_{k,\,k\neq i} \mu_{ki}\, e_{ki} \le (\lambda_{it}^{\mathrm{R}}/n_t^\mathrm{FD})\,a_i^{\mathrm{FD}} & \le \sum_{k,\,k\neq i} \mu_{ki}\,e_{ki} + \sum_{k} \mu_{ki}\,f_{ki} \label{eq:flow_cons2} \text{ }& \forall i \in \mathcal{M},  & \text{ }\\
	(\lambda_{it}^{\mathrm{R}}/n_t^\mathrm{FD})\, a_i^{\mathrm{FD}} + \sum_{j,\,j\neq i} \mu_{ij}\,e_{ij}& = \sum_{k,\,k\neq i} \mu_{ki}\,e_{ki} + \sum_{k} \mu_{ki}\,f_{ki} \label{eq:flow_cons3}\text{ }&  \forall i \in \mathcal{M}, & \text{ } \\
	a_i^{\mathrm{FD}}+\sum_j \left(a_{ij}^{\mathrm{GW}}+a_{ij}^{\mathrm{OD}}+a^\emptyset_{ij}\right) = 1 & \text{ }& \forall i \in \mathcal{M}, & \text{ }\label{eq:availabilities} \\
	0\le a_i^{\mathrm{\gls{FD}}} \le 1 &  \text{ } & \forall i,j \in\mathcal{M}, & \text{  } \label{eq:availability_fd} \\
	0\le a_{ij}^{\mathrm{\gls{GW}}} \le P_{ij}^{\mathrm{R}}; \quad 0\le a_{ij}^{\mathrm{\gls{OD}}} \le P_{ij}^{\mathrm{R}}  &  \text{ } & \forall i,j \in\mathcal{M}, & \text{  } \label{eq:availability_cd} \\ 
	0\le a_{ij}^{\emptyset} \le 1 &  \text{ } & \forall i,j \in\mathcal{M}, & \text{  } \label{eq:availability_empty}\\
	0\le e_{ij} \le 1,\quad 0 \le f_{ij}\le 1, \quad \sum_i\sum_j e_{ij}+ f_{ij} = 1 & \text{  } & \forall i,j \in \mathcal{M}. & \text{ } \label{eq:fluid_queue_lengths}
	\end{align}
	\label{eq:model_operational_level_fluid}
\end{subequations} 
The Objective \eqref{eq:objective_operational_level_fluid} describes the minimization of operational costs based on costs from serving requests and the empty routing costs. The term $\delta_{ij}$ denotes the Kronecker delta. Constraints \eqref{eq:little1} to \eqref{eq:little3} ensure flow conservation. Constraints \eqref{eq:flow_cons1} to \eqref{eq:flow_cons3} result from a linear relaxation. Constraints \eqref{eq:flow_cons1} reflect the relaxed Little's law, which states that the outgoing flow of relocating \glspl{FD} in one direction $j$ at one \location{}~$i$ cannot be higher than the incoming flow of \glspl{FD} that serve requests. Constraints \eqref{eq:flow_cons2} and \eqref{eq:flow_cons3} ensure that the total \gls{FD} flow leaving \location{} $i$ is equal to the total \gls{FD} flow entering \location{}~$i$. Constraints \eqref{eq:availabilities} ensure that a request is either matched or not matched in every \location{} $i$. Constraints \eqref{eq:availability_fd} to \eqref{eq:availability_empty} ensure that no more requests are matched to the corresponding delivery option than possible. Finally, Constraints~\eqref{eq:fluid_queue_lengths} ensure that the sum of \gls{FD} flows sums up to 1. We can readily compute the optimal solution to this LP with commercially available solvers.
\begin{proposition}\label{prop:fluid_model}
	The optimal objective of the \gls{LP} described by Equations \eqref{eq:objective_operational_level_fluid} to \eqref{eq:fluid_queue_lengths} is a lower bound on the operational costs, as $n_t^{\mathrm{\gls{FD}}}\to\infty$ and $\lambda_{it}^R(t) \to \infty$.
\end{proposition}
\textbf{Proof:} See Appendix \ref{app:fluid_model}.

We use the solution of the \gls{LP} \ref{eq:model_operational_level_fluid} to approximate the operational costs obtained in $\mathcal{\bar{T}}$.
\FloatBarrier
\subsection{Dynamic Programming on the Strategic Level} \label{sec:dynamic_programming}
In this section, we present a stochastic dynamic programming approach to solve the workforce planning problem on the strategic level. Section \ref{sec:exact_approach} describes a standard \gls{bdp} procedure to find the optimal policy $\pi$. As this approach becomes intractable for large fleet sizes, we present a \gls{PL-VFA} in Section \ref{sec:adp}, to compute a near-optimal $\pi$.
\subsubsection{Exact Approach: Backward Dynamic Programming} \label{sec:exact_approach}
\label{sec:bdp}
We start by describing a \gls{bdp} approach, which allows us to determine the optimal workforce planning policy~$\pi$.
Herein, we consider a deterministic policy $\pi$ and recall that the value function $V_t$ of being in pre-decision state $s_t$ reads
\begin{align}
V_t(s_t) = \underset{\pi\in\Pi}{\mathrm{min}}\mathop{\mathbb{E}}\left[\sum_{t'=t}^{T} \gamma^{t'-t}\cdot C_{t'}^{\mathrm{tot}}|s_t\right] = \underset{a_t}{\mathrm{min}}\left(C_{t}^{\mathrm{tot}}(s_{t}^a,\mathcal{R}^t)+\gamma\,\mathbb{E}\left[V_{t+1}(s_{t+1})|s_t^a\right]\right),\label{eq:bellman_strategic_lvl0}
\end{align}
where the value of being in state $s_{t+1}$ is defined as
\begin{equation}
V_{t+1}(s_{t+1}) = \underset{\pi\in\Pi}{\mathrm{min}}\mathop{\mathbb{E}}\left[\sum_{t'=t+1}^{T} \gamma^{t'-t-1}\cdot C_{t'}^{\mathrm{tot}}|s_{t+1}\right].
\end{equation}
Using the transition probabilities as defined in Equation \eqref{eq:transition_probabilities} we can rewrite Equation \eqref{eq:bellman_strategic_lvl0} as
\begin{equation}
V_t(s_t) = \underset{a_t}{\mathrm{min}}\left(C_{t}^{\mathrm{tot}}(s_{t}^a,\mathcal{R}^t)+\gamma\,\sum_{s_{t+1}}P(s_{t+1}|s_t^a)\,V_{t+1}(s_{t+1}|s_t^a)\right). \label{eq:bellman_strategic_lvl}
\end{equation}
Since the strategic level's time horizon $\mathcal{T}$ is finite, we can apply \gls{bdp}, as described in Algorithm~\ref{alg:BDP_pseudocode}, to solve~\eqref{eq:bellman_strategic_lvl}. First, for each fleet state $s_T$ in $t=T$, we set the value at the end of the time horizon to the minimum cost (l. 2), as the value of a state $t>T$ is zero in the final time step. We store the corresponding action in the policy (l. 3). Then, for each time step $t<T$, we obtain the value of state $s_t$ by solving Equation~\eqref{eq:bellman_strategic_lvl}~(l. 6) and store the corresponding action in $\pi$ (l. 7). Algorithm \ref{alg:BDP_pseudocode} requires the enumeration of all states and, consequently, the calculation of operational costs for every state and time step. We adapt the implementation of Brent search \citep{brent1973} from \cite{Limix2023} to our problem setting to solve Equation~\eqref{eq:bellman_strategic_lvl}. For LSPs operating with fleets of more than $1,000$ drivers per type, \gls{bdp} would require to solve the operational problem more than one billion times per time step and becomes intractable. Accordingly, we develop an approximate algorithm to solve \eqref{eq:bellman_strategic_lvl} in the following section, which allows us to study larger fleet sizes in reasonable computation times and to consider unbounded state spaces.
\begin{algorithm}
	\caption{\gls{bdp} algorithm}\label{alg:BDP_pseudocode}
	\begin{algorithmic}[1]
		\For{$s_T\in\mathcal{S}_T$}
		\State Solve $V_T(s_T) = \underset{a_T}{\mathrm{min}}(C_T(s_T^{a_T},\mathcal{R}^t)),\, \forall s_T\in\mathcal{S}_T$
		\State Set $\pi(s_T) = a_T^* = \underset{a_T}{\mathrm{argmin}}(C_T(s_T^{a_T},\mathcal{R}^t))$
		\EndFor
		\For{$t = (T-1),\ldots,0$}
		\For{$s_t \in \mathcal{S}_t$}
		\State Solve $V_t(s_t) = \underset{a_t}{\mathrm{min}}\left[C_t^{\mathrm{tot}}(s_t^{a_t})+\sum_{s_{t+1}}P(s_{t+1}|s_{t}^{a_t})V_{t+1}(s_{t+1})\right]$
		\State Set $\pi(s_t) = a_t^* =  \underset{a_t}{\mathrm{argmin}}\left[C_t^{\mathrm{tot}}(s_t^{a_t})+\sum_{s_{t+1}}P(s_{t+1}|s_{t}^{a_t})V_{t+1}(s_{t+1})\right]$
		\EndFor
		\EndFor
	\end{algorithmic}
\end{algorithm}
\FloatBarrier
\subsubsection{Piecewise Linear Value Function Approximation} \label{sec:adp}
In this section, we introduce an algorithm that approximates the value of being in post-decision state $s_t^{a_t}$, represented by $V_t^{a}(s_t^{a_t})$. Let us denote the set of all possible \gls{CD} combinations in $t$ by $\mathcal{W}_t$, and one \gls{CD} combination in time step $t$ by $w_t = (n_t^{\mathrm{\gls{GW}}},n_t^{\mathrm{\gls{OD}}})$. For each $w_t\in\mathcal{W}_t$, we seek for a piecewise linear approximation of $V_t^{a}(s_t^{a_t})$ along the \gls{FD} dimension. To efficiently obtain this approximation, we rely on Proposition~\ref{prop:convexity_total_costs}~\&~\ref{prop:convexity_value_function}, which state that $C_t^{\mathrm{tot}}$ and $V_t^{a}(s_t^{a_t})$ are piecewise-linear convex in $a_t$.
\begin{proposition}\label{prop:convexity_total_costs}
	$C_t^{tot}(n_t^{\mathrm{\gls{FD}}}+a_t,w_t)$ is piecewise-linear and convex in $a_t$.
\end{proposition}
\textbf{Proof:} See Appendix \ref{app:fluid_model}.
\begin{proposition}\label{prop:convexity_value_function}
	$V_t^{a}(n_t^{\mathrm{\gls{FD}}}+a_t,w_t)$ is piecewise-linear and convex in $a_t$.
\end{proposition}
\textbf{Proof:} See Appendix \ref{app:proof_convexity_value_function}.

\begin{figure}[htbp]
	\centering
	\includegraphics[scale=0.35]{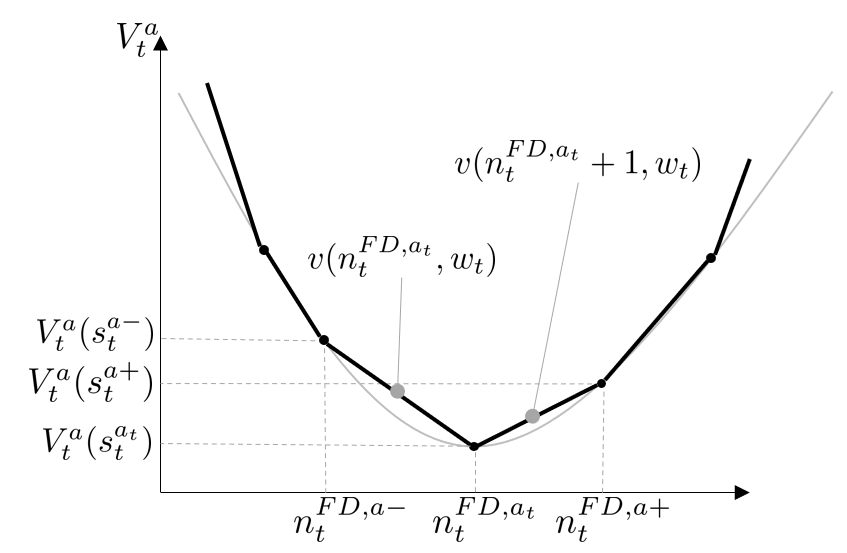}
	\caption{Piecewise linear approximation along the \gls{FD} dimension for a fixed $W_t$.}\label{fig:slope_definition}
\end{figure}

We denote the number of \glspl{FD} in the post-decision state by $n_t^{\mathrm{FD},a_t} = n_t^{\mathrm{FD}}+a_t$. Let $v_t(w_t)$ denote the set of slopes describing $V_t^{a}(s_t^{a_t})$ along the \gls{FD} dimension for fixed $w_t$, and let $v_t(n_t^{\mathrm{FD},a_t},w_t)$ be the slope to the "left" of $n_t^{\mathrm{FD},a_t}$, as illustrated in Figure \ref{fig:slope_definition}. The value of the post-decision state $s_t^{a_t}$, $V_t^{a}(s_t^{a_t})$, then reads
\begin{equation}
V_t^{a}(s_t^{a_t}) = V_t^{a}(0,w_t)+\sum_{k=0}^{n_t^{\mathrm{FD}}+a_t}v_t(k,w_t).
\end{equation}
From here on, we omit $V_t^{a}(0,w_t)$ as shifting the value function by a constant does not impact the optimal decision.
We obtain the value of being in a pre-decision state as
\begin{equation}
V_t^{a}(n_t^{\mathrm{FD},a_t},w_t) = \underset{a_t}{\mathrm{min}}\left(C_t^{\mathrm{tot}}(n_t^{\mathrm{FD},a_t},w_t)+\sum_{k=0}^{n_t^{\mathrm{FD},a_t}}v(k,w_t)\right)
\end{equation}
and the optimal number of \glspl{FD} as
\begin{equation}
a_t^* = \underset{a_t}{\mathrm{argmin}}\left(C_t^{\mathrm{tot}}(n_t^{\mathrm{FD},a_t},w_t)+\sum_{k=0}^{n_t^{\mathrm{FD},a_t}}v(k,w_t)\right). \label{eq:max_action_plvfa}
\end{equation}
We denote the post-decision states to the left and right of $s_t^a$ by $s_t^{a-} = (n_t^{\mathrm{FD}}+a_t-1,w_t)$ and~${s_t^{a+} = (n_t^{\mathrm{FD}}+a_t+1,w_t)}$ respectively, see Figure \ref{fig:slope_definition}.  
Moreover we denote by $s_{t+1}^-$ and $s_{t+1}^+$ the pre-decision states in $t+1$, to which we transition from $s_t^{a-}$ and $s_t^{a+}$ respectively. Accordingly, we denote by~$n_t^{\mathrm{FD},a-}$ and~$n_t^{\mathrm{FD},a+}$ the number of \glspl{FD} in $s_t^{a-}$ and $s_t^{a+}$, and by $n^{\mathrm{FD}-}_{t+1}$ and $n^{\mathrm{FD}+}_{t+1}$ the number of \glspl{FD} in $s_{t+1}^-$ and $s_{t+1}^+$. 
We use the following relations
\begin{align*}
V^{a}_t(s_t^{a_t}) = \mathbb{E}\left[V_{t+1}(s_{t+1})\right], \qquad V^{a}_t(s_t^{a-}) = \mathbb{E}\left[V_{t+1}(s^-_{t+1})\right], \qquad V^{a}_t(s_t^{a+}) = \mathbb{E}\left[V_{t+1}(s_{t+1}^+)\right],
\end{align*}
to obtain an explicit expression of the slopes to the "left" and "right" of the post-decision state $s_t^{a_t}$
\begin{align*}
v_t(s_t^{a-}) = V_t^{a}(s_t^{a_t})-V_t^{a}(s_t^{a-}) =  \mathbb{E}\left[V_{t+1}(s_{t+1})\right] - \mathbb{E}\left[V_{t+1}(s_{t+1}^-)\right], \label{eq:slope_neg} \\
v_t(s_t^{a+}) =V_t^{a}(s_t^{a+})-V_t^{a}(s_t^{a_t}) =  \mathbb{E}\left[V_{t+1}(s_{t+1}^+)\right] - \mathbb{E}\left[V_{t+1}(s_{t+1})\right].
\end{align*}
To approximate the slopes of the optimal value function, we adapt an iterative approach initially proposed by \cite{NascimentoPowell2009}. We denote the approximated slopes by $\bar{v}$. Moreover, we indicate sample information by $(\hat{\cdot})$.
Algorithm~\ref{alg:piecewise_linear_pseudocode} shows the procedure for calculating approximated value function slopes. We initialize the approximated slopes with zeros (l. 1). Then, we sample, for each episode, an initial state~(l.~3). Subsequently, we walk through the episode and first obtain a decision, $a_t$, by solving \eqref{eq:max_action_plvfa} (l. 5) and sample the transition to the next state according to our transition function described by the resignation and joining processes of \glspl{CD} (l. 7). 

Then, we observe samples of $s_{t+1}$ (l. 8), $s_{t+1}^-$ (l. 9), and $s_{t+1}^+$ (l. 10), and use these to calculate $v(s_t^{a-})$ (l. 11) and $v(s_t^{a+})$ (l. 12). We store the current slope approximations in a temporary vector $z$ (l. 13) and obtain a new slope based on a running mean update (l. 14 and 15), where $\alpha$ denotes the learning rate. These updates can lead to temporary convexity violations. We therefore preserve convexity by correcting the slopes to the "left" and "right" of $n_t^{\mathrm{\gls{FD}},a_t}$  as follows
\begin{equation}
\verb|Conv|(z(n,w)) = \begin{cases}
z(n_t^{\mathrm{\gls{FD}},a_t},w_t), \,\,\mathrm{if}\,\, n < n_t^{\mathrm{\gls{FD}},a_t}\,\,, w = w_t\,\,  \mathrm{and}\,\, z(n,w)>z(n_t^{\mathrm{\gls{FD}},a_t},w_t)\\
z(n_t^{\mathrm{\gls{FD}},a_t}+1,w_t), \,\,\mathrm{if}\,\, n > n_t^{\mathrm{\gls{FD}},a_t} + 1\,\,, w = w_t\,\,  \mathrm{and}\,\, z(n,w)<z(n_t^{\mathrm{\gls{FD}},a_t}+1,w_t) \\
z(n,w),\,\,\mathrm{else}
\end{cases}\label{eq:concavity_correction}
\end{equation}

Finally, learning the slopes for all $w_t\in\mathcal{W}_t$ requires many iterations to ensure that each \gls{CD} combination $w_t$ is sufficiently sampled. To reduce the number of samples required, we consider a homogeneous aggregation approach, in which we aggregate $\mathcal{W}_t$ \citep[cf.][p. 144]{Powell2011}. Let us denote the aggregation factors for the \gls{GW} and \gls{OD} dimension by $k^{\mathrm{\gls{GW}}}$ and $k^{\mathrm{\gls{OD}}}$ respectively, and the corresponding aggregated \gls{CD} fleet sizes by $\bar{n}_t^{\mathrm{\gls{GW}}} = \lfloor\frac{n_t^{\mathrm{\gls{GW}}}}{k^{\mathrm{\gls{GW}}}}\rfloor$ and $\bar{n}_t^{\mathrm{\gls{OD}}} = \lfloor\frac{n_t^{\mathrm{\gls{OD}}}}{k^{\mathrm{\gls{OD}}}}\rfloor$ respectively. Moreover, we let $\bar{w}_t = (\bar{n}_t^{\mathrm{\gls{GW}}},\bar{n}_t^{\mathrm{\gls{OD}}})$. Then, when considering homogeneous aggregation, the approximation of a slope for some $n_t^{\mathrm{\gls{FD}},a_t}$ and $w_t$ reads
\begin{equation}
\bar{v}_t(n_t^{\mathrm{\gls{FD}},a_t},w_t) = \bar{v}_t(n_t^{\mathrm{\gls{FD}},a_t},n_t^{\mathrm{\gls{GW}}},n_t^{\mathrm{\gls{OD}}}) \approx \bar{v}_t(n_t^{\mathrm{\gls{FD}},a_t},\bar{n}_t^{\mathrm{\gls{GW}}},\bar{n}_t^{\mathrm{\gls{OD}}}) = \bar{v}_t(n_t^{\mathrm{\gls{FD}},a_t},\bar{w}_t).
\end{equation}
We obtain the \gls{PL-VFA} algorithm with homogeneous aggregation by replacing all $w_t$ by $\bar{w}_t$ in Algorithm \ref{alg:piecewise_linear_pseudocode}.

\begin{algorithm}
	\caption{\gls{PL-VFA} algorithm.}\label{alg:piecewise_linear_pseudocode}
	\begin{algorithmic}[1]
		\State $\bar{v}_t(w_t) = 0, \,\,\forall t\in\mathcal{T},\,\, \forall w_t\in\mathcal{W}$
		\For{each training episode} 
		\State Sample $w_0$, $n_0^{\mathrm{\gls{FD}}}$
		\For{$t=0...T$}
		\State $a_t \leftarrow \underset{a}{\mathrm{argmin}}\left(C_t^{\mathrm{tot}}(n_t^{\mathrm{\gls{FD}}}+a,w_t)+\sum_{k=0}^{n_t^{\mathrm{\gls{FD}}}+a}\bar{v}(k,w_t)\right)$
		\State $n_t^{\mathrm{\gls{FD}},a_t}\leftarrow n_t^{\mathrm{\gls{FD}}} + a_t$
		\State $(n_{t+1}^{\mathrm{\gls{FD}}},w_{t+1}) \leftarrow \verb|transferFunction|(n_t^{\mathrm{\gls{FD}},a_t},w_{t})$
		\State $\hat{V}_{t+1}(n_{t+1}^{\mathrm{\gls{FD}}},w_{t+1}) \leftarrow \underset{a}{\mathrm{min}}\left(C_{t+1}^{\mathrm{tot}}(n_{t+1}^{\mathrm{\gls{FD}}}+ a,w_{t+1})+\sum_{k=0}^{n_{t+1}^{\mathrm{\gls{FD}}}+a}\bar{v}(k,w_{t+1})\right)$
		\State $\hat{V}_{t+1}(n_{t+1}^{\mathrm{\gls{FD}}-},w_{t+1})\leftarrow  \underset{a}{\mathrm{min}}\left(C_{t+1}^{\mathrm{tot}}(n_{t+1}^{\mathrm{\gls{FD}}-}+a,w_{t+1})+\sum_{k=0}^{n_{t+1}^{\mathrm{\gls{FD}}-}+a}\bar{v}(k,w_{t+1})\right)$
		\State $\hat{V}_{t+1}(n_{t+1}^{\mathrm{\gls{FD}}+},w_{t+1}) \leftarrow \underset{a}{\mathrm{min}}\left(C_{t+1}^{\mathrm{tot}}(n_{t+1}^{\mathrm{\gls{FD}}+}+a,w_{t+1})+\sum_{k=0}^{n_{t}^{\mathrm{\gls{FD}}+}+ a}\bar{v}(k,w_{t+1})\right)$
		\State $\hat{v}_t(n_t^{\mathrm{\gls{FD}}}+a_t,w_t) \leftarrow \hat{V}_{t+1}(n_{t+1}^{\mathrm{\gls{FD}}},w_{t+1})-\hat{V}_{t+1}(n_{t+1}^{\mathrm{\gls{FD}}-},w_{t+1})$
		\State $\hat{v}_t(n_t^{\mathrm{\gls{FD}}}+a_t+1,w_t) \leftarrow \hat{V}_{t+1}(n_{t+1}^{\mathrm{\gls{FD}}+},w_{t+1})-\hat{V}_{t+1}(n_{t+1}^{\mathrm{\gls{FD}}},w_{t+1})$ 
		\State $z\leftarrow\bar{v}(w_t)$
		\State $z(n_t^{\mathrm{\gls{FD}},a_t}) \leftarrow z(n_t^{\mathrm{\gls{FD}},a_t})\,(1-\alpha) + \alpha\,\hat{v}(n_t^{\mathrm{\gls{FD}},a_t},w_t)$
		\State $z(n_t^{\mathrm{\gls{FD}},a_t}+1) \leftarrow z(n_t^{\mathrm{\gls{FD}},a_t}+1)\,(1-\alpha) + \alpha\,\hat{v}(n_t^{\mathrm{\gls{FD}},a_t}+1,w_t)$  
		\State $\bar{v}(w_t) \leftarrow\verb|Conv|(z)$
		\EndFor
		\EndFor
	\end{algorithmic}
\end{algorithm} 
\FloatBarrier

%% file: contents/DesignOfExperiments.tex
\section{Design of Experiments} \label{sec:design_of_experiments}
This section describes our experimental setup for a subsequent managerial analysis. In the first part, we present the setup for the operational level's problem, which bases on a real-world data set describing spatial and temporal order patterns for on-demand food deliveries. In the second part, we discuss the parameter settings on the strategic level and sensitivities to be analyzed. 
\subsection{Experimental setup}
\begin{figure}[htbp]
	\centering
	\subfigure[Origin and destination pairs of orders.]{
		\vspace{10cm}\includegraphics[width=0.35\textwidth]{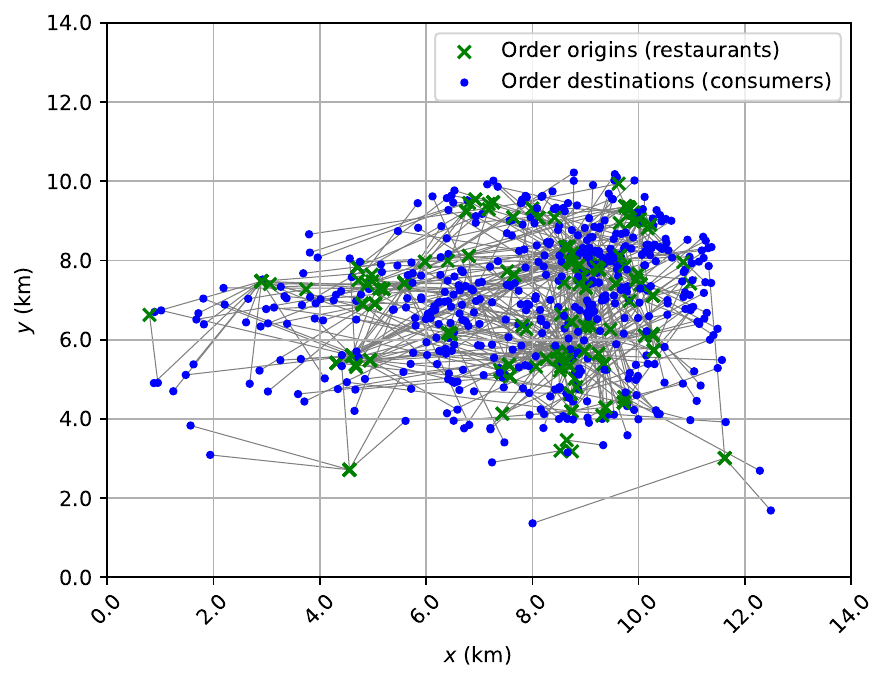}}
	\label{fig:orders_spatial_dist}
	\subfigure[Total number of orders between $t=0$ and $t=850$.]{
		\includegraphics[width=0.34\textwidth]{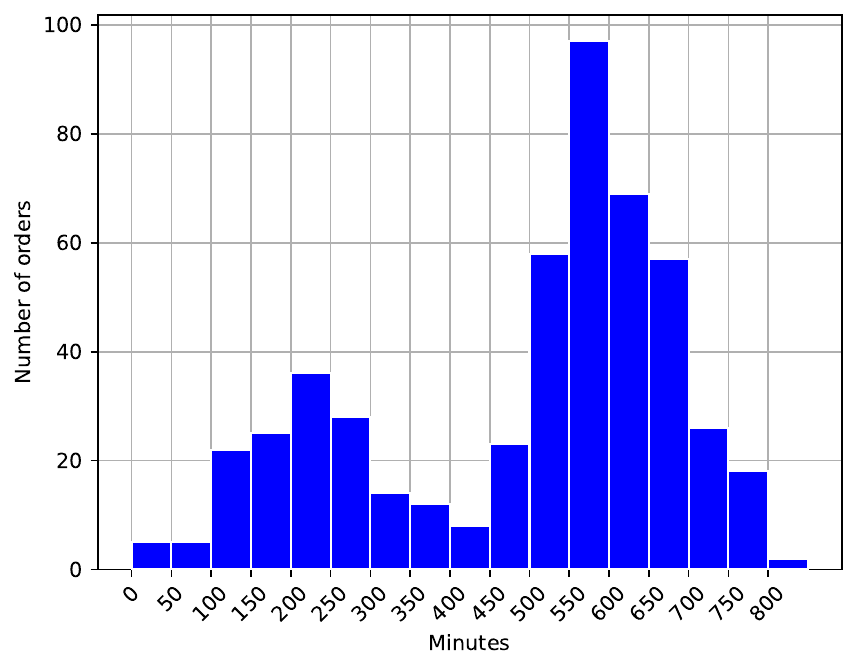}
		\label{fig:orders_temporal_dist}}
	\caption{Spatial and temporal request patterns in instance 0o100t75s1p100.}\label{fig:description_0o100t75s1p100}
\end{figure}

To account for a real-world scenario, we consider a data set provided by \cite{Grubhub2018}, which describes anonymized food delivery orders. The data set consists of ten different instances. Each instance represents one US metropolitan area. Each order is characterized by its origin and destination coordinates, placement, and ready time. The former is when a customer orders through the Grubhub platform, and the latter is when the order is ready to be delivered. In our base case, we use instance 0o100t75s1p100. Figure \ref{fig:description_0o100t75s1p100} highlights its spatial and temporal order distribution. Orders occur from minute $t=0$ to minute $t=850$. 

We use the provided data as base for deriving the origin-destination distance matrix $r_{ij}$, the order arrival rates per \location{} $i\in\mathcal{M}$, and the request pattern matrix $P_{ij}^{\mathrm{R}}$ as follows. First, to account for average zip code surfaces, we discretize the area into squares of $4\,\mathrm{km}^2$ (visualized by the gray grid in Figure \ref{fig:orders_spatial_dist}), each representing a \location{} $i\in\mathcal{M}$. We obtain $r_{ij}$, for $i \neq j$, as the Euclidean distance between centers of \location{}s $i$ and $j$. We assume that $r_{ii}$ corresponds to the half of the square's side length, which is the average distance between any two points within a square, in this case $1\,\mathrm{km}$. This results in 18 \location{}s and consequently in a $18\times18$ origin-destination matrix. Based on the request distribution and the aggregated \location{}s, we obtain the demand pattern matrix $P_{ij}^{\mathrm{R}}$ and demand arrival rates $\lambda_{it}^{\mathrm{R}}$. We briefly discuss the adequateness of the chosen discretization in Appendix \ref{app:adequateness_discretization}.

The data set provides information about courier locations, which are only comparable to \glspl{FD} and don't provide any information about \glspl{CD}' spatial availability. Moreover, since the data set is anonymized, the metropolitan area on which the data bases is unknown. Hence, we sample \gls{GW} and \gls{OD} arrivals. We let \gls{GW} arrivals depend on the arrival patterns of requests, and we set the \gls{GW} arrival intensity function $I_i^{\mathrm{\gls{GW}}}$ to $I^{\gls{GW}}_i = \frac{\lambda_{it}^{\mathrm{R}}}{\sum_j\lambda_{jt}^{\mathrm{R}}}$,
%\begin{equation}
%
%\end{equation}
and obtain $\lambda^{\mathrm{\gls{GW}}}_{it}$ via Equation \eqref{eq:lambda_rates}. This modeling approach reflects the main characteristic of \glspl{GW}, which financially depend on the work for \glspl{LSP} and therefore try to maximize their earnings by frequenting \location{}s with high demand for deliveries. We randomly generate \gls{OD} arrivals $I_i^{\mathrm{OD}}$ and mobility patterns $P_{ij}^\mathrm{\gls{OD}}$. We report \gls{CD} mobility and request patterns in Appendix \ref{app:patterns}. We consider an average travel speed of $v^{\mathrm{avg}}=19\,\mathrm{km/h}$ as reported in the Grubhub data set for all driver types.

We set the strategic time horizon to a year and divide it into $T=26$ two-week segments. We restrict our study on the operational level to the time window between minutes 550 and 600 (cf. Figure \ref{fig:orders_temporal_dist}) of instance~0o100t75s1p100 in the Grubhub data set, wherein we can assume steady-state conditions.

\subsection{Description of base case parameters and variations}
We now present the parameter settings required for the strategic level \gls{MDP} and the fluid approximation on the operational level. We start by describing a base case and then present parameter variations.

We motivate the base case resignation probability and joining rate by a statistical evaluation initially made for Uber drivers between 2012 and 2016 \citep{HallEtKrueger2016}. On the strategic level, we consider a constant resignation probability of $p^{\alpha} = 0.01$ for the base case. Hence, the number $\tilde{x}_t^{\alpha}$ out of $n_t^{\alpha}$ drivers that decide to leave the platform in time step $t$ follows a binomial distribution. Therefore, the probability of~$\tilde{x}_t^{\alpha}$ leaving, reads
\begin{equation}
\mathbb{P}(\tilde{x}^{\alpha}_t) = {n_t^{\alpha}\choose \tilde{x}^{\alpha}_t} \,{(p^{\alpha})}^{\tilde{x}_t^{\alpha}}\,\left(1-p^{\alpha}\right)^{n_t^{\alpha}-\tilde{x}_t^{\alpha}}, \alpha \in \left\{\mathrm{\gls{GW}},\mathrm{\gls{OD}}\right\}. \label{eq:resignation_process}
\end{equation}
In Section \ref{sec:results_driver_perspective}, we consider $p^{\alpha}$ to depend on the \glspl{CD}' matching sensitivity, i.e., their increased likelihood to leave the \gls{LSP}'s platform when they do not receive sufficiently many requests. Let us denote the slack variables from Equations \eqref{eq:little2} and \eqref{eq:little3}, which bound the number of \glspl{CD} being matched to requests on the operational level, by $s_{ij}^{\mathrm{\gls{GW}}}$ and $s_{ij}^{\mathrm{\gls{OD}}}$. Then, we can model $p^{\alpha}$ as
\begin{equation}
p^{\alpha} = p^{\alpha}_{\mathrm{high}}\cdot\frac{\sum_{ij} s_{ij}^{\alpha}}{n_t^{\alpha}} + p^{\alpha}_{\mathrm{low}}\cdot\left(1-\frac{\sum_{ij} s_{ij}^{\alpha}}{n_t^{\alpha}}\right), \alpha \in \left\{\mathrm{\gls{GW}},\mathrm{\gls{OD}}\right\}. \label{eq:matching_sensitive_resig}
\end{equation}
Herein $p^{\alpha}_{\mathrm{high}}$ and $p^{\alpha}_{\mathrm{low}}$ are upper and lower resignation probabilities, which we set to 1 (all \glspl{CD} resign when not being matched at all) and 0.01 (base case resignation probability) respectively. The term $\frac{\sum_{ij} s_{ij}^{\alpha}}{n_t^{\alpha}}$ describes the share of unmatched \glspl{CD}. When $\frac{\sum_{ij} s_{ij}^{\alpha}}{n_t^{\alpha}}$ is high $p^{\alpha}_{\mathrm{high}}$ receives a higher weight and resignations take place at a higher rate. When $\frac{\sum_{ij} s_{ij}^{\alpha}}{n_t^{\alpha}}$ is low, the opposite is the case. Note that if \gls{CD} resignation rates depend on the operational level's cost function and accordingly non-linearly on the number of \glspl{CD}, the convexity of the post-decision state's value function is not guaranteed anymore.

To model the joining process, we assume the number of newly joining \glspl{CD} $\tilde{y}_t^{\alpha}$ to also follow a binomial distribution. Hence, the probability of $\tilde{y}_t^{\alpha}$ newly joining \glspl{CD} reads
\begin{equation*}
\mathbb{P}(\tilde{y}^{\alpha}_t) = {n_t^{\alpha}\choose \tilde{y}^{\alpha}_t} \,{(q^{\alpha})}^{\tilde{y}_t^{\alpha}}\,\left(1-q^{\alpha}\right)^{n_t^{\alpha}-\tilde{y}_t^{\alpha}}, \alpha \in \left\{\mathrm{\gls{GW}},\mathrm{\gls{OD}}\right\}. \label{eq:joining_process}
\end{equation*}
where $n_t^{\alpha}$ is the number of \glspl{CD} currently active on the platform, and $q^{\alpha}$ is the average joining rate of \glspl{CD}. The number of \glspl{CD} in the next time step can be calculated based on Equations \eqref{eq:driver_transitions}. Making the number of newly joining \glspl{CD} dependent on the currently available ones allows us to account for network effects, i.e., platforms with more users/workers attract more users/workers. We set $q^{\alpha} = 0.09$ in the base case. Now, we describe the fraction of \glspl{CD} being active on the operational level and start by \glspl{GW}. As \glspl{GW} align their working times to the demand distribution, and the demand peaks in the $[550,600]$ time window, we assume that all \glspl{GW} are active. Hence, we set $\zeta^{\mathrm{\gls{GW}}} = 1$. \glspl{OD}' main working times are aligned with their primary occupation and lie mainly in the late afternoon/evening times, e.g., when returning home from work \citep{le2019,GalkinEtAl2021}. Accordingly, we assume \gls{OD} arrivals to occur uniformly within the~${\left[400,800\right]}$ time window. This results to a share of \gls{OD} arrivals within $\tilde{\mathcal{T}}$ of $\zeta^{\mathrm{\gls{OD}}} = \frac{50}{400} = 0.125$. 

We consider a homogeneous demand growth rate of roughly $0.6\,\%$ per strategic time step, leading to a compound annual growth rate of $20\,\%$. We consider an overall hourly demand comparable to New York City of approximately $24,000$ requests \citep{Washingtonpost2022}. Since the area we consider has a surface of approximately $100\,\mathrm{km}^2$ (cf. Figure \ref{fig:orders_spatial_dist}), which is smaller than New York City (approximately $800\,\mathrm{km}^2$), we divide the overall demand by a factor of 8 and obtain a total demand of $\sum_i\lambda_{iT}^{\mathrm{R}} = 3,000$ requests per hour in the final time-step. We assume that the spatial distribution of requests remains constant over $\mathcal{T}$. The request arrival rate reads $\lambda_{it}^{\mathrm{R}} = \frac{3,000}{1.006^{T-t}}\frac{\lambda_{it}^{\mathrm{R}}}{\sum_j^{|\mathcal{M}|}\lambda_{jt}^{\mathrm{R}}}$.

In the base case, we set wages for \glspl{FD} to $20\,\$/\mathrm{h}$ \citep{HallEtKrueger2016} and their variable
costs to~$0.34\,\$/\mathrm{km}$ \citep{BoeschBeckerEtAl2018,LanzettiSchifferEtAl2021}. For \glspl{GW}, we use route-based compensation schemes (see, e.g., Grubhub or Postmates), and for \glspl{OD}, we assume a constant compensation per request as \glspl{OD} are only paid for the detour from their private route. Moreover, we assume that \glspl{CD}' value of time corresponds to $0.225\,\$/\mathrm{min}$ in the base case \citep{Wadud2017}. Hence, if we take into account the instance's average velocity of $v=19\,\mathrm{km/h}$, we obtain a payment per km of  $c^{\mathrm{\gls{GW}}} = \frac{0.225\,\frac{\$}{\mathrm{min}}\cdot 60\,\frac{\mathrm{min}}{\mathrm{h}}}{19\,\frac{\mathrm{km}}{\mathrm{h}}}\approx 0.7\,\$/\mathrm{km}$ for \glspl{GW}. 
%\glspl{OD} perform a detour of, on average, $6\,\mathrm{min}$ (cf. Appendix \ref{app:adequateness_discretization}) from their private route when accepting a request. Hence, 
We set the \gls{OD} compensation to $c^{\mathrm{\gls{OD}}}_{ij} =c^{\mathrm{\gls{OD}}} = 5\,\text{\$/request}$ in the base case, as this corresponds to the minimum compensation expectation for \glspl{OD} when delivering a request, according to a representative study \citep[cf.][]{le2019}. In the base case, we set $C^{\mathrm{sev}} = \infty$ to account for a context wherein firing is impossible. Moreover, we consider a penalty of 10\$ per undelivered request. This ensures that it is always cheaper to outsource requests to \glspl{CD} or deliver them with \glspl{FD} than not delivering them in the base case. We set the initially available \glspl{CD} to 500 for both \glspl{GW} and \glspl{OD} in the base case. We perform sensitivity analyses for all parameters according to Table \ref{tab:parameter_settings}.

%and it ensures that $c_{ij}^{\mathrm{\gls{FD}}}<c_{ij}^{\mathrm{\gls{OD}}} = c^{\mathrm{\gls{OD}}},\, \forall i,j \in \mathcal{M}$
\begin{table}[htbp]
	\centering{\def\arraystretch{1.1}
		\caption{Base case parameters and their variations.}
		\begin{tabular}{llll}
			\toprule
			Quantity & Parameter & Base case & Variation range \\
			\midrule
			\gls{CD} joining rates (cf. Equation \eqref{eq:joining_process}) &  $q^{\alpha}$ & 0.09 & $\left[0.01,0.17\right]$\\
			Severance payment (cf. Equation \eqref{eq:total_costs}) & $C^{\mathrm{sev}}$& $\infty$ & $\left[0,60\right]$ \\
			\gls{FD} fix costs per hour (cf. Equation \eqref{eq:total_costs}) & $C^{\mathrm{fix}}$ & $20~\$/\mathrm{h}$ & $\left[4,34\right]$ \\
			\gls{GW} per km costs (cf. Equation \eqref{eq:objective_operational_level_fluid}) & $c^{\mathrm{\gls{GW}}}$ & $0.7~\$/\mathrm{km}$ & $\left[0.5,7.5\right]$  \\
			\gls{OD} costs per request (cf. Equation \eqref{eq:objective_operational_level_fluid}) & $c^{\mathrm{\gls{OD}}}$ & $5.5~\$/\mathrm{request}$ & $\left[1,8\right]$ \\
			\bottomrule
		\end{tabular}
		\label{tab:parameter_settings}}
\end{table}

To assess the results, we evaluate the quotient $h$(\%) between total cumulated costs in the final time step of a mixed fleet (i.e., consisting of \glspl{FD} and \glspl{CD}) and an \gls{FD}-only fleet, and the cost saving $\bar{h}(\%)$
\begin{equation}
h(\%)=100\,\frac{\sum_{t=0}^{T}C_{\mathrm{t}}^{\mathrm{tot}}(\text{mixed fleet of \glspl{FD} and \glspl{CD}})}{\sum_{t=0}^{T}C_{\mathrm{t}}^{\mathrm{tot}}(\text{\gls{FD}-only fleet})}; \qquad \bar{h}(\%) = 100(\%)-h(\%).
\end{equation}
\FloatBarrier

%% file: contents/Results.tex
\section{Results}\label{sec:results}
In the first part of this section, we validate our \glspl{PL-VFA} (Section \ref{sec:validation_plvfa}) before analyzing the structural properties of a policy derived by \gls{PL-VFA} in the base case (Section \ref{sec:strategic_level_impact}). In Section \ref{sec:lsp_perspective}, we study the policies' and parameter variations' impact on total costs from an \gls{LSP} perspective. Finally, we take the \glspl{CD}' perspective and compare different behavioral assumptions. We implemented the strategic level's \gls{MDP} in Python and used Gurobi 9.1.2 to solve the operational problem. We performed all experiments on a workstation with a GHz i9-9900 CPU at 16x3.10 GHz and 16GB RAM. If not mentioned otherwise, reported results are average values based on executing the respective policy 50 consecutive times. %We report the technical parameters we chose for the \gls{PL-VFA} in Appendix \ref{app:technical_parameters}.
\subsection{Validation of \gls{PL-VFA}}\label{sec:validation_plvfa}
To validate the \gls{PL-VFA} approach, we evaluate \gls{PL-VFA} on smaller instances and consider only \glspl{GW}. In these instances, we can compute a solution with \gls{bdp}. We study three demand scenarios: constant demand, growing demand, and peak demand. Moreover, we vary the initially available numbers of \glspl{GW}. We compare the results obtained by \gls{PL-VFA} and a \gls{MY}, which always hires enough \glspl{FD} to serve the demand in the current time step $t$, to our \gls{bdp}, which yields the optimal solution. The total cumulated costs in the final time step $T$ read
%We report the difference (in \%) of total cumulated costs in the final time step of \gls{PL-VFA} and \gls{bdp}, and \gls{MY} and \gls{bdp} respectively. The total cumulated costs in the final time step result to
\begin{equation*}
\bar{C}_{T} = \sum_{t=0}^{T} C_t^{\mathrm{tot}}. \label{eq:total_cumulated_costs}
\end{equation*}
Moreover, we define the gap to the optimal solution as 
\begin{equation*}
\delta(\%) = 100\cdot \frac{\bar{C}_{T}-\bar{C}_{T}^{\mathrm{\gls{bdp}}}}{\bar{C}_{T}^{\mathrm{\gls{bdp}}}}
\end{equation*}
We summarize the average gaps $\delta$ in Table \ref{tab:summary_convergence}. The \gls{PL-VFA} algorithm converged after $10\,\mathrm{k}$ iterations for the constant and growth demand case and after $5\,\mathrm{k}$ iterations for the peak demand case. We refer the interested reader to Appendix \ref{app:boxplots_validation} for a more detailed comparison of this section's results and the instance's characteristics. \gls{PL-VFA} (almost) matches \gls{bdp} results in all cases. In contrast, results based on a myopic hiring policy show deviations of up to $5.79\,\%$.
\begin{table}[h!]
	\centering{\def\arraystretch{1.1}
		\caption{Deviation from optimal solution, $\delta(\%)$, of policies obtained with \gls{PL-VFA} and \gls{MY} for different initial \gls{GW} fleet sizes and demand scenarios.}
		\begin{tabular}{ccccccc} 
			\toprule
			& \multicolumn{2}{c}{Constant} & \multicolumn{2}{c}{Growth} & \multicolumn{2}{c}{Peak}\\
			\cmidrule(r){2-3} 	\cmidrule(r){4-5} \cmidrule(r){6-7}
			$(n_0^\mathrm{\gls{FD}},n_0^\mathrm{\gls{GW}},n_0^\mathrm{\gls{OD}})$ & \gls{PL-VFA} & \gls{MY} & \gls{PL-VFA} & \gls{MY} & \gls{PL-VFA} & \gls{MY} \\
			\midrule
			(0,6,0)  &  0.00 & 4.18 & 0.02 & 1.63 & 0.02 & 4.25 \\
			(0,9,0)  &  0.05 &  5.79 & 0.53 & 2.92 & 0.40 & 4.69 \\
			(0,12,0)  &  0.15 & 3.38 & 0.51 & 3.12 & 0.28  & 4.17 \\
			\bottomrule
		\end{tabular}
		\label{tab:summary_convergence}}
\end{table}
Given that \gls{PL-VFA}'s costs consistently stay within an error margin of a maximum of $0.53\,\%$ compared to the costs derived from \gls{bdp}, it can serve as a reliable and effective tool for obtaining hiring policies. We obtained this section's results using a constant learning rate $\alpha = 0.01$ and an aggregation factor $k^{\mathrm{\gls{GW}}} = 10$. In the following we use a learning rate of~${\alpha = 0.001}$ to ensure stable convergence of the \gls{PL-VFA} algorithm and aggregation factors~${k^{\mathrm{\gls{GW}}} = k^{\mathrm{\gls{OD}}} = 100}$ as they yield the best solution after $100\,\mathrm{k}$ iterations of Algorithm \ref{alg:piecewise_linear_pseudocode}. In Appendix, \ref{app:parameter_tuning} we detail the hyperparameter tuning process.
\FloatBarrier
\subsection{Hiring policy comparison in the base case} \label{sec:strategic_level_impact}
\begin{figure}[htbp]
	\centering
	\subfigure[Number of \glspl{FD} hired with \gls{PL-VFA}.]{
		\includegraphics[width=0.4\textwidth]{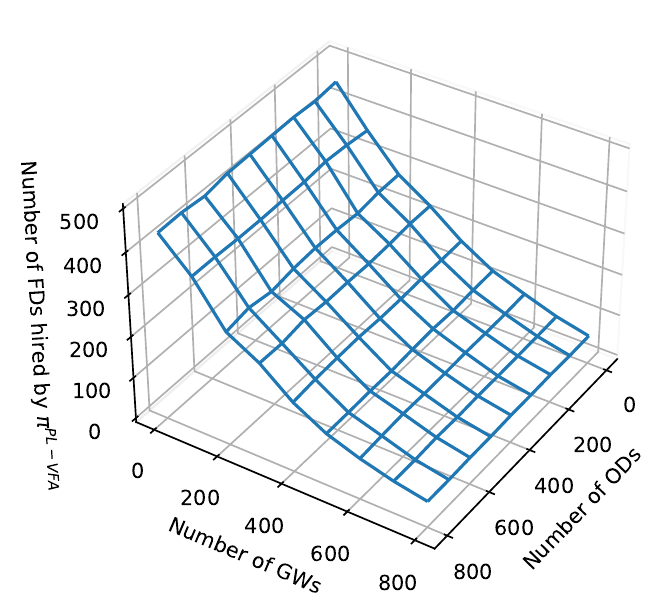}
		\label{fig:policy_comparison_PLVFA}}
	\subfigure[Number of \glspl{FD} hired with \gls{MY}.]{
		\includegraphics[width=0.4 \textwidth]{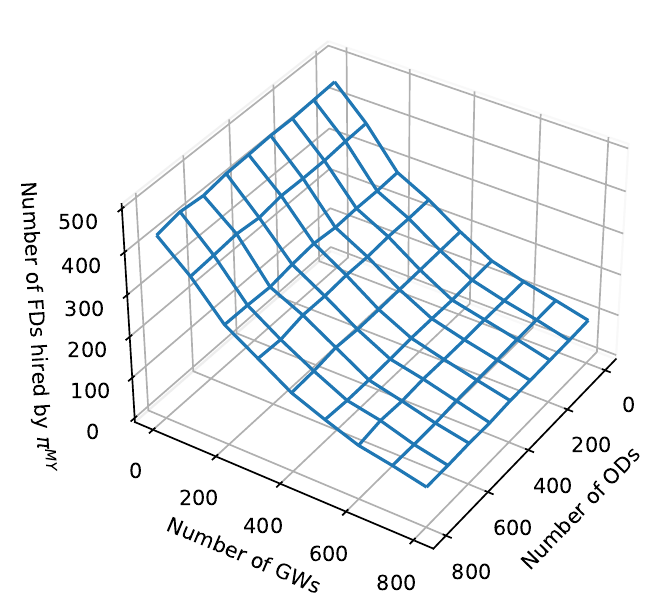}
		\label{fig:policy_comparison_myopic}}
	\caption{Number of \glspl{FD} hired in $t=0$ if $n_0^{\gls{FD}} = 0$.}\label{fig:policy_comparison_base_case}
\end{figure}

\begin{figure}[htbp]
	\centering
	\subfigure[$\pi^{\mathrm{\gls{MY}}}-\pi^{\mathrm{\gls{PL-VFA}}}$ in $t=0$.]{
		\includegraphics[width=0.4\textwidth]{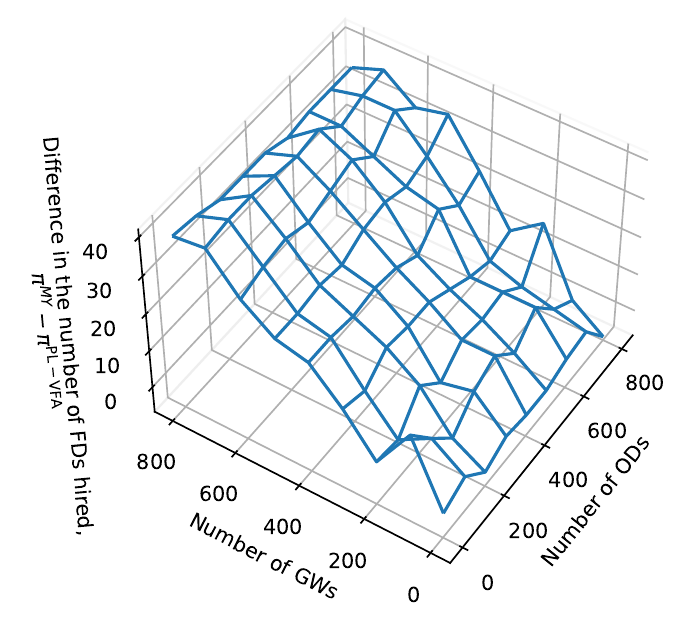}
		\label{fig:diff_t0}}
	\subfigure[Over- vs. underhiring over time.]{	\includegraphics[width=0.4\textwidth]{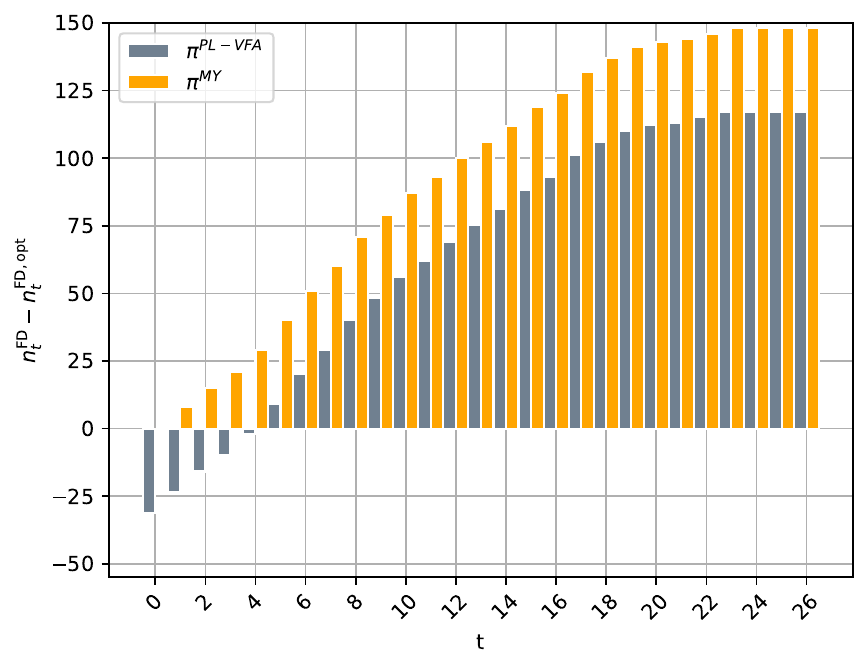}
		\label{fig:base_case_underoverhiring}} 	
	\caption{Difference between $\pi^{\mathrm{\gls{PL-VFA}}}$ and $\pi^{\mathrm{\gls{MY}}}$ and over- vs. underhiring over time.}\label{fig:policy_comparison_diff}
\end{figure}

We begin this section by studying the difference in the number of \glspl{FD} hired by a policy obtained from \gls{PL-VFA} and \gls{MY}, which we denote by $\pi^{\mathrm{\gls{PL-VFA}}}$ and $\pi^{\mathrm{\gls{MY}}}$ respectively. To this end, Figure \ref{fig:policy_comparison_base_case} shows the number \glspl{FD} hired by  $\pi^{\mathrm{\gls{PL-VFA}}}$ (cf. Figure \ref{fig:policy_comparison_PLVFA}) and  $\pi^{\mathrm{\gls{MY}}}$ (cf. Figure \ref{fig:policy_comparison_myopic}) in $t=0$ as a function of the number of \glspl{CD} with initially zero \glspl{FD}, i.e., $n_0^{\mathrm{\gls{FD}}} = 0$. Both policies hire more \glspl{FD} when fewer \glspl{CD} are available. The number of \glspl{FD} hired shows a low sensitivity concerning the number of \glspl{OD} compared to the number of \glspl{GW}. 
In Figure \ref{fig:diff_t0}, we show the difference in the number of \glspl{FD} hired between $\pi^{\mathrm{\gls{MY}}}$ and $\pi^{\mathrm{\gls{PL-VFA}}}$ in $t=0$. Firstly, we observe that the difference is always positive, hence $\pi^{\mathrm{\gls{MY}}}$ always hires more \glspl{FD} than $\pi^{\mathrm{\gls{PL-VFA}}}$. The difference increases with the number of \glspl{GW}. This is plausible as if \gls{CD} supply is initially not high,~$\pi^{\mathrm{\gls{MY}}}$ hires enough \glspl{FD} to minimize total costs in $t$, whereas $\pi^{\mathrm{\gls{PL-VFA}}}$ hires less \glspl{FD} than required to minimize total costs in $t$ to prevent an oversupply of \glspl{FD} in later time steps, where it cannot fire \glspl{FD} anymore. When \gls{CD} supply is initially very low,~$\pi^{\mathrm{\gls{PL-VFA}}}$ hires similarly many \glspl{FD} as $\pi^{\mathrm{\gls{MY}}}$, because the \gls{CD} fleet will not become large enough over time to contribute to request deliveries. 
Next we study the difference between~$\pi^{\mathrm{\gls{PL-VFA}}}$ and $\pi^{\mathrm{\gls{MY}}}$ from a temporal point of view. Let $n_t^{\mathrm{\gls{FD},opt}}$ denote the number of \glspl{FD} required to cover the entire demand in a time step $t$, given the current \gls{CD} fleet composition and demand. Figure \ref{fig:base_case_underoverhiring} shows the difference between~$n_t^{\mathrm{\gls{FD},opt}}$ and $n_t^{\mathrm{\gls{FD}}}$ as a function of $t$ for both $\pi^{\mathrm{\gls{PL-VFA}}}$ and $\pi^{\mathrm{\gls{MY}}}$. When using $\pi^{\mathrm{\gls{PL-VFA}}}$ the \gls{LSP} underhires, i.e., does not have sufficient drivers to serve all requests, in the first time steps of the time horizon and overhires, i.e., has more drivers than they require, in the second half. When using $\pi^{\mathrm{\gls{MY}}}$ the \gls{LSP} overhires from the first time steps on. The number of overhired \glspl{FD} is approximately $20\,\%$ lower in the final time step $T$ when using $\pi^{\mathrm{\gls{PL-VFA}}}$ than when using $\pi^{\mathrm{\gls{MY}}}$.
The results are plausible, as $\pi^{\mathrm{\gls{PL-VFA}}}$ anticipates future \gls{CD} supply, it refrains from hiring many \glspl{FD} which might become obsolete as the number of \glspl{CD} grows. The myopic policy $\pi^{\mathrm{\gls{MY}}}$ does not take into account future \gls{CD} supply and therefore always tries to fulfill the demand in the current time step, which causes increased overhiring over time.
\begin{result}
	\textit{The number of overhired \glspl{FD} in $T$ is approximately $20\,\%$ lower when using $\pi^{\mathrm{\gls{PL-VFA}}}$ than when using $\pi^{\mathrm{\gls{MY}}}$. The myopic nature of $\pi^{\mathrm{\gls{MY}}}$ causes it to overhire \glspl{FD} from early time steps on.}
\end{result}
\FloatBarrier
\subsection{Sensitivity analysis} \label{sec:lsp_perspective}
\begin{figure}[htbp]
	\centering
	\subfigure[Variation of the \gls{GW} joining rate $q^{\mathrm{\gls{GW}}}$.]{
		\includegraphics[width=0.4\textwidth]{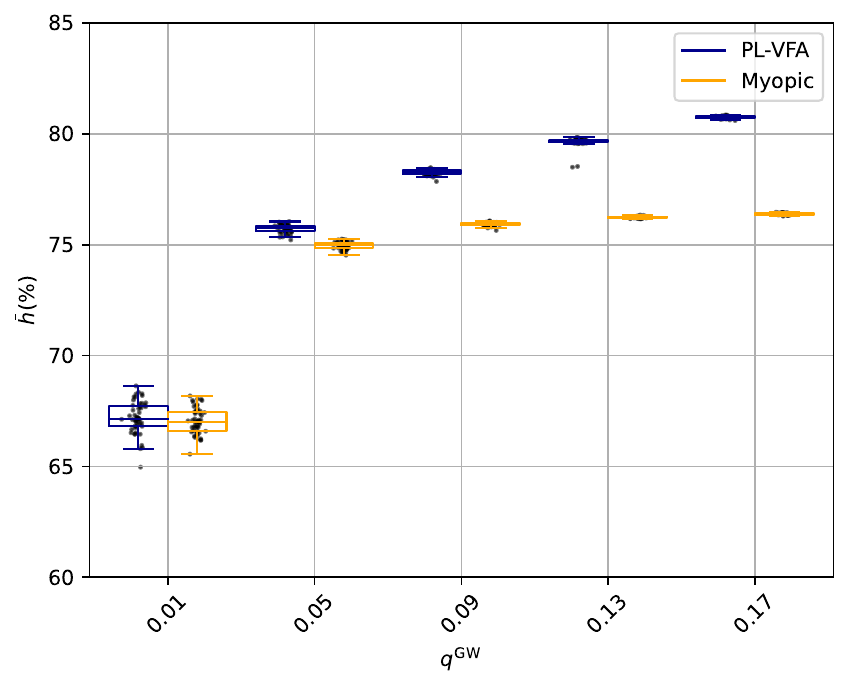}
		\label{fig:add_gw}}
	\subfigure[Variation of \gls{OD} joining rate $q^{\mathrm{\gls{OD}}}$.]{
		\includegraphics[width=0.4 \textwidth]{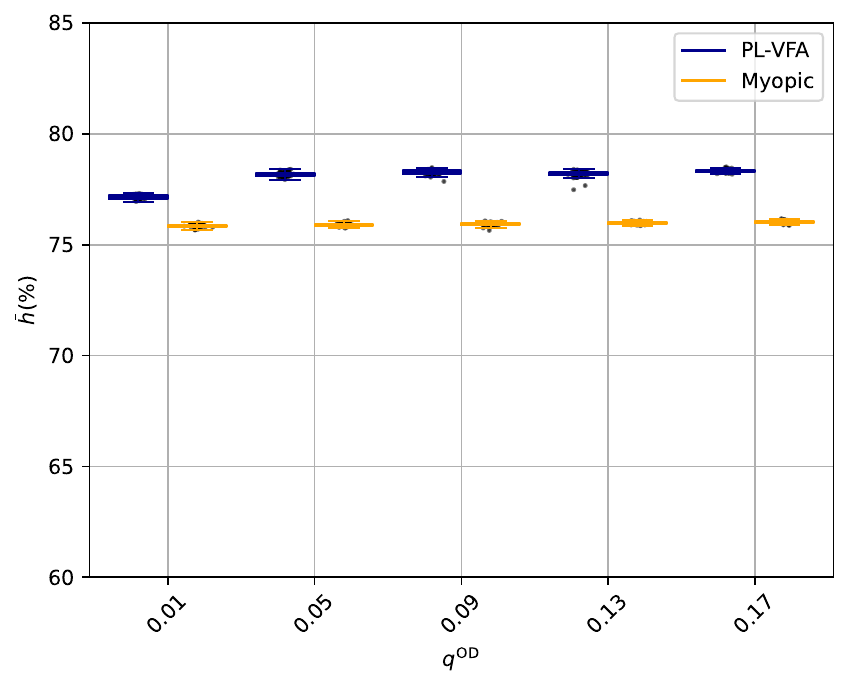}
		\label{fig:add_od}}
	\caption{Variation of \gls{CD} joining rate.}\label{fig:add_variation_strategic}
\end{figure}

Figure \ref{fig:add_variation_strategic} shows $\bar{h}$ as a function of different \gls{GW} and \gls{OD} joining rates. With increasing \gls{CD} joining rates, we observe higher cost savings of up to $81\,\%$, when increasing $q^{\mathrm{\gls{GW}}}$, and of up to $78\,\%$, when increasing $q^{\mathrm{\gls{OD}}}$. Moreover, we observe that for increasing $q^{\alpha}$, $\pi^{\mathrm{\gls{PL-VFA}}}$ outperforms $\pi^{\mathrm{\gls{MY}}}$ by up to five percentage points when varying $q^{\mathrm{\gls{GW}}}$ (cf. Figure \ref{fig:add_gw}), and two percentage points when varying $q^{\mathrm{\gls{OD}}}$ (cf. Figure \ref{fig:add_od}). A five percentage points higher $\bar{h}$ corresponds to $19\,\%$ lower total costs when using $\pi^{\mathrm{\gls{PL-VFA}}}$.

The total cost decrease's sensitivity concerning the variation of $q^{\mathrm{\gls{OD}}}$ is low compared to the sensitivity concerning $q^{\mathrm{\gls{GW}}}$. This is plausible, as the \glspl{OD}' likelihood of being matched to requests is lower than that of \glspl{GW}. Opposed to \glspl{GW}, \glspl{OD} only accept requests whose origin and destination correspond to their origin and destination. Moreover, the share of \glspl{OD} arriving within $\mathcal{\bar{T}}$ is lower, i.e., $\zeta^{\mathrm{\gls{OD}}}\ll\zeta^{\mathrm{\gls{GW}}}$.

\begin{result}
	\textit{The advantage of $\pi^{\mathrm{\gls{PL-VFA}}}$ grows with increasing $q^{\alpha}$ and can lead to up to five percentage points higher cost savings when using $\pi^{\mathrm{\gls{PL-VFA}}}$, which corresponds to $19\,\%$ lower total costs when using $\pi^{\mathrm{\gls{PL-VFA}}}$.} \label{result:cost_savings_plvfa}
\end{result}

\begin{figure}[htbp]
	\centering
	\subfigure[$q^{\mathrm{\gls{GW}}} = 0.05$]{
		\includegraphics[width=0.3\textwidth]{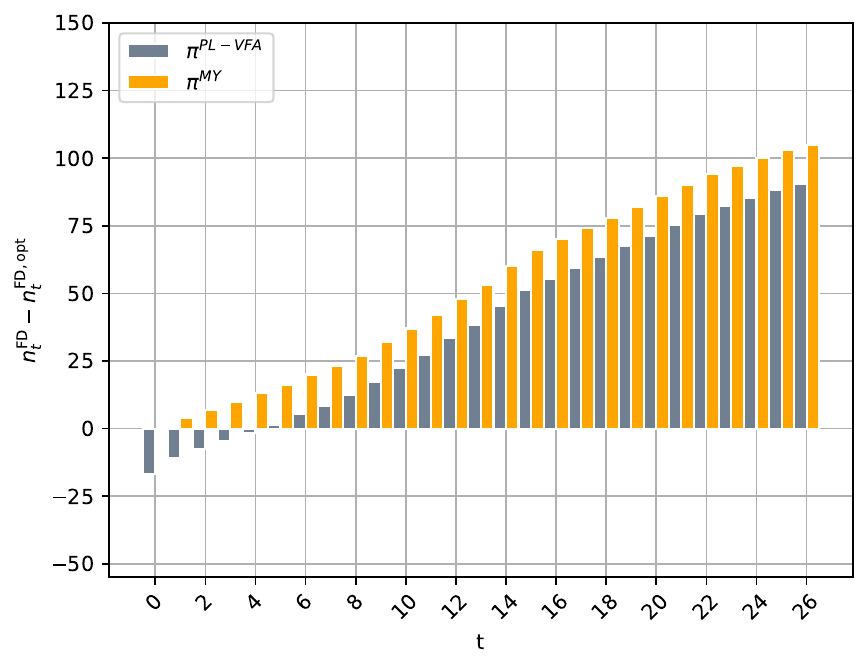}
		\label{fig:under_over_hiring_low_saving_potential}}
	\subfigure[$q^{\mathrm{\gls{GW}}} = 0.09$]{
		\includegraphics[width=0.3\textwidth]{figures/Results/over_under_hiring_009.pdf}
		\label{fig:under_over_hiring_medium_saving_potential}}
	\subfigure[$q^{\mathrm{\gls{GW}}} = 0.17$]{
		\includegraphics[width=0.3\textwidth]{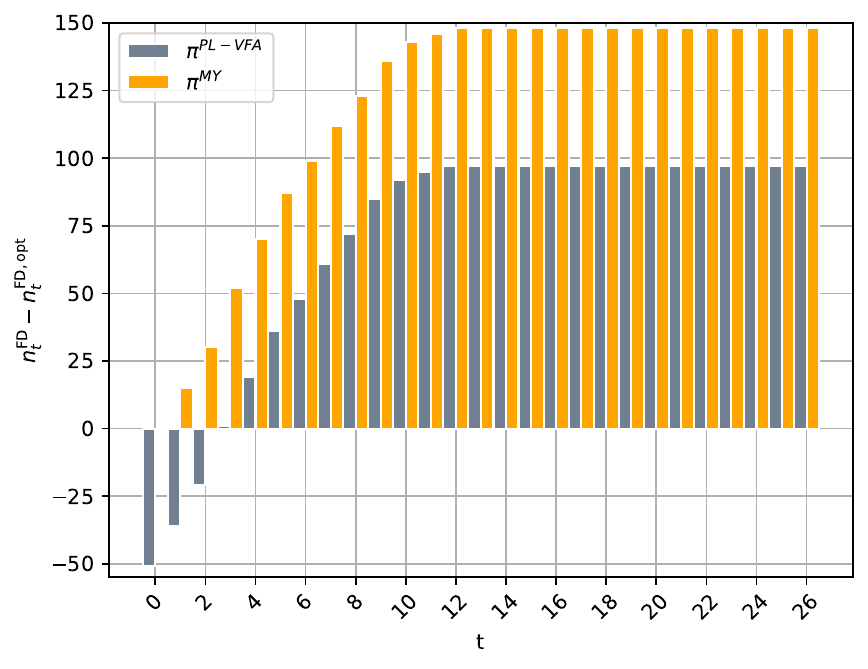}
		\label{fig:under_over_hiring_high_saving_potential}}
	\caption{$n_t^{\mathrm{\gls{FD}}}-n_t^{\mathrm{\gls{FD},opt}}$ for each time step $t$.}\label{fig:variation_joining_rates_underoverhiring} 	
\end{figure}

To provide a better intuition on how $\pi^{\mathrm{\gls{PL-VFA}}}$ achieves smaller total costs than $\pi^{\mathrm{\gls{MY}}}$, we study the temporal development of the number of \glspl{FD}. To this end, Figure \ref{fig:variation_joining_rates_underoverhiring} shows the difference between $n_t^{\mathrm{\gls{FD},opt}}$ and $n_t^{\mathrm{\gls{FD}}}$ as a function of $t$ for~$q^{\mathrm{\gls{GW}}} = 0.05$, $q^{\mathrm{\gls{GW}}} = 0.09$ (base case), and $q^{\mathrm{\gls{GW}}} = 0.17$. For $q^{\mathrm{\gls{GW}}} = 0.05$ (cf. Figure \ref{fig:under_over_hiring_low_saving_potential}), the difference between $n_t^{\mathrm{\gls{FD},opt}}$ and $n_t^{\mathrm{\gls{FD}}}$ grows over the time horizon. While $\pi^{\mathrm{\gls{PL-VFA}}}$ underhires in the first time steps, $\pi^{\mathrm{\gls{MY}}}$ overhires over the entire time horizon. For $q^{\mathrm{\gls{GW}}} = 0.09$ (cf. Figure \ref{fig:under_over_hiring_medium_saving_potential}) the behavior is similar to~$q^{\mathrm{\gls{GW}}} = 0.05$, but the number of under- and overhired \gls{FD} increases. Moreover, the difference between $\pi^{\mathrm{\gls{MY}}}$ and $\pi^{\mathrm{\gls{PL-VFA}}}$ also increases. Figure \ref{fig:under_over_hiring_high_saving_potential} explores under- and overhiring for $q^{\mathrm{\gls{GW}}} = 0.17$. When using $\pi^{\mathrm{\gls{PL-VFA}}}$, we observe the same behavior as for $q^{\mathrm{\gls{GW}}} = 0.05$ and $q^{\mathrm{\gls{GW}}} = 0.09$, however with even stronger under- and overhiring as well as a stronger difference between $\pi^{\mathrm{\gls{MY}}}$ and $\pi^{\mathrm{\gls{PL-VFA}}}$.

\begin{figure}[htbp]
	\centering
	\subfigure[Cost saving $\bar{h}$ in $t=T$ for varying $q^{\mathrm{\gls{GW}}}$ and $q^{\mathrm{\gls{OD}}}$.]{\includegraphics[width=0.4\textwidth]{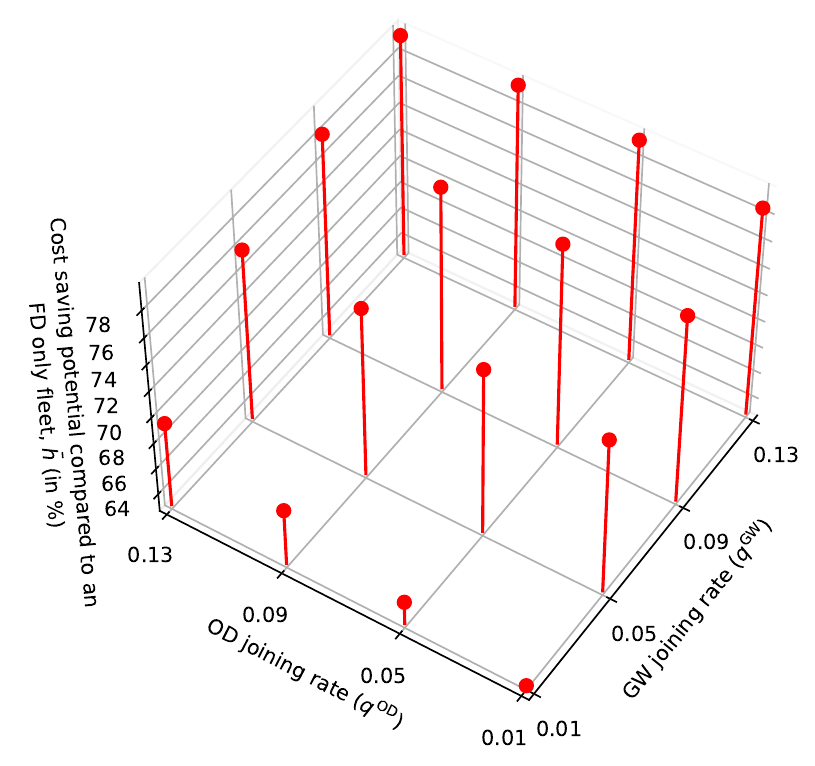}
		\label{fig:total_costs_joining_rates_2d}}
	\subfigure[Average share of driver and penalty costs in total cumulated costs mix in $t=T$.]{		\includegraphics[width=0.4\textwidth]{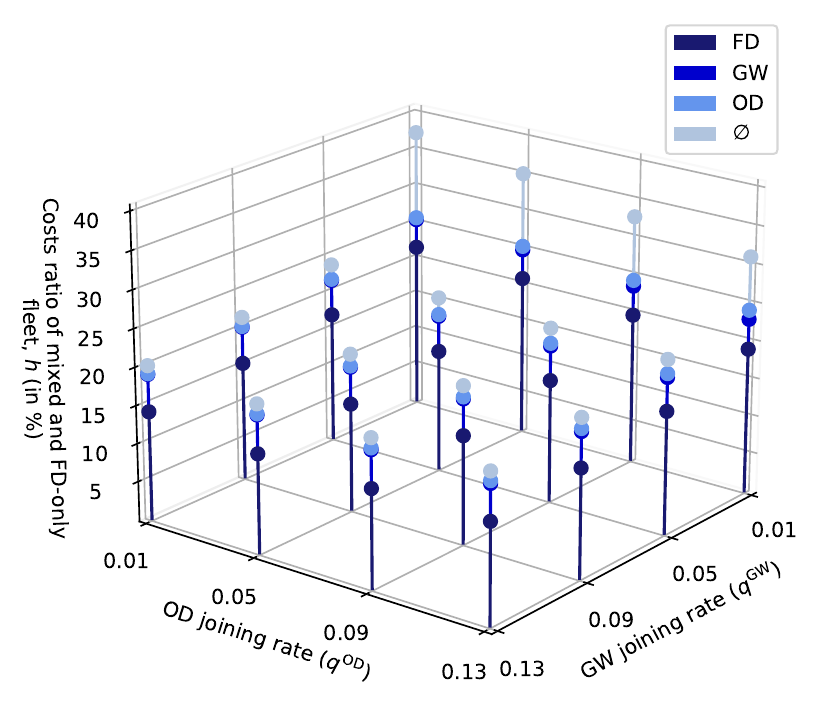}
		\label{fig:cost_splits}}
	\caption{Combined impact of varying joining rates on total costs.}
	
\end{figure}

When \gls{CD} joining rates are low, $\pi^{\mathrm{\gls{PL-VFA}}}$ cannot leverage potential future \gls{CD} supply in its decision-making and the need for \glspl{FD} remains higher over the entire time horizon. Hence, the difference between~$\pi^{\mathrm{\gls{MY}}}$ and~$\pi^{\mathrm{\gls{PL-VFA}}}$ is lower. When \gls{CD} joining rates are higher, \glspl{FD} risk becoming obsolete. Hence, $\pi^{\mathrm{\gls{PL-VFA}}}$ underhires in early time steps to prevent an obsolete \gls{FD} pool in later time steps. As total cumulated costs are smaller for $q^{\mathrm{\gls{GW}}} = 0.17$ than for $q^{\mathrm{\gls{GW}}} = 0.09$, we interpret the previous result as follows: By underhiring in early time steps $\pi^{\mathrm{\gls{PL-VFA}}}$ hedges against overhiring in later time steps, i.e., against having to remunerate a large amount of \glspl{FD} that are not required anymore as the number of \glspl{CD} increased over time. We provide the impact of underhiring on the \gls{LSP}'s service level in Appendix \ref{app:sevice_level_analysis}.
\begin{result}
	\textit{$\pi^{\mathrm{\gls{PL-VFA}}}$ hedges against overhiring in later time steps by hiring up to 50 \glspl{FD} less than required in early time steps and therefore investing in penalties for requests not delivered.} \label{result:service_levels}
\end{result}
We now study the effect of $q^{\mathrm{\gls{GW}}}$ and $q^{\mathrm{\gls{OD}}}$ jointly. To this end, Figure \ref{fig:total_costs_joining_rates_2d} shows the cost saving compared to an \gls{FD}-only fleet, $\bar{h}$. We observe that increasing joining rates lead to higher cost savings. Moreover, we can see that cost savings are more sensitive to $q^{\mathrm{\gls{GW}}}$ than to $q^{\mathrm{\gls{OD}}}$. For example, when $q^{\mathrm{\gls{GW}}} = 0.13$, varying $q^{\mathrm{\gls{OD}}}$ has no impact on cost savings. The \gls{LSP} achieves the highest cost saving, i.e., $78\,\%$, when $q^{\mathrm{\gls{GW}}}$ is highest.
\begin{result}
	\textit{The increase of $q^{\mathrm{\gls{GW}}}$ reduces costs by up to $78\,\%$ and therefore more than the increase of $q^{\mathrm{\gls{OD}}}$. Hence, \glspl{GW} are the main driver for cost savings within the \gls{CD} fleet.}
\end{result}

\begin{figure}[htbp]
	\centering
	\subfigure[Variation of \glspl{GW}' costs per km $c^{\mathrm{\gls{GW}}}$.]{\includegraphics[width=0.4\textwidth]{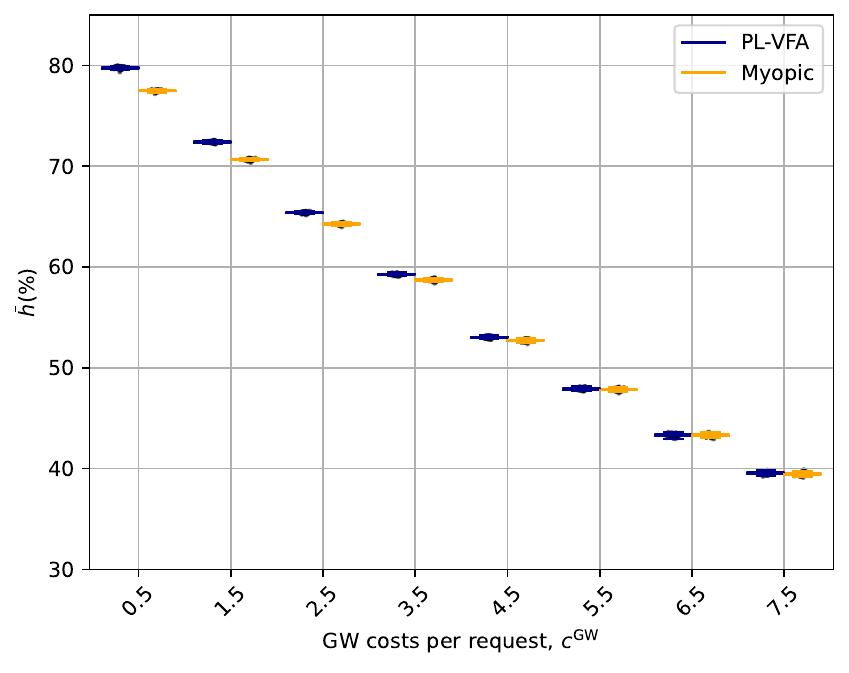}\label{fig:c_g_total_utility}}
	\subfigure[Variation of \glspl{OD}' costs per request $c^{\mathrm{\gls{OD}}}$.]{\includegraphics[width=0.4\textwidth]{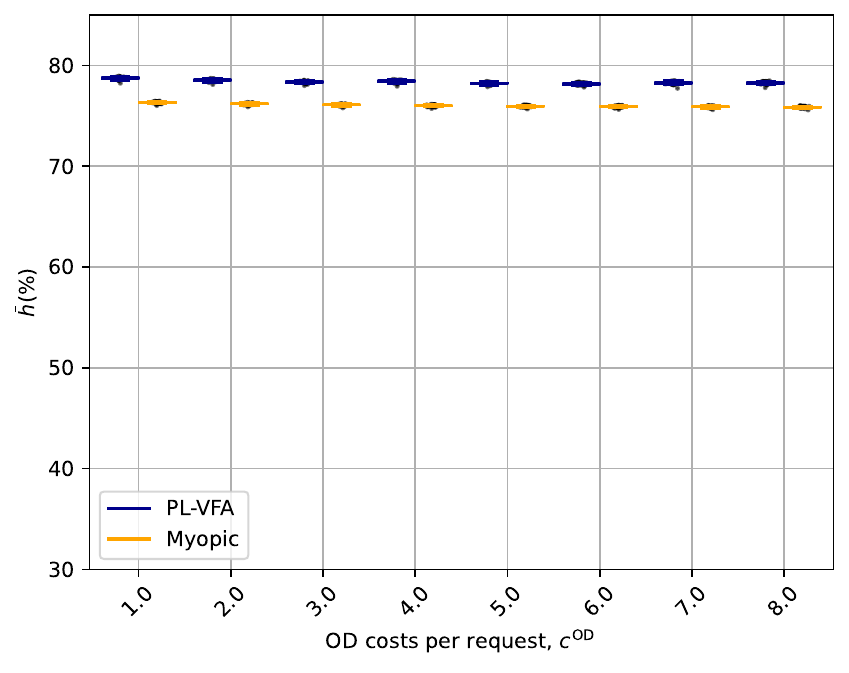}	\label{fig:c_o_total_utility}}
	\caption{Variation of \gls{CD} costs.}
\end{figure}

Figure \ref{fig:cost_splits} shows the driver and penalty cost split as a percentage of total costs for varying $q^{\mathrm{\gls{GW}}}$ and~$q^{\mathrm{\gls{OD}}}$. We observe that \glspl{FD} have, overall, the highest cost share, with up to $50\,\%$ for $q^{\mathrm{\gls{GW}}} = q^{\mathrm{\gls{OD}}} = 0.01$. \gls{GW} costs have the overall second highest cost share with up to $30\,\%$ when $q^{\mathrm{\gls{GW}}}=0.13$, while \glspl{OD} have lower importance in the cost mix. Penalty costs are only high when the \gls{GW} joining rate is low and highest for~${(q^{\mathrm{\gls{GW}}} = 0.01,q^{\mathrm{\gls{OD}}}=0.01)}$ with a share of up to $40\,\%$.
\begin{result}
	\textit{In mixed fleets, \glspl{FD} are the main total costs driver with a cost share of up to $50\,\%$, followed by \glspl{GW} with up to $30\,\%$.}
\end{result}
In Figure \ref{fig:c_g_total_utility}, we report \gls{GW} costs per km. 
The cost-saving potential decreases with increasing $c^{\mathrm{\gls{GW}}}$. For low $c^{\mathrm{\gls{GW}}}$ the advantage of using $\pi^{\mathrm{\gls{PL-VFA}}}$ is highest and leads to higher cost savings compared to $\pi^{\mathrm{\gls{MY}}}$ of up to 5 percentage points. In Figure \ref{fig:c_g_total_utility}, we report \gls{OD} costs per request. We observe that $\bar{h}$ remains constant over the entire range of $c^{\mathrm{\gls{OD}}}$.
The advantage of using $\pi^{\mathrm{\gls{PL-VFA}}}$ observed for low $c^{\mathrm{\gls{GW}}}$ in Figure \ref{fig:c_g_total_utility} is plausible since the \gls{LSP} can increasingly leverage \glspl{GW} when their costs are low.
%and whose potential $\pi^{\mathrm{\gls{PL-VFA}}}$ exploits better than by $\pi^{\mathrm{\gls{MY}}}$.

\begin{result}
	\textit{The cost saving potential is more sensitive to \gls{GW} costs than to \gls{OD} costs. The advantage of~$\pi^{\mathrm{\gls{PL-VFA}}}$ is highest when \gls{CD} costs are low, for which~$\pi^{\mathrm{\gls{PL-VFA}}}$ yields more than five percentage points higher cost savings than $\pi^{\mathrm{\gls{MY}}}$.}
\end{result}

\begin{figure}[htbp]
	\centering
	\includegraphics[width=0.4\textwidth]{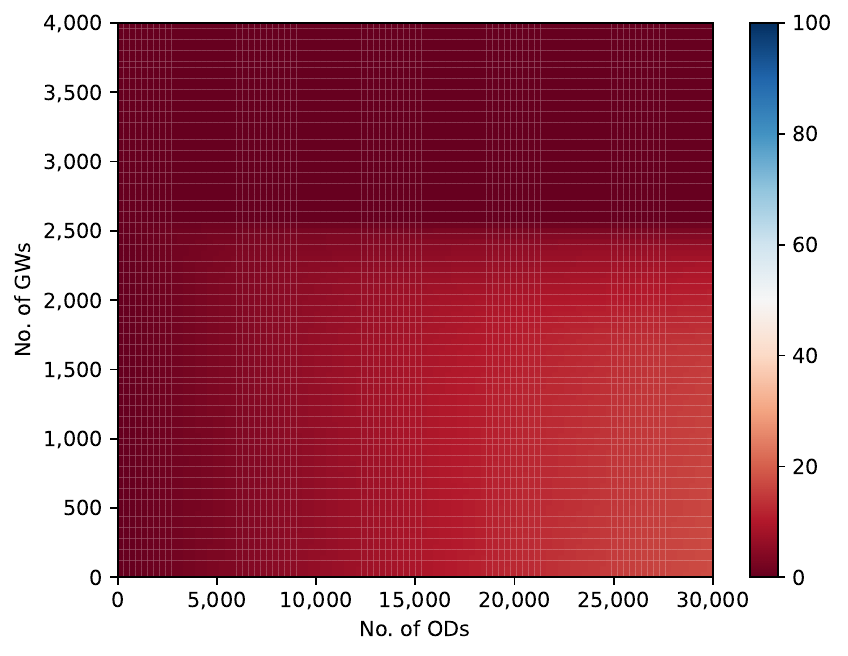}
	\caption{Share of requests delivered (in \%) by \gls{OD} in $t=0$ when $n_0^{\mathrm{\gls{FD}}}=0$ for different fleet compositions.}
	\label{fig:requests_delivered_base}
\end{figure}

In the previous analyses, we observed that total costs are less sensitive to \gls{OD} than to \gls{GW} specific parameters, thereby indicating that \glspl{OD} play an insignificant role in the request delivery process. In Figure \ref{fig:requests_delivered_base}, we investigate the share of requests delivered by \glspl{OD} in the first time step for different fleet compositions and zero \glspl{FD}. We observe that the share of requests delivered corresponds to approximately $5\,\%$ for most fleet compositions. Only when the number of \glspl{GW} is low and the number of \glspl{OD} very high ($\ge15,000$ \glspl{OD}), \glspl{OD}' share in requests delivered increases to above $20\,\%$. 

\begin{figure}[htbp]
	\centering
	\subfigure{
	\includegraphics[width=0.0425\textwidth,height=0.2\textheight]{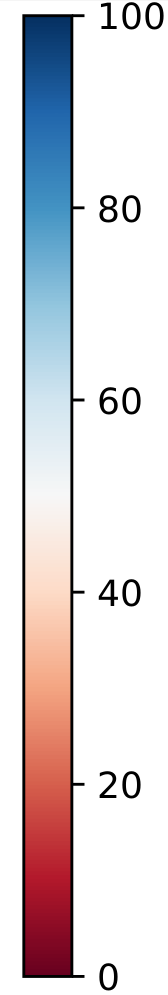}}
	\subfigure[Reduction of \gls{OD} costs to $2.5\,\$$.]{
		\includegraphics[width=0.3\textwidth]{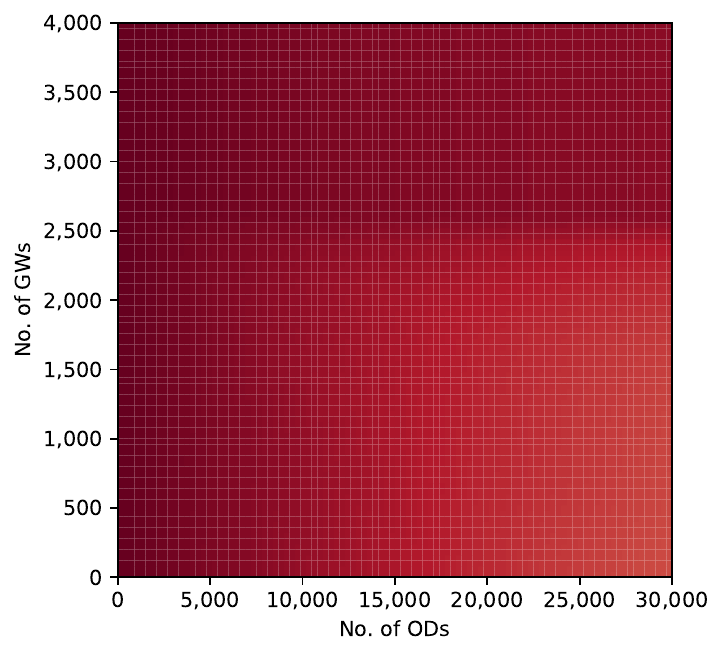}
		\label{fig:requests_delivered_c275}\setcounter{subfigure}{1}}
	\subfigure[Increase of $\zeta^{\mathrm{\gls{OD}}}$ to 0.5.]{
		\includegraphics[width=0.3\textwidth]{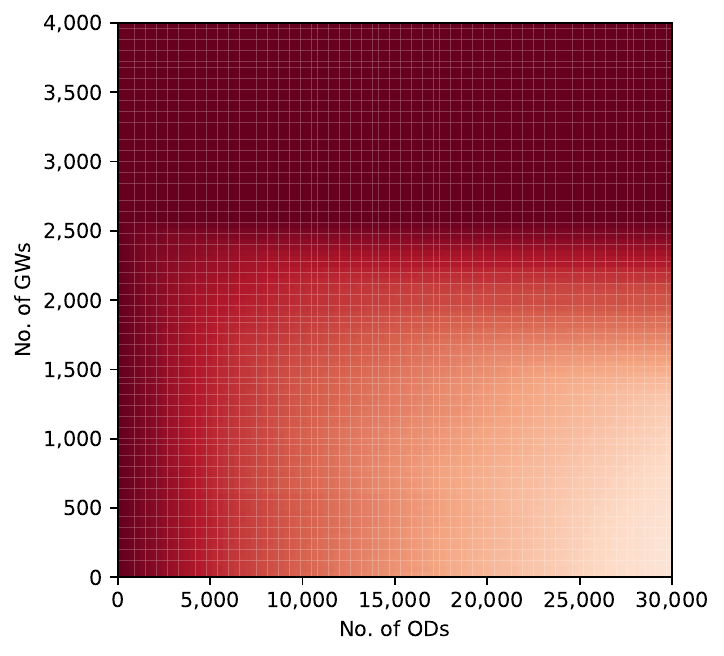}
		\label{fig:requests_delivered_capacity05}}
	\subfigure[Snychronizing \gls{OD} and request routes.]{
		\includegraphics[width=0.3\textwidth]{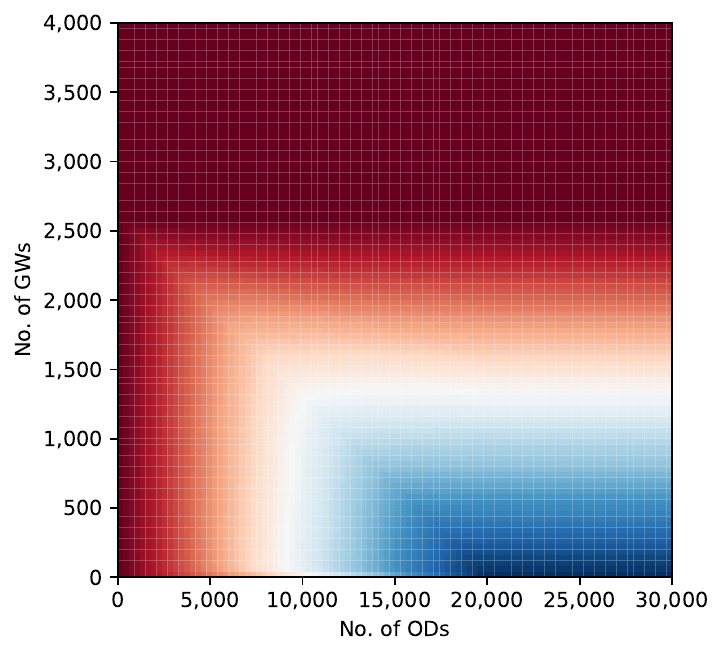}
		\label{fig:requests_delivered_odpatterns}}
	\caption{Share of requests delivered (in \%) in $t=0$ and $n_0^{\mathrm{\gls{FD}}} = 0$ for different fleet compositions across \glspl{OD}' costs, temporal patterns, and spatial patterns.} \label{fig:requests_delivered_variations}
\end{figure}

\begin{figure}[htbp]
	\centering
	\includegraphics[width=0.4\textwidth]{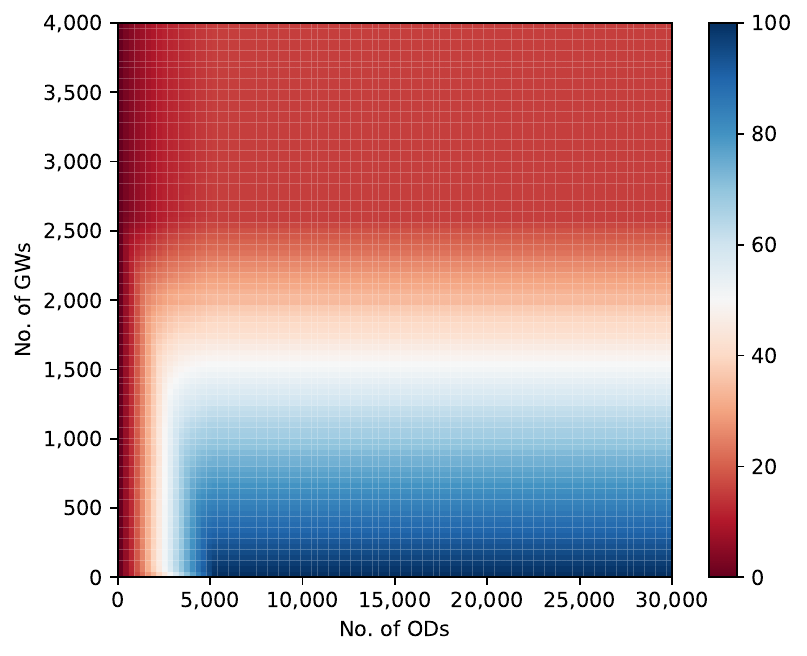}
	\caption{Share of requests delivered (in \%) by \gls{OD} in $t=0$ when $n_0^{\mathrm{\gls{FD}}}=0$ for different fleet compositions considering all variations in Figure \ref{fig:requests_delivered_variations} simultaneously.}
	\label{fig:requests_delivered_allvar}
\end{figure}

Three factors influence \glspl{OD}' share in request matches. Firstly, \glspl{OD}' costs per request play a significant role when deciding whether to outsource a request to \glspl{GW} or \glspl{OD}. To understand its impact, we reduce \glspl{OD}' costs per request from $5\,\$$ to $2.5\,\$$ and report this variation's result in Figure \ref{fig:requests_delivered_c275}. Overall, the share of delivered requests rises to around $10\,\%$, even when more \glspl{GW} are present. Secondly, we change \glspl{OD}' temporal patterns, represented by $\zeta^{\mathrm{\gls{OD}}}$, from 0.13 to 0.5, thereby increasing the number of \glspl{OD} being active within $\mathcal{\bar{T}}$ (cf. Figure \ref{fig:requests_delivered_capacity05}). The share of requests delivered by \glspl{OD} reaches around $50\,\%$ when the number of \glspl{OD} is high, and only a few \glspl{GW} are available. However, when the number of \glspl{GW} is high, the share of requests delivered by \glspl{OD} remains as low as in the base case. Thirdly, we synchronize \glspl{OD}' spatial driving patterns, $P^{\mathrm{\gls{OD}}}_{ij}$, with request route patterns, i.e., $P^{\mathrm{\gls{OD}}}_{ij} = P^{\mathrm{R}}_{ij}$ (cf. Figure \ref{fig:requests_delivered_odpatterns}). This has the strongest impact on the share of requests delivered by \glspl{OD} as it reaches values of above $80\,\%$ when the number of \glspl{OD} is high and \glspl{GW} is low.
In all three cases, the share of requests delivered by \glspl{OD} remained low as the number of \glspl{GW} increased. Hence, we study the impact of varying all three parameters simultaneously in Figure~\ref{fig:requests_delivered_allvar}. Now, the share of requests delivered by \glspl{OD}, even for high numbers of \glspl{GW}, mounts from $5\,\%$ in the base case to roughly $35\,\%$. This bears a significant potential for municipalities to ameliorate sustainable logistics. Unlike \glspl{GW}, \glspl{OD} do not induce traffic, and municipalities could motivate \glspl{LSP} to implement measures increasing the share of \gls{OD} deliveries. \glspl{LSP} can, for example, provide incentives for customers to buy during \gls{OD} peak times by offering rebates during these times, or synchronize request and \gls{OD} routes by pooling requests at micro-hubs which coincide with \location{}s frequently visited by \glspl{OD}.
\begin{result}
	\textit{\glspl{OD}' share in requests delivered increases to more than $90\,\%$ when synchronizing \glspl{OD}' temporal and spatial patterns with those of requests, \gls{OD} costs are reduced, and the number of \glspl{GW} is low.}
\end{result}

\begin{figure}[htbp]
	\centering
	\subfigure[Variation of severance payment $C^{\mathrm{sev}}$.]{\includegraphics[width=0.4\textwidth]{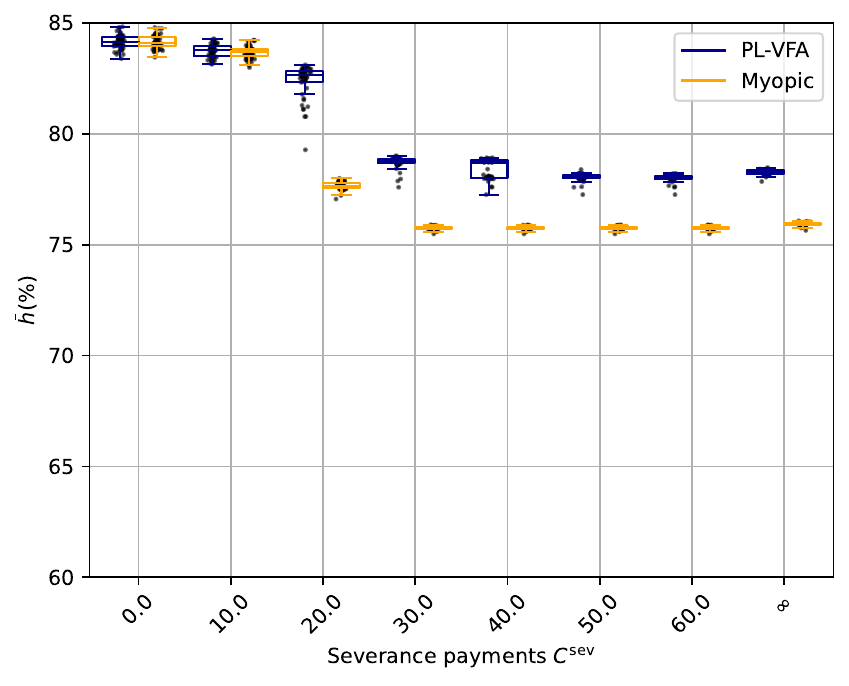}\label{fig:firing_request_total_utility}
	}
	\subfigure[Variation of \glspl{FD}' fix costs $C^{\mathrm{fix}}$.]{\includegraphics[width=0.4\textwidth]{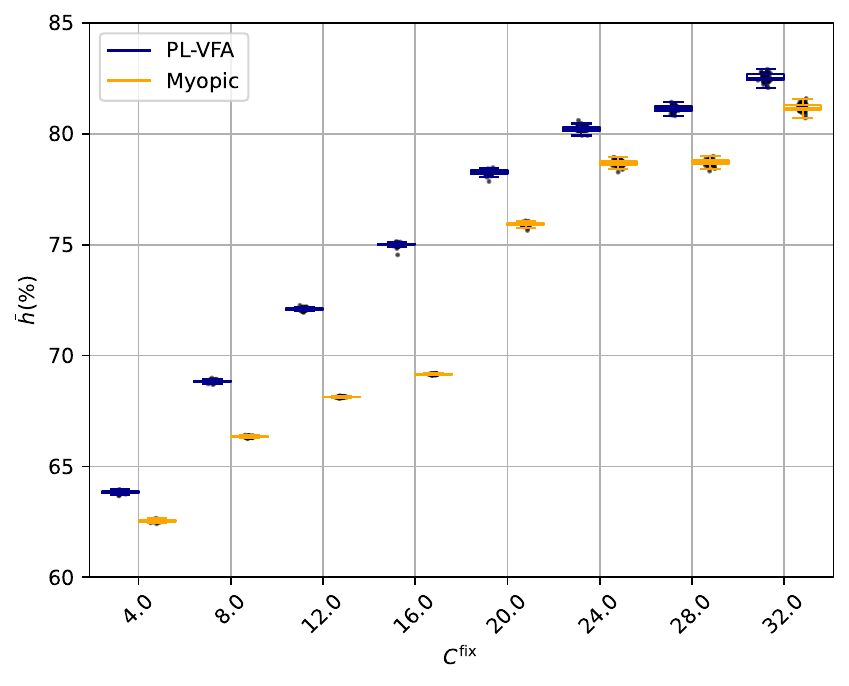}\label{fig:c_fd_total_utility}}
	\caption{Variation of severance payment and \gls{FD} fix costs.}
\end{figure}

Finally, to understand the impact of firing flexibility in workforce planning, we vary the severance payment $C^{\mathrm{sev}}$ and report the results in Figure \ref{fig:firing_request_total_utility}. We observe that with increasing $C^{\mathrm{sev}}$, the saving potential compared to an \gls{FD}-only fleet decreases. The cost advantage of using $\pi^{\mathrm{\gls{PL-VFA}}}$ increases with increasing $C^{\mathrm{sev}}$ and remains approximately constant from $C^{\mathrm{sev}}=40\,\$$ on. When $C^{\mathrm{sev}}$ is low, the \gls{LSP} can lay off \glspl{FD} at any time without additional costs. Hence, the advantage of using \gls{PL-VFA} is negligible. When $C^{\mathrm{sev}}$ is higher, the value of not hiring \glspl{FD} is potentially higher as \glspl{FD} can become obsolete in later time steps. Firing them then results in penalty costs that could have been avoided if they were not hired in the first place. This trade-off is only made by $\pi^{\mathrm{\gls{PL-VFA}}}$.
In Figure \ref{fig:c_fd_total_utility} we explore the impact of $C^{\mathrm{{fix}}}$ on $\bar{h}$. For small $C^{\mathrm{fix}}$ $\bar{h}$ is smaller and the difference between $\pi^{\mathrm{\gls{PL-VFA}}}$ and $\pi^{\mathrm{\gls{MY}}}$ is negligible. The cost saving potential increases up to $82.5\,\%$ for $\pi^{\mathrm{\gls{PL-VFA}}}$ and $81\,\%$ for $\pi^{\mathrm{\gls{MY}}}$. The difference between $\pi^{\mathrm{\gls{PL-VFA}}}$ and $\pi^{\mathrm{\gls{MY}}}$ increases up to seven percentage points when $C^{\mathrm{fix}}=16\,\$$, which corresponds to $19\,\%$ lower total costs. When $C^{\mathrm{fix}}$ increases further, the difference between both policies decreases again. For low fixed \gls{FD} costs, the decision of hiring or not hiring \glspl{FD} does not significantly impact total costs. As $C^{\mathrm{fix}}$ increases, this decision needs to be traded off more carefully. Interestingly, the advantage of using \gls{PL-VFA} vanishes for even higher fixed costs, as both policies refrain from hiring costly \glspl{FD}.
\begin{result}
	\textit{Compared to $\pi^{\mathrm{\gls{MY}}}$, $\pi^{\mathrm{\gls{PL-VFA}}}$ achieves three percentage points higher cost savings when severance payments become infinite (base case) and seven percentage points higher cost savings, i.e., $19\,\%$ lower total costs, when \gls{FD} fix costs are set to $16\,\$/\mathrm{h}$.}
\end{result}
%This section stressed the importance of \gls{CD} supply in reducing the \gls{LSP}'s total costs and the total costs' sensitivities concerning \gls{CD} and \gls{FD} costs. 
So far, we assumed that \glspl{CD} leave the \gls{LSP} according to a fixed resignation rate. In the next section, we study the effect of a resignation probability that depends on the number of unmatched \glspl{CD}.
\FloatBarrier
\subsection{Analysis of unmatched \glspl{CD}} \label{sec:results_driver_perspective}
\begin{figure}[htbp]
	\centering
	\subfigure[Percentage of unmatched \glspl{GW}.]{
		\includegraphics[width=0.4\textwidth]{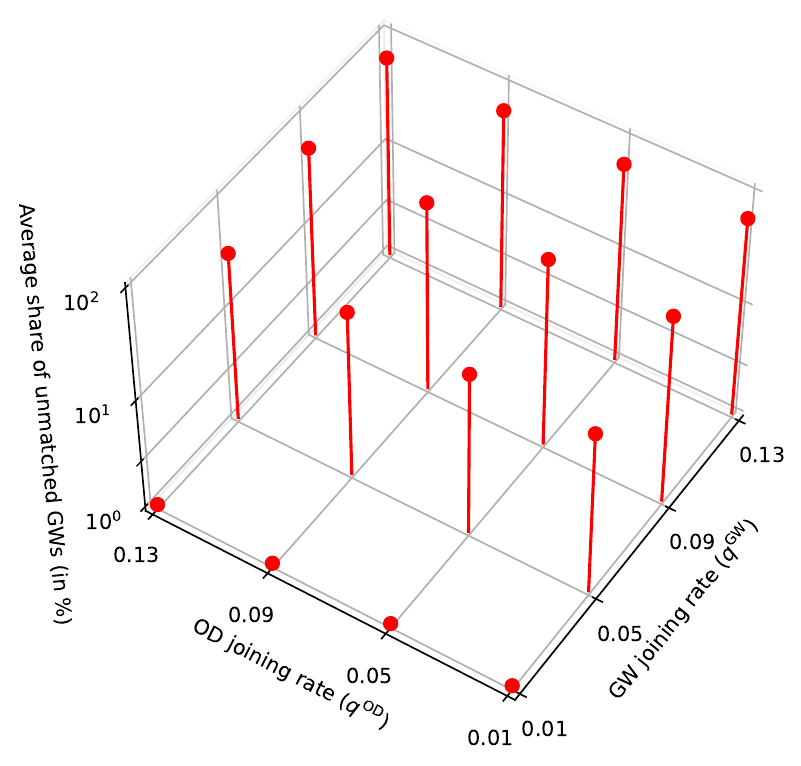}
		\label{fig:add_gw_add_od_s_g}}
	\subfigure[Percentage of unmatched \glspl{OD}.]{
		\includegraphics[width=0.4\textwidth]{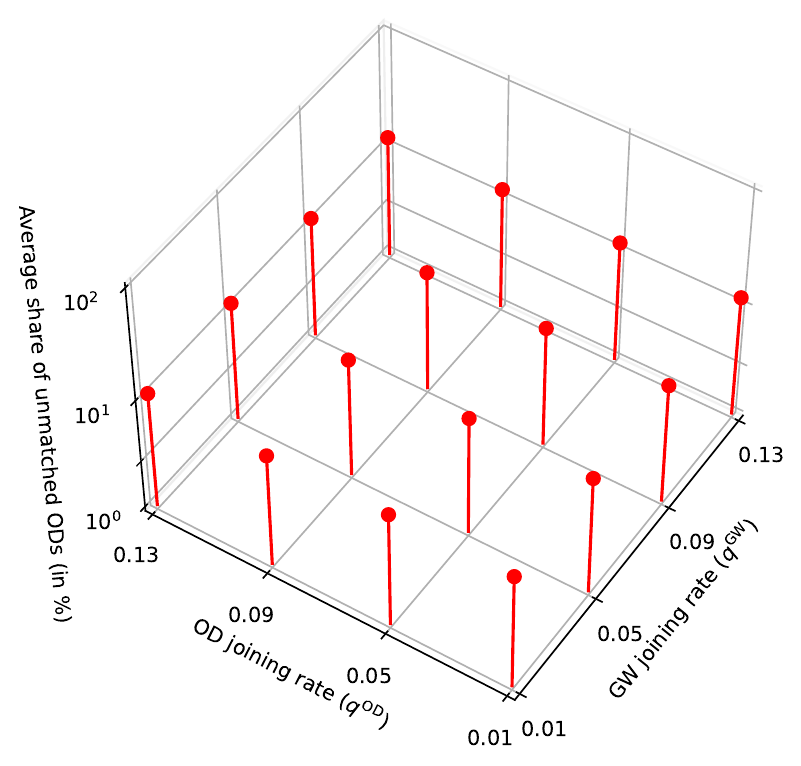}
		\label{fig:add_gw_add_od_s_o}}
	\caption{Percentage of unmatched \glspl{CD} when the resignation probability does not depend on the number of unmatched \glspl{CD}. The $z$-axis is logarithmically scaled. } \label{fig:unmatched_CDs}
\end{figure}

Figure \ref{fig:unmatched_CDs} reports the average percentage of unmatched \glspl{GW} and \glspl{OD} for different joining rates over the entire time horizon. We observe that the percentage of unmatched \glspl{GW} (cf. Figure \ref{fig:add_gw_add_od_s_g}) is higher than the percentage of unmatched \glspl{OD}, especially when $q^{\mathrm{\gls{GW}}}$ is high. For high $q^{\mathrm{\gls{GW}}}$ the percentage of unmatched \glspl{GW} amounts to $60\,\%$. The percentage of unmatched \glspl{OD} is lower and amounts to a constant value of $10\,\%$ across different joining rates (cf. Figure \ref{fig:add_gw_add_od_s_o}).
As the joining rate grows, \gls{CD} supply surpasses demand, and the \gls{LSP} can no longer outsource demand to \glspl{CD}. Furthermore, the higher quotient of unmatched \glspl{GW} is plausible since more \glspl{GW} are active on the operational level than \glspl{OD} due to their higher $\zeta^{\mathrm{\gls{GW}}}$. Hence, they are more likely to be active when there is no demand.

\begin{figure}[htbp]
	\centering
	\subfigure[Percentage of unmatched \glspl{GW}.]{
		\includegraphics[width=0.4\textwidth]{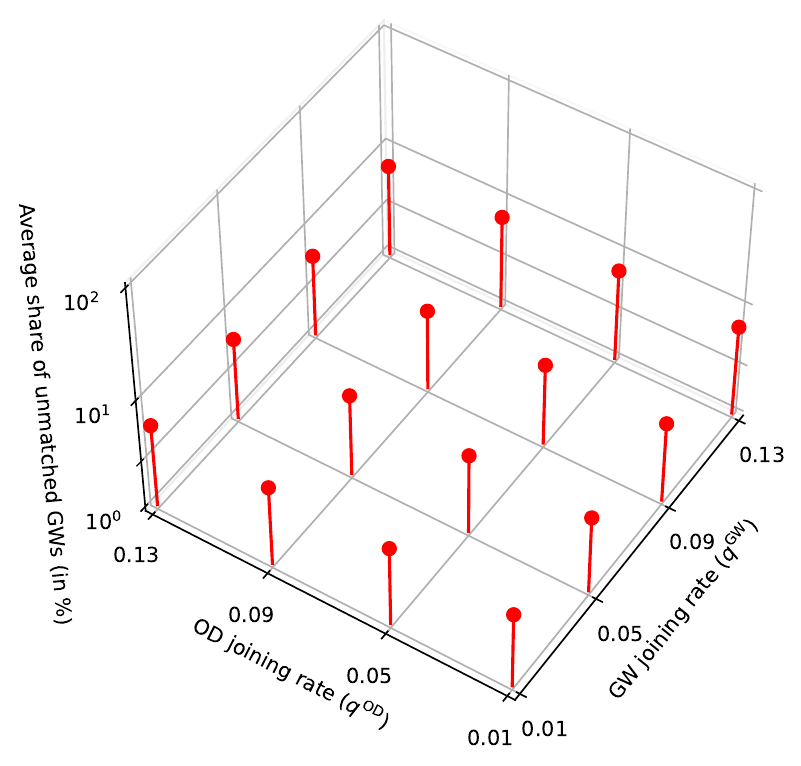}
		\label{fig:add_gw_add_od_s_g_ms}}
	\subfigure[Percentage of unmatched \glspl{OD}.]{
		\includegraphics[width=0.4\textwidth]{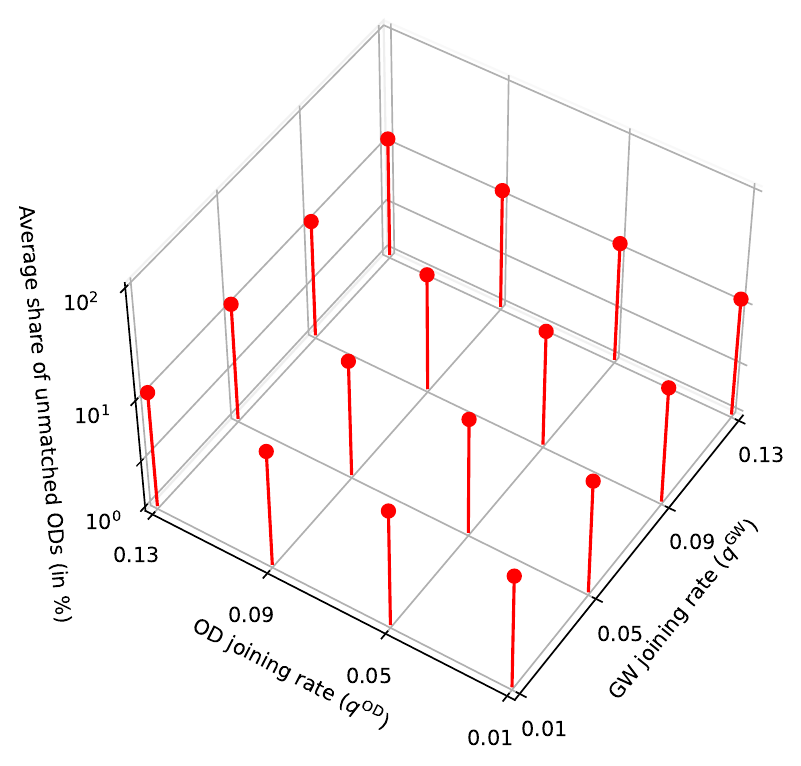}
		\label{fig:add_gw_add_od_s_o_ms}}
	\caption{Percentage of unmatched \glspl{CD} when \glspl{CD}' resignation probability depends on number of unmatched \glspl{CD}. The $z$-axis is logarithmically scaled.} \label{fig:add_gw_add_od_s}
\end{figure}

Figure \ref{fig:add_gw_add_od_s} shows the number of unmatched \glspl{CD} now assuming that \glspl{CD}' resignation probability depends on $s_{ij}^{\alpha}$ (cf. Equations \eqref{eq:matching_sensitive_resig}). Overall, the quotient of unmatched CDs is significantly lower than
in Figure~\ref{fig:unmatched_CDs}. Moreover, the difference between \glspl{GW} and \glspl{OD} is small. Both have a relatively constant ratio of unmatched drivers across joining rates of around $10\,\%$. The lower ratio of unmatched \glspl{GW} is plausible, as the \glspl{GW}' resignation probability depends on the number of \glspl{GW} unmatched. Hence, it can significantly surpass the base case resignation probability of 0.01 and, therefore, the joining rate. When \glspl{GW} leave the \gls{LSP}'s platform at a higher rate than joining it, the \gls{LSP} accumulates fewer \glspl{GW} than in the case when the resignation probability has a constant value of 0.01. Fewer \glspl{GW} imply less unmatched \glspl{GW}. 

\begin{figure}[htbp]
	\subfigure[Difference for \glspl{FD}.]{
		\includegraphics[width=0.3\textwidth]{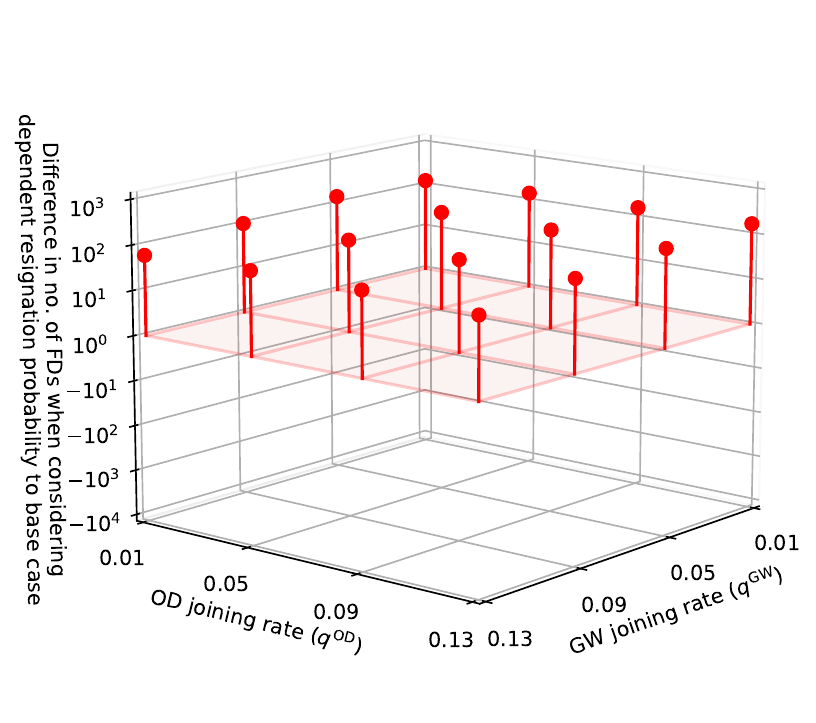}
		\label{fig:add_gw_add_od_diff_fd}}
	\subfigure[Difference for \glspl{GW}.]{
		\includegraphics[width=0.3\textwidth]{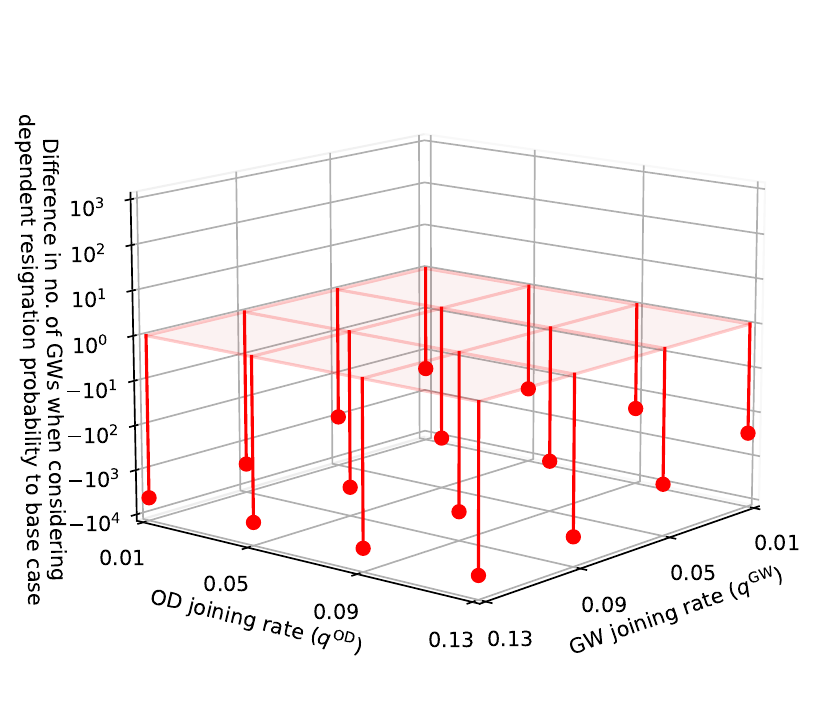}
		\label{fig:add_gw_add_od_diff_gw}}
	\subfigure[Difference for \glspl{OD}.]{
		\includegraphics[width=0.3\textwidth]{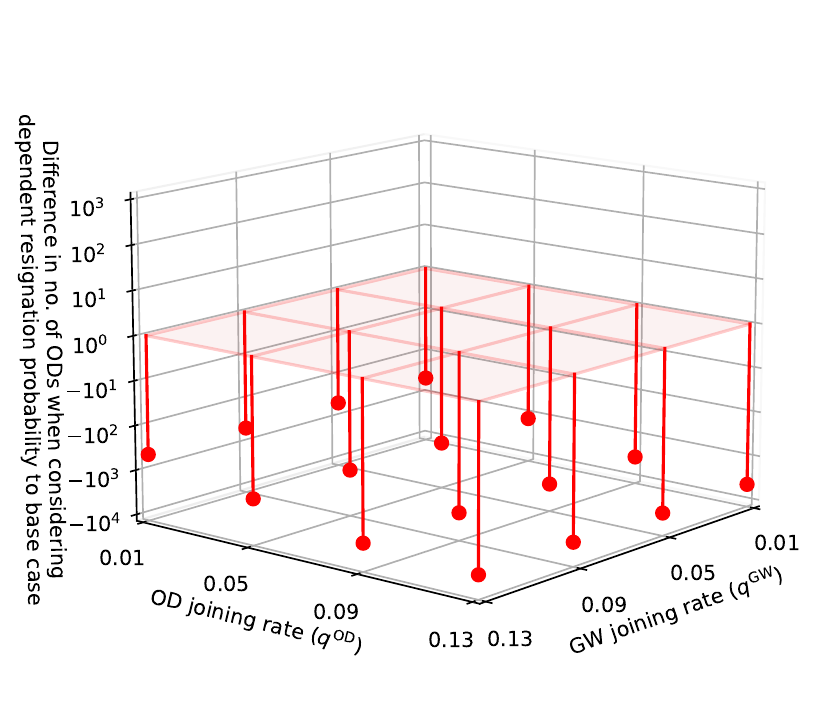}
		\label{fig:add_gw_add_od_diff_od}}
	\caption{Difference between the total number of drivers when resignation probability depends on the number of unmatched \glspl{CD} and when it does not. The $z$-axis is logarithmically scaled.}
	\label{fig:add_gw_add_od_drivers_difference}
\end{figure}

In Figure \ref{fig:add_gw_add_od_drivers_difference}, we show the difference in the number of \glspl{FD} and \glspl{CD} in the final time step $t=T$ between the case wherein the resignation probability depends and wherein the resignation probability does not depend on the percentage of unmatched \glspl{CD}. We observe that the \gls{LSP} hires more than 100 additional \glspl{FD} and has significantly less \glspl{CD} at their disposition when the resignation probability depends on the number of unmatched \glspl{CD}. The lower \gls{CD} supply urges the \gls{LSP} to hire more \glspl{FD} to ensure high service levels.
\begin{result}
	\textit{The \gls{LSP} has to hire $\ge 100$ more \glspl{FD}, when the resignation probability depends on the number of unmatched \glspl{CD}, as the effective \gls{CD} supply is significantly lower.}
\end{result}
\FloatBarrier

%% file: contents/Conclusion.tex
\section{Conclusion} \label{sec:conclusion}
\glsresetall
In this paper, we studied the strategic workforce planning of a \gls{LSP} providing on-demand delivery services with a mixed fleet of couriers consisting of \glspl{FD} and \glspl{CD}. We integrated long-term strategic \gls{FD} hiring decisions and short-term operational decisions regarding driver dispatching. We formalized the strategic hiring and firing problem as a \gls{MDP} and solved it with approximate dynamic programming based on piecewise linear value function approximation, which allows us to study large-scale instances. We incorporated operational costs in the \gls{MDP}'s cost function using a fluid approximation to account for delivery operations.

We conducted a case study based on a real-world data set from Grubhub for food delivery in a metropolitan area located in the US. Herein, our studies led to several findings, which we synthesize in the following. 

\textbf{Total costs obtained with \gls{PL-VFA} are either equal to or up to $19\,\%$ lower than total costs obtained with a myopic hiring policy.} \gls{PL-VFA} achieved this cost saving by hiring less \glspl{FD} than required to serve the demand and consequently accepts lower service levels in early time steps. It does so to hedge against remunerating obsolete \glspl{FD} in later time steps as the amount of \glspl{CD} grew.

\textbf{\glspl{FD} and \glspl{GW} are the main cost drivers in the total cost mix with up to a 50\% and a 30\% share in total costs, respectively.} The significance of \gls{FD} costs in the total cost mix stresses the importance of finding good \gls{FD} hiring policies, which minimize the number of \glspl{FD} hired. \glspl{GW} have the second highest contribution to total costs, when $q^{\mathrm{\gls{GW}}}$ is high. The \glspl{GW}' contribution is more significant than the one of \glspl{OD} because \glspl{OD} only accept requests coinciding with their origin and destination. Moreover, their share of arrivals within $\mathcal{\bar{T}}$ is significantly lower than the one of \glspl{GW}.

\textbf{The \gls{LSP} has to hire more than $100$ additional \glspl{FD} when \glspl{CD} are matching sensitive} as an oversupply of \glspl{CD}, i.e., a higher number of \glspl{CD} than required to serve the demand, does not arise anymore. 

This work opens up a promising new research avenue in the field of crowdsourced deliveries, by combining the study of crowdsourced delivery fleets with long-term workforce planning. Specifically, this work provides a foundation for follow-up studies. Firstly, the \gls{FD} workforce planning problem could be extended by accounting for \glspl{FD} with different contract durations or working schedules. Moreover, one could introduce uncertainty in the demand dimension on the strategic workforce planning level. Finally, one could implement behavioral components for \glspl{CD}, e.g., discrete choice models based on real-world data, to more accurately represent the \glspl{CD}' behavior, e.g., regarding resignation processes.

%% file: contents/Appendices.tex
%\begin{APPENDIX}{Multiple Operational Time Horizons}
%\end{APPENDIX}
%\begin{APPENDIX}{}
\setcounter{page}{1}
		\section{Multiple Operational Time Horizons}\label{app:multiple_ops_time_horizons}	
	If we consider~$K$ different time horizons, denoted by~$\mathcal{\bar{T}}_k,\, k\in\left\{1,\ldots,K\right\}$, we need to solve an optimization problem to compute accurate~$C_t^{\mathrm{ops}}$.
	Let us denote the total number of \glspl{FD} by~$n_t^{\mathrm{\gls{FD}}}$. Moreover, we denote the \glspl{FTE} with ~$\bar{n}_t^{\mathrm{\gls{FD}}}$, which we obtain from~$\bar{n}_t^{\mathrm{\gls{FD}}} = \zeta^{\mathrm{\gls{FD}}}\,n_t^{\mathrm{\gls{FD}}}$, where~$\zeta^{\mathrm{\gls{FD}}}$ denotes the share of time \glspl{FD} are willing to work based on a full-time working schedule of an \gls{LSP}. For conciseness, we consider this share of time to be homogeneous, i.e., all \glspl{FD} either accept full-time or part-time contracts with the same amount of hours. We denote the number of \gls{FD} \glspl{FTE} per~$\bar{\mathcal{T}}_k$ with~$\bar{n}_{tk}^{\mathrm{\gls{FD}}}$.  Since each operational level problem potentially requires a different number of \glspl{FD} to serve the demand, we can not distribute~$\bar{n}_t^{\mathrm{\gls{FD}}}$ homogeneously among~$\mathcal{\bar{T}}_k$. Instead, we need to solve the following problem, which yields an optimal distribution of~$\bar{n}_{tk}$ over all~$\mathcal{T}_k$
	\begin{align}
	C_t^{\mathrm{ops}}(n_t^{\mathrm{\gls{FD}}},\cdot) & = \underset{\bar{n}_{tk}^{\mathrm{\gls{FD}}}}{\mathrm{min}}\,\sum_k^K\,\bar{C}_{tk}^{\mathrm{ops}}(\bar{n}_{tk}^\mathrm{\gls{FD}},\cdot), \\
	\text{s.t. } & \qquad \sum_k^K \bar{n}_{tk}^{\mathrm{\gls{FD}}} = \bar{n}_t^{\mathrm{\gls{FD}}},
	\end{align}
	where~$\cdot$ denotes the current \gls{CD} fleet.
	\section{Fluid Model}\label{app:fluid_model}
In this section, we first describe how the operational level's problem (cf. Equations \eqref{eq:objective_operational_level_fluid} to \eqref{eq:availabilities}) is obtained. Then, we prove Proposition \ref{prop:fluid_model}, which allows us to state that the operational problem's objective value is a lower bound on the operational costs we would obtain if using a state-independent or dependent policy in a finite-sized system and, therefore a suitable approximation of the operational level's costs.
%	\subsection{Origin of Equations \eqref{eq:objective_operational_level_fluid} to \eqref{eq:availabilities}} \label{app:fluid_model_derivation}

In the first step, we consider the closed queuing network introduced in Section \ref{sec:ops_problem} without restricting ourselves to the equilibrium state. Let us define the cumulative idle time corresponding to a single server queue~$e_{ii}(\bar{t})$ as follows
\begin{equation}
u_i(\bar{t}) = \int_{0}^{\bar{t}} \mathbbm{1}(e_{ii}(s) = 0)\,ds, \qquad \forall\,i\in\mathcal{M},
\end{equation}
where~$\mathbbm{1}(\cdot)$ represents the indicator function. 

In the first step, we neglect the presence of \glspl{CD}. We can then state the flow conservation equations that the fluid queue lengths~$e_{ij}(\bar{t})$ and~$f_{ij}(\bar{t})$ ($\forall i,j\in\mathcal{M}$) need to fulfil, which are analogous to Equations 19 to 21 in \cite{BravermanDaiEtAl2019}.
%\begin{equation}
\begin{alignat}{1}
f_{ij}(\bar{t}) &= f_{ij}(0)+\frac{\lambda_{it}^{\mathrm{R}}}{n_t^{\mathrm{\gls{FD}}}}\,P_{ij}^{\mathrm{R}}(\bar{t}-u_i(\bar{t}))-\mu_{ij}\,\int_0^{\bar{t}}f_{ij}(s)\,ds, \qquad\qquad\qquad\qquad\qquad\,\,\,\,\,\,\,\,\,\,\,\, \forall i,j \in \mathcal{M}, \label{eq:f_queue} \\
e_{ij}(\bar{t}) &= e_{ij}(0)-\mu_{ij}\,\int_{0}^{\bar{t}} e_{ij}(s)\,ds+Q_{ij}\,\sum_{k}\mu_{ki}\,\int_{0}^{\bar{t}} f_{ki}(s)\,ds,\qquad\qquad\qquad\,\,\,\,\,\,\,\,  \forall i,j \in \mathcal{M}, i\neq j, \label{eq:e_ij_queue}\\
e_{ii}(\bar{t}) &= e_{ii}(0)-\frac{\lambda_{it}^{\mathrm{R}}}{n_t^{\mathrm{\gls{FD}}}}\left(\bar{t}-u_i(\bar{t})\right)+\sum_{j,j\neq i}\mu_{ji}\int_0^{\bar{t}}e_{ji}(s)ds+Q_{ii}\sum_{j}\mu_{ji}\int_0^{\bar{t}}f_{ji}(s)ds, \qquad \forall i \in \mathcal{M}. \label{eq:e_ii_queue}
\end{alignat}
%\end{equation}
Equation \eqref{eq:f_queue} states that all queues describing \glspl{FD} that serve requests, i.e.,~$f_{ij}(\bar{t})$, are equal to the sum of \glspl{FD} initially serving requests, i.e.,~$f_{ij}(0)$, the number of requests per \gls{FD} in~$(i,j)$-direction up until~$\bar{t}$, i.e.,~$\frac{\lambda_{it}^{\mathrm{R}}\,P_{ij}^{\mathrm{R}}}{n_t^{\mathrm{\gls{FD}}}}(\bar{t}-u_i(\bar{t}))$, and the outgoing \glspl{FD} that serve requests, i.e.,~$-\mu_{ij}\,\int_0^{\bar{t}}f_{ij}(s)\,ds$. Equation \eqref{eq:e_ij_queue} describes~$e_{ij}(\bar{t})$ as the sum of initially relocating \glspl{FD}, i.e.,~$e_{ij}(0)$, relocating \glspl{FD} up until~$\bar{t}$, i.e.,~$-\mu_{ij}\int_{0}^{\bar{t}} e_{ij}(s)\,ds$, and \glspl{FD} arriving at~$i$ being relocated to~$j$, i.e.,~$Q_{ij}\,\sum_{k}\mu_{ki}\,\int_{0}^{\bar{t}} f_{ki}(s)\,ds$. Finally, Equation \eqref{eq:e_ii_queue} describes \glspl{FD} idling at~$i$ as the sum of initially idling \glspl{FD} at~$i$ ($e_{ii}(0)$), outgoing relocated \glspl{FD} ($-\frac{\lambda_{it}^{\mathrm{R}}}{n_t^{\mathrm{\gls{FD}}}}\left(\bar{t}-u_i(\bar{t})\right)$), incoming relocated \glspl{FD} (~$\sum_{j,j\neq i}\mu_{ji}\int_0^{\bar{t}}e_{ji}(s)ds$), and incoming \glspl{FD} serving a request from~$j$ to~$i$, which, upon arrival, are relocated from~$i$ to~$i$ ($Q_{ii}\sum_{j}\mu_{ji}\int_0^{\bar{t}}f_{ji}(s)ds$). 

Since we are interested in the queues' equilibrium behaviour, we set derivatives with respect to~$\bar{t}$ to zero, yielding
\begin{alignat}{2}
\dot{f}_{ij}(\bar{t}) & = 0 = \frac{\lambda_{it}^{R}\,P_{ij}^{\mathrm{R}}}{n_t^{\mathrm{\gls{FD}}}}(1-\dot{u}_i(\bar{t}))-\mu_{ij}\,f_{ij}(\bar{t}), & \forall i,j \in \mathcal{M}, \\
\dot{e}_{ij}(\bar{t}) & =0 = -\mu_{ij}\,e_{ij}(\bar{t})+Q_{ij}\sum_{k}\mu_{ki}\,f_{ki}(\bar{t}),& \forall i,j \in \mathcal{M},i\neq j,\\
\dot{e}_{ii}(\bar{t}) & = 0 = -\frac{\lambda_{it}^{R}}{n_t^{\mathrm{\gls{FD}}}}(1-\dot{u}_i(\bar{t}))+\sum_{ji,j\neq i}\mu_{ji}\,e_{ji}(\bar{t})+Q_{ii}\sum_{j}\mu_{ji}\,f_{ji}(\bar{t}), & \forall i \in \mathcal{M}.
\end{alignat}

The term~$\dot{u}_i(\bar{t})$ is either 1, which means that the queue is idle or 0, which means that the queue is not idle. We relax the requirement of~$\dot{u}(\bar{t})$ being binary and replace it with the server utilization~$1-a^{\mathrm{\gls{FD}}}_i$ and therefore allow partial utilization. Moreover, we remove the dependence of the variables on the time~$\bar{t}$, as, in the equilibrium state, the system of equations is not time-dependent anymore. This leads to the following set of equations
\begin{alignat}{2}
\frac{\lambda_{it}^{\mathrm{R}}\,P_{ij}^{\mathrm{R}}}{n_t^{\mathrm{\gls{FD}}}}\,a_i^{\mathrm{\gls{FD}}} &= \mu_{ij}\,f_{ij}, & \forall i \in \mathcal{M}, \label{eq:continuity_0} \\
\mu_{ij}\,e_{ij} &= Q_{ij}\,\sum_{ki}\mu_{ki}\,f_{ki}, & \forall i,j\in\mathcal{M},i\neq j, \label{eq:continuity_1}  \\
\frac{\lambda_{it}^{\mathrm{R}}}{n_t^{\mathrm{\gls{FD}}}}\, a_i^{\mathrm{FD}} &= \sum_{ji,j\neq i}\mu_{ji}\,e_{ji} + Q_{ii}\sum_{j}\mu_{ji}\,f_{ji}, & \forall i \in \mathcal{M}. \label{eq:continuity_2} 
\end{alignat}
Proof of Lemma 1 in \cite{BravermanDaiEtAl2019} shows how this set of equations can be transformed to Equations \eqref{eq:flow_cons1} to \eqref{eq:flow_cons3} (cf. Section \ref{sec:ops_problem}), by relying on~$0\le Q_{ij}\le 1$. 

We now incorporate \glspl{CD} by letting them serve demand flow, described by~$a_{ij}^{\mathrm{\gls{GW}}}$ and~$a_{ij}^{\mathrm{\gls{OD}}}$ respectively (cf. Equations \eqref{eq:little2} to \eqref{eq:little3}). Requests can also be rejected, denoted by~$a_{ij}^{\mathrm{\emptyset}}$. Taking these steps into account and stating our optimization objective (described by Equation \eqref{eq:objective_operational_level_fluid}), we obtain the operational level's problem as described by Equations \eqref{eq:objective_operational_level_fluid} to \eqref{eq:availabilities}. We are ready to prove Proposition \ref{prop:fluid_model}.

\textbf{Proof of Proposition \ref{prop:fluid_model}:} In the first step, we reformulate our problem such that our constraints equal the ones described in \cite{BravermanDaiEtAl2019}. In the second step, we show that our problem's objective fulfils certain conditions formulated by \cite{BravermanDaiEtAl2019}, which we detail in the following section. These conditions allow us to state that the solution to the \gls{LP} defined by Equations \eqref{eq:objective_operational_level_fluid} to~\eqref{eq:availabilities}, is a lower bound on the operational costs obtained when considering a finite-sized system and using a state-dependent or state-independent routing policy~$Q_{ij}$.

Firstly, we define the following two sets describing origin-destination pairs~$(i,j)$
\begin{align}
(i,j) \in 
\begin{cases}
\mathcal{\bar{M}}^1, & \text{if } \left(c_{ij}^{\mathrm{\gls{GW}}} < c_{ij}^{\mathrm{\gls{OD}}}\right), \\	
\mathcal{\bar{M}}^2, & \text{if } \left(c_{ij}^{\mathrm{\gls{OD}}}<  c_{ij}^{\mathrm{\gls{GW}}}\right), \\	
\end{cases} \qquad \forall i,j \in\mathcal{M}.
\end{align}
Note that, as already highlighted in the model description, we assume that~$c_{ij}^{\mathrm{\gls{FD}}}< c_{ij}^{\mathrm{\beta}}, \, \beta \in \left\{\mathrm{\gls{GW}},\mathrm{\gls{OD}},\emptyset\right\}$ on all arcs. 
\begin{remark} \label{rem:outsourcing_to_cds}
	We can easily see that, for each~$i\in\mathcal{M}$, and any choice of~$a_i^{\mathrm{FD}}$, obtaining~$a^{\mathrm{\gls{GW}}}_{ij}$ and ~$a^{\mathrm{\gls{OD}}}_{ij}$ is trivial. In fact, if~$(i,j)\in\mathcal{\bar{M}}^1$, for any choice of~$a_i^{\mathrm{\gls{FD}}}$, one would first try to outsource the remaining demand,~$\lambda_{it}^{\mathrm{R}}\,P_{ij}^{\mathrm{R}}\,\left(1-a_i^{\mathrm{FD}}\right)$, to \glspl{GW}, as they are cheaper, entailing~$a_{ij}^{\mathrm{\gls{GW}}}$ being set to its upper bound. If, after outsourcing, some demand~$\lambda_{it}^{\mathrm{R}}\,P_{ij}^{\mathrm{R}}\,\left(1-a_i^{\mathrm{FD}}\right)\,\left(1-a_{ij}^{\mathrm{\gls{GW}}}\right)$ remains, it is outsourced to \glspl{OD} and therefore~$a_{ij}^{\mathrm{\gls{OD}}}$ is set to its upper bound. If \glspl{OD} don't yet cover all the demand in~$(i,j)$-direction, the remaining amount, namely~$\lambda_{it}^{\mathrm{R}}\,P_{ij}^{\mathrm{R}}\,\left(1-a_i^{\mathrm{FD}}\right)\,\left(1-a_{ij}^{\mathrm{\gls{GW}}}\right)\,\left(1-a_{ij}^{\mathrm{\gls{OD}}}\right)$, is penalized. Proceeding in a different order would not be cost-optimal, as on the route~$(i,j)\in\bar{\mathcal{M}}^1$,~$c_{ij}^{\mathrm{\gls{GW}}}<c_{ij}^{\mathrm{\gls{OD}}}<c_{ij}^{\emptyset}$. The same procedure applies if~$(i,j)\in\bar{\mathcal{M}}^2$. Then, requests are first outsourced to \glspl{OD} and then to \glspl{GW}. If costs for both are equal, the attribution to one of the two is indifferent.
\end{remark}

Based on Remark \ref{rem:outsourcing_to_cds}, we can gather \gls{GW}, \gls{OD}, and penalty costs to obtain alternative penalty costs~$c_{ij}^{\emptyset,\bar{\mathcal{M}}^1}$, when~$(i,j)\in\bar{\mathcal{M}}^1$, yielding
\begin{align}
c_{ij}^{\emptyset,\bar{\mathcal{M}}^1}(a_i^{\mathrm{\gls{FD}}}) = & \mathrm{min}\left[\lambda_{it}^{\mathrm{R}}\,P_{ij}^{\mathrm{R}}\,(1-a_i^{\mathrm{\gls{FD}}}),\lambda_{it}^{\mathrm{\gls{GW}}}\,P_{ij}^{\mathrm{\gls{GW}}}\right]\,c_{ij}^{\mathrm{\gls{GW}}} &\notag \\
+ & \mathrm{min}\left[\mathrm{max}\left[\lambda_{it}^{\mathrm{R}}\,P_{ij}^{\mathrm{R}}(1-a_i^{\mathrm{\gls{FD}}})-\lambda_{it}^{\mathrm{\gls{GW}}}\,P_{ij}^{\mathrm{\gls{GW}}},0\right],\lambda_{it}^{\mathrm{\gls{OD}}}\,P_{ij}^{\mathrm{\gls{OD}}}\right]\,c_{ij}^{\mathrm{\gls{OD}}} & \notag\\
+ &\mathrm{max}\left[\lambda_{it}^{\mathrm{R}}\,P_{ij}^{\mathrm{R}}(1-a_i^{\mathrm{\gls{FD}}})-\lambda_{it}^{\mathrm{\gls{GW}}}\,P_{ij}^{\mathrm{\gls{GW}}}-\lambda_{it}^{\mathrm{\gls{OD}}}\,P_{ij}^{\mathrm{\gls{OD}}},0\right]\,c_{ij}^{\emptyset},& \forall (i,j)\in\bar{\mathcal{M}}^1
\end{align}
This definition reflects the outsourcing procedure explained in Remark \ref{rem:outsourcing_to_cds} for~$(i,j)\in\bar{\mathcal{M}}^1$. Moreover, it implicitly fulfils Constraints \eqref{eq:little2} and \eqref{eq:little3}, as~$a_{ij}^{\mathrm{\gls{GW}}}$ and~$a_{ij}^{\mathrm{\gls{OD}}}$ are set, at maximum, to their upper bound.

Analogously, when~$(i,j)\in\bar{\mathcal{M}}^2$, we obtain~$c_{ij}^{\emptyset,\bar{\mathcal{M}}^2}$ as
\begin{align}
c_{ij}^{\emptyset,\bar{\mathcal{M}}^2}(a_i^{\mathrm{\gls{FD}}}) = & \mathrm{min}\left[\lambda_{it}^{\mathrm{R}}\,P_{ij}^{\mathrm{R}}\,(1-a_i^{\mathrm{\gls{FD}}}),\lambda_{it}^{\mathrm{\gls{OD}}}\,P_{ij}^{\mathrm{\gls{OD}}}\right]\,c_{ij}^{\mathrm{\gls{OD}}}& \notag  \\
+ & \mathrm{min}\left[\mathrm{max}\left[\lambda_{it}^{\mathrm{R}}\,P_{ij}^{\mathrm{R}}(1-a_i^{\mathrm{\gls{FD}}})-\lambda_{it}^{\mathrm{\gls{OD}}}\,P_{ij}^{\mathrm{\gls{OD}}},0\right],\lambda_{it}^{\mathrm{\gls{GW}}}\,P_{ij}^{\mathrm{\gls{GW}}}\right]\,c_{ij}^{\mathrm{\gls{GW}}} & \notag \\
+ &\mathrm{max}\left[\lambda_{it}^{\mathrm{R}}\,P_{ij}^{\mathrm{R}}(1-a_i^{\mathrm{\gls{FD}}})-\lambda_{it}^{\mathrm{\gls{GW}}}\,P_{ij}^{\mathrm{\gls{GW}}}-\lambda_{it}^{\mathrm{\gls{OD}}}\,P_{ij}^{\mathrm{\gls{OD}}},0\right]\,c_{ij}^{\emptyset},&  \forall (i,j)\in\bar{\mathcal{M}}^2. \label{eq:alternative_penalty_1}
\end{align}
Consequently, we denote an alternative definition of penalty costs by 
\begin{align}
c_{ij}^{\emptyset'}(a_i^{\mathrm{\gls{FD}}}) = \begin{cases}
c_{ij}^{\emptyset,\bar{\mathcal{M}}^1}(a_i^{\mathrm{\gls{FD}}}), \text{if } (i,j)\in\bar{\mathcal{M}}^1 \\
c_{ij}^{\emptyset,\bar{\mathcal{M}}^2}(a_i^{\mathrm{\gls{FD}}}), \text{if } (i,j)\in\bar{\mathcal{M}}^2.  \label{eq:alternative_penalty_2}
\end{cases}
\end{align}
We can now state the optimization problem (\eqref{eq:objective_operational_level_fluid} to \eqref{eq:fluid_queue_lengths}) as

\begin{subequations}
	\begin{equation}
	\underset{e,f,a^{\mathrm{\gls{FD}}}_i}{\mathrm{min}} \sum_{i}\sum_{j} \left[a_i^{\mathrm{\gls{FD}}}\,c_{ij}^{\mathrm{\gls{FD}}}\,\lambda_{it}^{\mathrm{R}}\,P_{ij}^{\mathrm{R}}+c_{ij}^{\emptyset'}(a_i^{\mathrm{\gls{FD}}}) + c_{ij}^\mathrm{FD}\,(1-\delta_{ij})\,n^{\mathrm{FD}}_t\,e_{ij}\right] \label{eq:objective_operational_level_fluid_alt}
	\end{equation}
	\begin{align}
	(\lambda_{it}^{\mathrm{R}}/n_t^\mathrm{FD})\cdot a_i^{\mathrm{FD}}\cdot P_{ij}^{\mathrm{R}} & = \mu_{ij}\cdot f_{ij}\text{ } & \forall i,j \in \mathcal{M}  & \text{ } \tag*{\eqref{eq:little1}}\\
	\mu_{ij}\,e_{ij} \le \sum_{k}\mu_{ki} f_{ki}, \text{ }& \text{  } i \neq j & \forall i,j \in \mathcal{M}  &\text{ } \tag*{\eqref{eq:flow_cons1}}\\
	\sum_{k,\,k\neq i} \mu_{ki}\, e_{ki} \le (\lambda_{it}^{\mathrm{R}}/n_t^\mathrm{FD})\,a_i^{\mathrm{FD}} \le \sum_{k,\,k\neq i} \mu_{ki}\,e_{ki} + \sum_{k} \mu_{ki}\,f_{ki} \tag*{\eqref{eq:flow_cons2}}& \text{ }& \forall i \in \mathcal{M} & \text{ }\\
	(\lambda_{it}^{\mathrm{R}}/n_t^\mathrm{FD})\, a_i^{\mathrm{FD}} + \sum_{j,\,j\neq i} \mu_{ij}\,e_{ij} = \sum_{k,\,k\neq i} \mu_{ki}\,e_{ki} + \sum_{k} \mu_{ki}\,f_{ki} \tag*{\eqref{eq:flow_cons3}}&\text{ }&  \forall i \in \mathcal{M} & \text{ }, \\
	0\le a_i^{\mathrm{\gls{FD}}} \le 1,\quad 0\le e_{ij} \le 1,\quad 0 \le f_{ij}\le 1, \quad \sum_i\sum_j e_{ij}+ f_{ij} = 1 \label{eq:bounding2} & \text{  } & \forall i,j \in \mathcal{M} & \text{ },
	\end{align}
\end{subequations}
where the Constraints \eqref{eq:little2} and \eqref{eq:little3} are not required anymore since they are implicitly fulfilled by Equations \eqref{eq:alternative_penalty_1} and \eqref{eq:alternative_penalty_2}.

These constraints are of same type as in \cite{BravermanDaiEtAl2019} (cf. Lemma 3), with~$P_{ij} = P_{ij}^{\mathrm{R}}$ and~${\lambda_{it} = \frac{\lambda_{it}^{\mathrm{R}}}{n_t^{\mathrm{\gls{FD}}}}}$. \cite{BravermanDaiEtAl2019} add the constraints~$e_{ii}\,(1-a_i^{\mathrm{\gls{FD}}}) = 0,  \forall i \in \mathcal{M}$
to account for~$e_{ii} = 0$, if~$a_i^{\mathrm{\gls{FD}}}<1$ (queue fully utilized), or~$a_{i}^{\mathrm{\gls{FD}}} = 1$, if~$e_{ii}>0$ (queue not utilized, \glspl{FD} idling in~$i$). Adding constraint~$e_{ii}\,(1-a_i^{\mathrm{\gls{FD}}}) = 0$ does not influence the objective and is only required when we need to retrieve an empty vehicle routing policy. As our problem setting does not require retrieving the empty vehicle routing policy, we only consider Equations \eqref{eq:continuity_0} to \eqref{eq:continuity_2}.

\cite{BravermanDaiEtAl2019} show that any problem with boundary conditions \eqref{eq:little1} to \eqref{eq:availabilities} and an objective, which is 1) non-increasing in~$a_i^{\mathrm{\gls{FD}}}$, 2) non-decreasing in~$f_{ij}$, 3) non-decreasing in~$e_{ij}, i\neq j$, 4) independent of~$e_{ii}$, and 5) convex in~$(e,f)$, is a lower bound on the operational costs one would obtain using a state-dependent or independent routing policy~$Q_{ij}$. Thus, we show in the following that these five conditions hold.

\begin{enumerate}
	\item[] \underline{\textbf{Condition 1} -- Non-increasingness in~$a_i^{\mathrm{FD}}$,~$\forall i \in \mathcal{M}$}: By assumption, ${c_{ij}^{\mathrm{\gls{FD}}} < c_{ij}^{\mathrm{\gls{OD}}}}$, ${c_{ij}^{\mathrm{\gls{FD}}} < c_{ij}^{\mathrm{\gls{GW}}}}$, ${c_{ij}^{\mathrm{\gls{FD}}} < c_{ij}^{\emptyset}}$, $\forall (i,j) \in \mathcal{M}$. Therefore, any increase in~$a_i^{\mathrm{FD}}$ implies an increase in demand delivered by \glspl{FD} and consequently a decrease of more costly \gls{CD} or penalty costs. Hence, operational costs decrease. Thus, Condition 1 holds. 
	\item[] \underline{\textbf{Condition 2} -- Non-increasingness in~$f_{ij}$,~$\forall i,j \in \mathcal{M}$}: The objective function is implicitly depending on~$f_{ij}$. From Equation \eqref{eq:little1} we know that~$f_{ij}$ varies according to~$a_i^{\mathrm{\gls{FD}}}$. Moreover, we know from Condition 1 that the objective is non-increasing in~$a_i^{\mathrm{FD}}$. Consequently, the objective is non-increasing in~$f_{ij}$ and Condition 2 holds.
	\item[]  \underline{\textbf{Condition 3} -- Non-decreasingness in~$e_{ij}$,~$\forall i,j \in \mathcal{M}, i\neq j$}: The Objective \eqref{eq:objective_operational_level_fluid_alt} linearly increases with~$e_{ij}$, when~$i\neq j$. Therefore, it is non-decreasing in~$e_{ij}$.
	\item[] \underline{\textbf{Condition 4} -- Independence from~$e_{ii}$,~$\forall i \in \mathcal{M}$}: By definition,~$c_{ij}^{\mathrm{\gls{FD}}}\left(1-\delta_{ij}\right)$ is zero, when~$i=j$, therefore making the objective independent of~$e_{ii}$.
	\item[] \underline{\textbf{Condition 5} -- Convexity in~$(e,f)$}: The objective is linear in~$e$. Moreover, due to Equation \eqref{eq:little1}, if the objective is convex in~$a_i^{\mathrm{\gls{FD}}}$, it is also convex in~$f$. The objective is convex in~$a_i^{\mathrm{\gls{FD}}}$, as it has a linear~$a_i^{\mathrm{\gls{FD}}}$ term and~$c_{ij}^{\emptyset,\bar{\mathcal{M}}^1}(a_i^{\mathrm{\gls{FD}}})$ and ~$c_{ij}^{\emptyset,\bar{\mathcal{M}}^2}(a_i^{\mathrm{\gls{FD}}})$ are convex in~$a_{i}^{\mathrm{\gls{FD}}}$. Thus, Condition 5 holds.
\end{enumerate}
\hfill~$\square$
	\section{Proof of Proposition \ref{prop:convexity_total_costs}} \label{app:proof_convexity_total_costs}
Recall the total cost function~$C^{\mathrm{tot}}_t(s_t,a_t)$
\begin{equation}
C_t^{\mathrm{tot}}(s_t,a_t) = C_t^{\mathrm{ops}}(s_t,a_t)+\underset{C_t^{\mathrm{fix}}}{\underbrace{C^{\mathrm{fix}}\,\left(n_t^{\mathrm{\gls{FD}}}+a_t\right)}}+\underset{C_t^{\mathrm{sev}}}{\underbrace{C^{\mathrm{sev}}\,\mathrm{min}(a_t,0)}}.
\end{equation}
Clearly,~$C_t^{\mathrm{fix}}$ is linear. The severance payments,~$-C^{\mathrm{sev}}\,\mathrm{min}(a_t,0)$, consist of a piecewise-linear convex curve with negative slope for~$a_t<0$ with and~$0$ for~$a_t\ge0$. 

It remains to show that~$C_t^{\mathrm{ops}}$ is convex in~$n_t^{\mathrm{\gls{FD}}}$. We do so by showing that~$C_t^{\mathrm{ops}}$ has an increasing slope in~$n_t^{\mathrm{\gls{FD}}}$. First, note, that~$C_t^{\mathrm{ops}}(n_t^{\mathrm{\gls{FD}}})$ decreases monotonically in~$n_t^{\mathrm{\gls{FD}}}$. As we minimize in Equation \eqref{eq:objective_operational_level_fluid_alt}, a solution for a fleet with~$n_t^{\mathrm{\gls{FD}}}+1$ \glspl{FD} is optimal either if~$C_t^{\mathrm{ops}}(n_t^{\mathrm{\gls{FD}}}+1)<C_t^{\mathrm{ops}}(n_t^{\mathrm{\gls{FD}}})$ or if~$C_t^{\mathrm{ops}}(n_t^{\mathrm{\gls{FD}}}+1)=C_t^{\mathrm{ops}}(n_t^{\mathrm{\gls{FD}}})$. In the latter case, the additional \gls{FD} fluid is entirely attributed to some~$e_{ii}$.

Let us now denote the cost reduction from adding one \gls{FD} to a fleet of~$n_t^{\mathrm{\gls{FD}}}$ \glspl{FD} by
\begin{equation}
\Delta_{n_t^{\mathrm{\gls{FD}}}} C_t^{\mathrm{ops}}(1) = C_t^{\mathrm{ops}}(n_t^{\mathrm{\gls{FD}}}+1)-C_t^{\mathrm{ops}}(n_t^{\mathrm{\gls{FD}}}).
\end{equation} 
As Equation \eqref{eq:objective_operational_level_fluid_alt} yields the minimum cost for a fleet with~$n_t^{\mathrm{\gls{FD}}}+1$ \glspl{FD}, the incremental cost reduction when adding one \gls{FD} to a fleet with~$n_t^{\mathrm{\gls{FD}}}$ \glspl{FD},~$\Delta_{n_t^{\mathrm{\gls{FD}}}} C_t^{\mathrm{ops}}(1)$, is the maximum cost reduction one can achieve with one additional \gls{FD}. To prove that~$C_t^{\mathrm{ops}}$ is convex in the number of \glspl{FD}, we need to show that
\begin{equation}
\Delta_{n_t^{\mathrm{\gls{FD}}}} C_t^{\mathrm{ops}}(1)\ge \Delta_{n_t^{\mathrm{\gls{FD}}}+1} C_t^{\mathrm{ops}}(1).
\end{equation}

We do so by showing that~$\Delta_{n_t^{\mathrm{\gls{FD}}}} C_t^{\mathrm{ops}}(1)< \Delta_{n_t^{\mathrm{\gls{FD}}}+1} C_t^{\mathrm{ops}}(1)$ leads to a contradiction. First, note that all feasible cost reductions,~$\Delta_{n_t^{\mathrm{\gls{FD}}}+m}C^{\mathrm{ops}}_t(1)$ obtainable by adding one \gls{FD} to a fleet of~$n_t^{\mathrm{\gls{FD}}}+m$ \glspl{FD} ($m\in\mathbb{N}$), are also feasible for a fleet of~$n_t^{\mathrm{\gls{FD}}}$ \glspl{FD}. If~$\Delta_{n_t^{\mathrm{FD}}+1} C_t^{\mathrm{ops}}(1)>\Delta_{n_t^{\mathrm{\gls{FD}}}} C_t^{\mathrm{ops}}(1)$ were possible, the following would not be optimal
\begin{equation}
{C_t^{\mathrm{ops}}(n_t^{\mathrm{\gls{FD}}}+1) = C_t^{\mathrm{ops}}(n_t^{\mathrm{\gls{FD}}})+\Delta_{n_t^{\mathrm{\gls{FD}}}} C_t^{\mathrm{ops}}(1)}.
\end{equation} 
As $\Delta_{n_t^{\mathrm{\gls{FD}}}+1} C_t^{\mathrm{ops}}(1)$ is feasible as well, applying it would lead to lower~$C_t^{\mathrm{ops}}(n_t^{\mathrm{\gls{FD}}}+1)$. This contradicts the minimum objective in Equation \eqref{eq:objective_operational_level_fluid_alt}. Therefore~$\Delta_{n_t^{\mathrm{\gls{FD}}}} C_t^{\mathrm{ops}}(1)\ge \Delta_{n_t^{\mathrm{\gls{FD}}}+1} C_t^{\mathrm{ops}}(1)$.

Consequently,~$C_t^{\mathrm{ops}}$ has a decreasing slope in~$n_t^{\mathrm{\gls{FD}}}$ and is, thus, convex in~$n_t^{\mathrm{\gls{FD}}}$. The summation of convex terms remains convex, therefore~$C_t^{\mathrm{tot}}$ is convex in~$n_t^{\mathrm{\gls{FD}}}$.
\hfill~$\square$
	
	\section{Proof of Proposition \ref{prop:convexity_value_function}} \label{app:proof_convexity_value_function}
	Let us recall the definition of the post-decision state's value function,
	\begin{equation}
	V_t^{a}(n_t^{\mathrm{\gls{FD}}}+a_t,\cdot) = \mathbb{E}_{\tilde{x}^{\alpha} \sim \mathcal{X}^{\alpha}, \tilde{y}^{\alpha}\sim \mathcal{Y}^{\alpha}}\left[V(n_{t+1}^{\mathrm{\gls{FD}}},\cdot)|s_t^{a_t}\right],
	\end{equation}
	where~$s_t^{a_t} = (n_t^{\mathrm{\gls{FD}}}+a_t,\cdot)$ and~$\cdot$ denotes the current \gls{CD} fleet composition, i.e., $V_t^{a}(n_t^{\mathrm{\gls{FD}}}+a_t,\cdot) = V_t^{a}(n_t^{\mathrm{\gls{FD}}}+a_t,n_t^{\mathrm{\gls{GW}}},n_t^{\mathrm{\gls{OD}}})$.
	
	We seek to prove that the value function of the post-decision state is convex in~$a_t$ by induction.
	
	Let us consider time step~$t=T+1$ in the induction basis. Here, the pre-decision state~$V_{T+1}(n_{T+1}^{\mathrm{\gls{FD}}},\cdot)$ and the post-decision state~$V_{T+1}^{a_{T+1}}(n_{T+1}^{\mathrm{\gls{FD}}}+a_{T+1},\cdot)$ are both equal to zero, as the \gls{MDP} is finite. Hence, they are trivially convex in the number of \glspl{FD}. Now, we suppose~$V_{t+1}(n_{t+1}^{\mathrm{\gls{FD}}},\cdot)$ and~$V_{t+1}^{a}(n_{t+1}^{\mathrm{\gls{FD}}}+a_{t+1},\cdot)$ are convex in the number of \glspl{FD}. Then, by showing that they are also convex in~$t$, convexity along the \gls{FD} dimension holds for every~$t$. The post-decision state's value in time step~$t$ reads
	\begin{align}
	V_{t}^{a_t}(n_{t}^{\mathrm{\gls{FD}}}+{a_{t}},\cdot) &= \mathbb{E}_{\tilde{x}^{\alpha} \sim \mathcal{X}^{\alpha}, \tilde{y}^{\alpha}\sim \mathcal{Y}^{\alpha}}\left[V_{t}(n_{t+1}^{\mathrm{\gls{FD}}},\cdot)|s_{t}^{a_t}\right] \\
	&=\mathbb{E}_{\tilde{x}^{\alpha} \sim \mathcal{X}^{\alpha}, \tilde{y}^{\alpha}\sim \mathcal{Y}^{\alpha}}\left[\underset{a_{t+1}}{\mathrm{min}}\left(C_{t+1}^{\mathrm{tot}}(n_{t+1}^{\mathrm{\gls{FD}}},a_{t+1},\cdot)+\gamma\mathbb{E}_{\tilde{y}^{\alpha}\sim \mathcal{Y}^{\alpha}, \tilde{x}^{\alpha}\sim \mathcal{X}^{\alpha}}\left[V_{t+2}(n_{t+2}^{\mathrm{\gls{FD}}},\cdot)|s_{t+1}^{a_{t+1}}\right]\right)|s_{t}^{a_t}\right]\\
	& =  \mathbb{E}_{\tilde{x}^{\alpha} \sim \mathcal{X}^{\alpha}, \tilde{y}^{\alpha}\sim \mathcal{Y}^{\alpha}}\left[\underset{a_{t+1}}{\mathrm{min}}\left(C_{t+1}^{\mathrm{tot}}(n_{t+1}^{\mathrm{\gls{FD}}},a_{t+1},\cdot)+\gamma\,V_{t+1}^{a_{t+1}}(n_{t+1}^{\mathrm{\gls{FD}}}+a_{t+1},\cdot)\right)|s_{t}^{a_t}\right].
	\end{align}
	
	Due to Proposition \ref{prop:convexity_total_costs}, we know that~$C_{t+1}^{\mathrm{tot}}$ is convex in the \gls{FD} dimension. Hence, we know that~$C_{t+1}^{\mathrm{tot}}(n_{t+1}^{\mathrm{\gls{FD}}},a_{t+1},\cdot)$ is convex in both~$n_{t+1}^{\mathrm{\gls{FD}}}$ and~$a_{t+1}$, as~$C_{t+1}^{\mathrm{tot}}(n_{t+1}^{\mathrm{\gls{FD}}},a_{t+1},\cdot) = C_{t+1}^{\mathrm{tot}}(n_{t+1}^{\mathrm{\gls{FD}}}+a_{t+1},\cdot)$. The same holds for~$V_{t+1}^{a_{t+1}}$, which we know is convex in the \gls{FD} dimension (induction basis).
	
	Since the sum of two convex functions is again convex,~$\underset{a_{t+1}}{\mathrm{min}}\left(C_{t+1}^{\mathrm{tot}}(n_{t+1}^{\mathrm{\gls{FD}}},a_{t+1},\cdot)+\gamma\,V_{t+1}^{a_{t+1}}(n_{t+1}^{\mathrm{\gls{FD}}},a_{t+1},\cdot)\right)$ is convex in~$n_{t+1}^{\mathrm{\gls{FD}}}$ \citep[Theorem 5.3]{Rockafellar1970}. As~$n_{t+1}^{\mathrm{\gls{FD}}} = n_{t}^{\mathrm{\gls{FD}}} +a_{t}$, which is linear in~$a_{t}$, it follows that~$\underset{a_{t+1}}{\mathrm{min}}\left(C_{t+1}^{\mathrm{tot}}(n_{t+1}^{\mathrm{\gls{FD}}},a_{t+1},\cdot)+\gamma\,V_{t+1}^{a_{t+1}}(n_{t+1}^{\mathrm{\gls{FD}}},a_{t+1},\cdot)\right)$ is convex in~$a_{t}$ \citep[Theorem 5.7]{Rockafellar1970}. 
	
	Moreover, Proposition  \ref{prop:convexity_total_costs} holds for every outcome of~$\tilde{y}^{\alpha}$ and~$\tilde{x}^{\alpha}$ and the outcome of~$\tilde{y}^{\alpha}$ and~$\tilde{x}^{\alpha}$ does not depend on~$a_t$. Finally, since the integral of a convex function is again convex,~$V_{t}^{a_{t}}(n_{t}^{\mathrm{\gls{FD}}}+a_{t},\cdot)$ is convex in~$a_{t}$. Consequently~$V_{t}^{a_{t}}(s_{t}^{a_t})$ is convex in the action~$a_{t}$ for all~$t$.
	\hfill~$\square$
	\FloatBarrier
\section{Adequateness of discretization.}\label{app:adequateness_discretization}
\glspl{OD} starting their journey within a \location{} need to perform a detour of on average three minutes both at the origin and destination \location{} to pick up the request. We obtain this detour duration by dividing the average distance between any two points within a square by the average travel speed. The average distance between any two points within a square can be approximated by half of the square's side length. A representative survey suggests that private individuals willing to act as \glspl{OD} are eager to perform a maximum detour of roughly 10 minutes when participating in a crowdshipping platform \citep{le2019}. In fact, for a discretization with squares of~$4\,\mathrm{km}^2$ surface, the average required detour for \glspl{OD} remains below~10 minutes up to a travel speed of~$12\,\mathrm{km/h}$, which would allow us to consider more congested settings such as New York City where average travel speeds across multiple transportation modes are instead in between~5-15$\,$km/h \citep{NYCDT2019}.
\section{Request patterns, \gls{OD} patterns, origin-destination matrix}\label{app:patterns}
This section provides the data we extracted from the Grubhub data set. We start by describing the route patterns of the requests,~$P_{ij}^{\mathrm{R}}$
\begin{tiny}
\begin{equation}
P_{ij}^{\mathrm{R}} = \begin{bmatrix}
1.00 & 0.00 & 0.00 & 0.00 & 0.00 & 0.00 & 0.00 & 0.00 & 0.00 & 0.00 & 0.00 & 0.00 & 0.00 & 0.00 & 0.00 & 0.00 & 0.00 & 0.00 \\
0.00 & 0.00 & 1.00 & 0.00 & 0.00 & 0.00 & 0.00 & 0.00 & 0.00 & 0.00 & 0.00 & 0.00 & 0.00 & 0.00 & 0.00 & 0.00 & 0.00 & 0.00 \\
0.00 & 0.20 & 0.40 & 0.40 & 0.00 & 0.00 & 0.00 & 0.00 & 0.00 & 0.00 & 0.00 & 0.00 & 0.00 & 0.00 & 0.00 & 0.00 & 0.00 & 0.00 \\
0.33 & 0.00 & 0.33 & 0.33 & 0.00 & 0.00 & 0.00 & 0.00 & 0.00 & 0.00 & 0.00 & 0.00 & 0.00 & 0.00 & 0.00 & 0.00 & 0.00 & 0.00 \\
0.00 & 0.00 & 0.10 & 0.00 & 0.00 & 0.00 & 0.00 & 0.30 & 0.40 & 0.20 & 0.00 & 0.00 & 0.00 & 0.00 & 0.00 & 0.00 & 0.00 & 0.00 \\
0.00 & 0.00 & 0.00 & 0.00 & 0.50 & 0.00 & 0.50 & 0.00 & 0.00 & 0.00 & 0.00 & 0.00 & 0.00 & 0.00 & 0.00 & 0.00 & 0.00 & 0.00 \\
0.00 & 0.18 & 0.00 & 0.18 & 0.09 & 0.00 & 0.00 & 0.27 & 0.18 & 0.09 & 0.00 & 0.00 & 0.00 & 0.00 & 0.00 & 0.00 & 0.00 & 0.00 \\
0.00 & 0.09 & 0.18 & 0.00 & 0.00 & 0.00 & 0.00 & 0.09 & 0.36 & 0.18 & 0.09 & 0.00 & 0.00 & 0.00 & 0.00 & 0.00 & 0.00 & 0.00 \\
0.00 & 0.00 & 0.00 & 0.00 & 0.14 & 0.29 & 0.00 & 0.00 & 0.57 & 0.00 & 0.00 & 0.00 & 0.00 & 0.00 & 0.00 & 0.00 & 0.00 & 0.00 \\
0.00 & 0.00 & 0.00 & 0.00 & 0.00 & 0.00 & 0.00 & 0.00 & 0.00 & 1.00 & 0.00 & 0.00 & 0.00 & 0.00 & 0.00 & 0.00 & 0.00 & 0.00 \\
0.00 & 0.00 & 0.00 & 0.00 & 0.00 & 0.00 & 0.00 & 1.00 & 0.00 & 0.00 & 0.00 & 0.00 & 0.00 & 0.00 & 0.00 & 0.00 & 0.00 & 0.00 \\
0.00 & 0.00 & 0.00 & 0.00 & 0.00 & 0.00 & 0.00 & 0.17 & 0.83 & 0.00 & 0.00 & 0.00 & 0.00 & 0.00 & 0.00 & 0.00 & 0.00 & 0.00 \\
0.00 & 0.07 & 0.00 & 0.07 & 0.07 & 0.00 & 0.00 & 0.40 & 0.13 & 0.13 & 0.07 & 0.07 & 0.00 & 0.00 & 0.00 & 0.00 & 0.00 & 0.00 \\
0.00 & 0.00 & 0.11 & 0.00 & 0.00 & 0.11 & 0.00 & 0.11 & 0.22 & 0.33 & 0.11 & 0.00 & 0.00 & 0.00 & 0.00 & 0.00 & 0.00 & 0.00 \\
0.00 & 0.00 & 0.00 & 0.00 & 0.00 & 0.00 & 0.00 & 0.00 & 0.00 & 1.00 & 0.00 & 0.00 & 0.00 & 0.00 & 0.00 & 0.00 & 0.00 & 0.00 \\
0.00 & 0.00 & 0.00 & 0.00 & 0.00 & 0.00 & 0.00 & 0.60 & 0.20 & 0.00 & 0.20 & 0.00 & 0.00 & 0.00 & 0.00 & 0.00 & 0.00 & 0.00 \\
0.00 & 0.00 & 0.00 & 0.00 & 0.00 & 0.00 & 0.00 & 0.33 & 0.67 & 0.00 & 0.00 & 0.00 & 0.00 & 0.00 & 0.00 & 0.00 & 0.00 & 0.00 \\
0.00 & 0.00 & 0.00 & 0.00 & 0.00 & 0.29 & 0.00 & 0.14 & 0.29 & 0.29 & 0.00 & 0.00 & 0.00 & 0.00 & 0.00 & 0.00 & 0.00 & 0.00  
\end{bmatrix}.
\end{equation}
\end{tiny}
Next, we describe the spatial distribution of request arrivals via normalized request arrival rates.
\begin{tiny}
\begin{equation}
\frac{\lambda_{it}^{\mathrm{R}}}{\sum_i^{|\mathcal{M}|}\lambda_{it}^{\mathrm{R}}} = \begin{bmatrix}
0.02 &
0.01 &
0.05 &
0.03 &
0.10 &
0.02 &
0.11 &
0.11 &
0.07 &
0.01 &
0.01 &
0.06 &
0.15 &
0.09 &
0.01 &
0.05 &
0.06 &
0.07 
\end{bmatrix}.
\end{equation}
\end{tiny}
We randomly sample \gls{OD} route patterns,~$P^{\mathrm{\gls{OD}}}_{ij}$, randomly, which results in
\begin{tiny}
\begin{equation}
P_{ij}^{\gls{OD}} = \begin{bmatrix}
0.02 & 0.02 & 0.09 & 0.11 & 0.03 & 0.08 & 0.10 & 0.10 & 0.01 & 0.00 & 0.02 & 0.10 & 0.01 & 0.05 & 0.11 & 0.06 & 0.08 & 0.03 \\
0.08 & 0.10 & 0.00 & 0.09 & 0.12 & 0.09 & 0.03 & 0.09 & 0.01 & 0.05 & 0.11 & 0.03 & 0.03 & 0.02 & 0.00 & 0.08 & 0.03 & 0.03 \\
0.06 & 0.01 & 0.07 & 0.02 & 0.07 & 0.08 & 0.01 & 0.05 & 0.08 & 0.05 & 0.01 & 0.06 & 0.08 & 0.06 & 0.11 & 0.07 & 0.11 & 0.02 \\
0.01 & 0.08 & 0.04 & 0.02 & 0.09 & 0.03 & 0.07 & 0.07 & 0.08 & 0.06 & 0.07 & 0.03 & 0.03 & 0.08 & 0.04 & 0.09 & 0.06 & 0.06 \\
0.01 & 0.10 & 0.05 & 0.06 & 0.04 & 0.03 & 0.10 & 0.06 & 0.00 & 0.07 & 0.03 & 0.06 & 0.09 & 0.04 & 0.10 & 0.07 & 0.00 & 0.10 \\
0.08 & 0.11 & 0.02 & 0.02 & 0.11 & 0.08 & 0.01 & 0.09 & 0.09 & 0.11 & 0.08 & 0.01 & 0.00 & 0.00 & 0.00 & 0.03 & 0.10 & 0.06 \\
0.06 & 0.10 & 0.01 & 0.03 & 0.07 & 0.11 & 0.06 & 0.00 & 0.09 & 0.03 & 0.09 & 0.04 & 0.10 & 0.09 & 0.06 & 0.02 & 0.01 & 0.01 \\
0.01 & 0.01 & 0.03 & 0.09 & 0.07 & 0.00 & 0.01 & 0.12 & 0.07 & 0.03 & 0.03 & 0.10 & 0.03 & 0.07 & 0.12 & 0.11 & 0.03 & 0.06 \\
0.07 & 0.10 & 0.02 & 0.00 & 0.01 & 0.06 & 0.07 & 0.07 & 0.04 & 0.12 & 0.07 & 0.05 & 0.07 & 0.09 & 0.08 & 0.03 & 0.01 & 0.04 \\
0.07 & 0.02 & 0.08 & 0.01 & 0.03 & 0.09 & 0.02 & 0.07 & 0.06 & 0.10 & 0.03 & 0.01 & 0.08 & 0.09 & 0.10 & 0.10 & 0.00 & 0.03 \\
0.06 & 0.09 & 0.09 & 0.05 & 0.08 & 0.06 & 0.04 & 0.04 & 0.06 & 0.05 & 0.09 & 0.09 & 0.04 & 0.09 & 0.02 & 0.01 & 0.01 & 0.05 \\
0.00 & 0.10 & 0.08 & 0.00 & 0.02 & 0.03 & 0.01 & 0.08 & 0.04 & 0.10 & 0.06 & 0.09 & 0.09 & 0.09 & 0.05 & 0.06 & 0.08 & 0.03 \\
0.05 & 0.06 & 0.00 & 0.06 & 0.05 & 0.09 & 0.03 & 0.10 & 0.06 & 0.02 & 0.08 & 0.07 & 0.01 & 0.04 & 0.07 & 0.10 & 0.00 & 0.10 \\
0.03 & 0.09 & 0.05 & 0.08 & 0.05 & 0.06 & 0.03 & 0.05 & 0.07 & 0.08 & 0.07 & 0.06 & 0.08 & 0.03 & 0.06 & 0.04 & 0.03 & 0.04 \\
0.04 & 0.03 & 0.06 & 0.04 & 0.09 & 0.07 & 0.02 & 0.04 & 0.03 & 0.08 & 0.09 & 0.09 & 0.06 & 0.03 & 0.02 & 0.08 & 0.05 & 0.08 \\
0.06 & 0.08 & 0.06 & 0.08 & 0.06 & 0.05 & 0.04 & 0.01 & 0.04 & 0.01 & 0.11 & 0.02 & 0.09 & 0.09 & 0.07 & 0.06 & 0.02 & 0.05 \\
0.04 & 0.03 & 0.08 & 0.04 & 0.12 & 0.12 & 0.03 & 0.01 & 0.02 & 0.06 & 0.09 & 0.03 & 0.03 & 0.11 & 0.05 & 0.08 & 0.03 & 0.01 \\
0.05 & 0.05 & 0.02 & 0.06 & 0.06 & 0.07 & 0.01 & 0.08 & 0.02 & 0.05 & 0.08 & 0.00 & 0.03 & 0.09 & 0.09 & 0.05 & 0.09 & 0.10
\end{bmatrix}
\end{equation}
\end{tiny}
Similarly, we randomly sample \gls{OD} arrival patterns~$I^{\mathrm{\gls{OD}}}$, which results in
\begin{tiny}
	\begin{equation}
I^{\mathrm{\gls{OD}}}_i = \begin{bmatrix}
0.06 &
0.10 &
0.00 &
0.04 &
0.02 &
0.01 &
0.03 &
0.05 &
0.06 &
0.08 &
0.06 &
0.10 &
0.03 &
0.13 &
0.00 &
0.10 &
0.06 &
0.08
	\end{bmatrix} 
	\end{equation}
\end{tiny}
Finally, we detail the origin-destination matrix resulting from the discretization of the Grubhub data set
\begin{tiny}
\begin{equation}
v^{\mathrm{avg}}\cdot \mu_{ij} = r_{ij} = \begin{bmatrix}
2.0 & 4.5 & 4.0 & 6.3 & 6.0 & 6.3 & 8.9 & 8.2 & 8.0 & 8.2 & 10.0 & 10.2 & 12.6 & 12.2 & 12.0 & 12.2 & 12.6 & 13.4 \\
2.0 & 2.0 & 2.8 & 4.0 & 4.5 & 5.7 & 6.3 & 6.0 & 6.3 & 7.2 & 8.2 & 8.9 & 10.2 & 10.0 & 10.2 & 10.8 & 11.7 & 12.8 \\
1.0 & 2.8 & 2.0 & 4.5 & 4.0 & 4.5 & 7.2 & 6.3 & 6.0 & 6.3 & 8.0 & 8.2 & 10.8 & 10.2 & 10.0 & 10.2 & 10.8 & 11.7 \\
2.8 & 1.0 & 2.0 & 2.0 & 2.8 & 4.5 & 4.5 & 4.0 & 4.5 & 5.7 & 6.3 & 7.2 & 8.2 & 8.0 & 8.2 & 8.9 & 10.0 & 11.3 \\
2.0 & 2.0 & 1.0 & 2.8 & 2.0 & 2.8 & 5.7 & 4.5 & 4.0 & 4.5 & 6.0 & 6.3 & 8.9 & 8.2 & 8.0 & 8.2 & 8.9 & 10.0 \\
5.7 & 2.8 & 4.5 & 2.0 & 4.0 & 6.0 & 2.0 & 2.8 & 4.5 & 6.3 & 5.7 & 7.2 & 6.0 & 6.3 & 7.2 & 8.5 & 10.0 & 11.7 \\
4.5 & 2.0 & 2.8 & 1.0 & 2.0 & 4.0 & 2.8 & 2.0 & 2.8 & 4.5 & 4.5 & 5.7 & 6.3 & 6.0 & 6.3 & 7.2 & 8.5 & 10.0 \\
4.0 & 2.8 & 2.0 & 2.0 & 1.0 & 2.0 & 4.5 & 2.8 & 2.0 & 2.8 & 4.0 & 4.5 & 7.2 & 6.3 & 6.0 & 6.3 & 7.2 & 8.5 \\
4.5 & 4.5 & 2.8 & 4.0 & 2.0 & 1.0 & 6.3 & 4.5 & 2.8 & 2.0 & 4.5 & 4.0 & 8.5 & 7.2 & 6.3 & 6.0 & 6.3 & 7.2 \\
5.7 & 6.3 & 4.5 & 6.0 & 4.0 & 2.0 & 8.2 & 6.3 & 4.5 & 2.8 & 5.7 & 4.5 & 10.0 & 8.5 & 7.2 & 6.3 & 6.0 & 6.3 \\
7.2 & 4.5 & 5.7 & 2.8 & 4.5 & 6.3 & 1.0 & 2.0 & 4.0 & 6.0 & 4.5 & 6.3 & 4.0 & 4.5 & 5.7 & 7.2 & 8.9 & 10.8 \\
6.3 & 4.0 & 4.5 & 2.0 & 2.8 & 4.5 & 2.0 & 1.0 & 2.0 & 4.0 & 2.8 & 4.5 & 4.5 & 4.0 & 4.5 & 5.7 & 7.2 & 8.9 \\
6.0 & 4.5 & 4.0 & 2.8 & 2.0 & 2.8 & 4.0 & 2.0 & 1.0 & 2.0 & 2.0 & 2.8 & 5.7 & 4.5 & 4.0 & 4.5 & 5.7 & 7.2 \\
6.3 & 5.7 & 4.5 & 4.5 & 2.8 & 2.0 & 6.0 & 4.0 & 2.0 & 1.0 & 2.8 & 2.0 & 7.2 & 5.7 & 4.5 & 4.0 & 4.5 & 5.7 \\
7.2 & 7.2 & 5.7 & 6.3 & 4.5 & 2.8 & 8.0 & 6.0 & 4.0 & 2.0 & 4.5 & 2.8 & 8.9 & 7.2 & 5.7 & 4.5 & 4.0 & 4.5 \\
8.2 & 6.0 & 6.3 & 4.0 & 4.5 & 5.7 & 2.8 & 2.0 & 2.8 & 4.5 & 2.0 & 4.0 & 2.8 & 2.0 & 2.8 & 4.5 & 6.3 & 8.2 \\
8.0 & 6.3 & 6.0 & 4.5 & 4.0 & 4.5 & 4.5 & 2.8 & 2.0 & 2.8 & 1.0 & 2.0 & 4.5 & 2.8 & 2.0 & 2.8 & 4.5 & 6.3 \\
8.2 & 7.2 & 6.3 & 5.7 & 4.5 & 4.0 & 6.3 & 4.5 & 2.8 & 2.0 & 2.0 & 1.0 & 6.3 & 4.5 & 2.8 & 2.0 & 2.8 & 4.5 
\end{bmatrix}
\end{equation}
\end{tiny}
\FloatBarrier
\section{Detailed validation results for \gls{PL-VFA} and smaller instance description} \label{app:boxplots_validation}
\begin{figure}[htbp]
	\subfigure[$(n_0^{\mathrm{\gls{FD}}},n_0^{\mathrm{\gls{GW}}},n_0^{\mathrm{\gls{OD}}}) = (0,6,0)$.]{
		\includegraphics[width=0.3\textwidth]{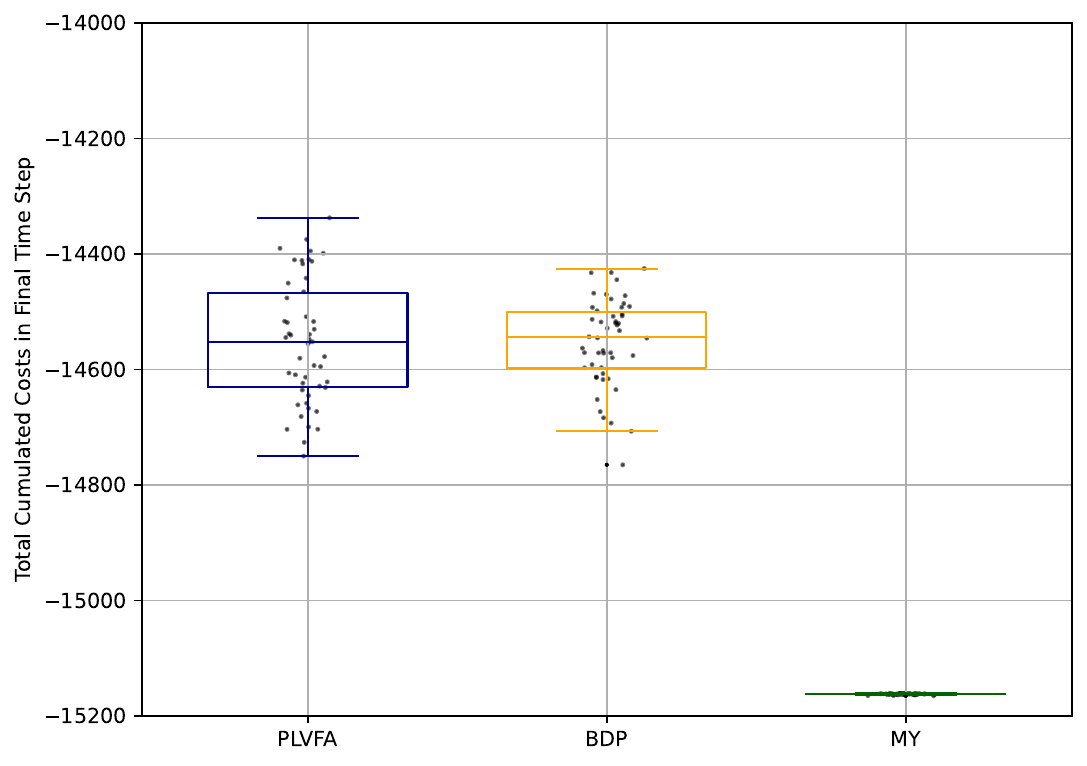}}
	\subfigure[$(n_0^{\mathrm{\gls{FD}}},n_0^{\mathrm{\gls{GW}}},n_0^{\mathrm{\gls{OD}}}) = (0,9,0)$.]{
		\includegraphics[width=0.3\textwidth]{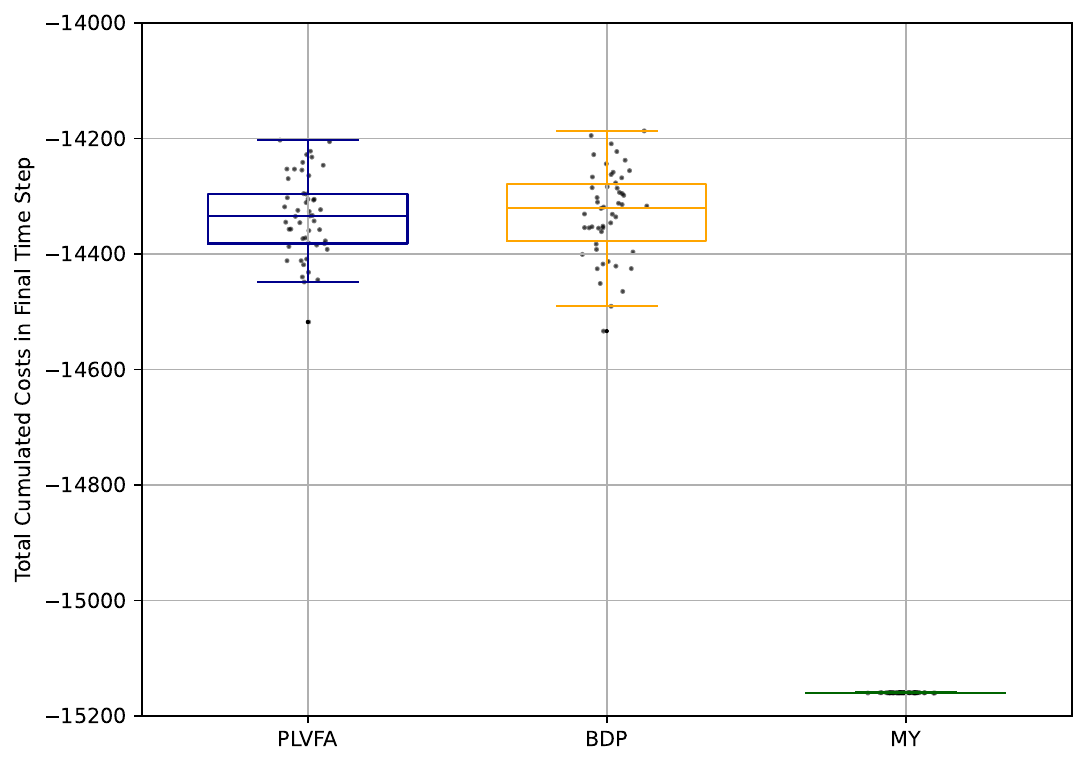}}
	\subfigure[$(n_0^{\mathrm{\gls{FD}}},n_0^{\mathrm{\gls{GW}}},n_0^{\mathrm{\gls{OD}}}) = (0,12,0)$.]{
		\includegraphics[width=0.3\textwidth]{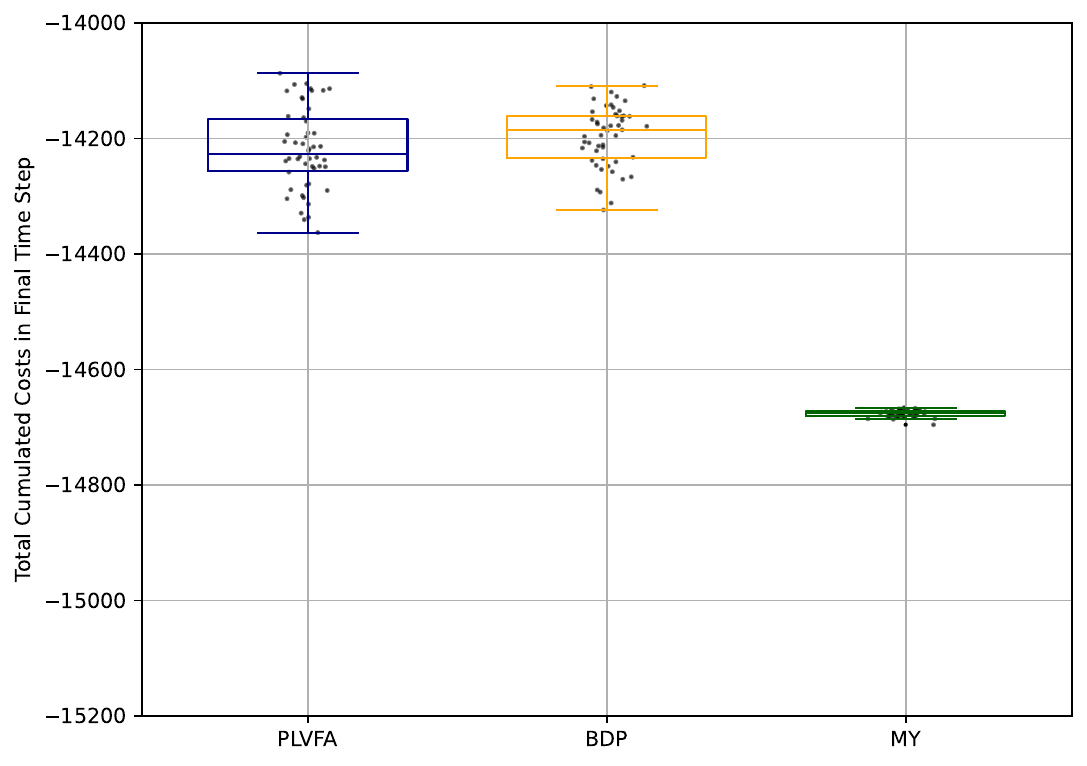}}
	\caption{Constant demand scenario: Total cumulated costs in~$t=T$ obtained by \gls{PL-VFA}, \gls{bdp}, and \gls{MY}. The \gls{PL-VFA} algorithm required~$10\,\mathrm{k}$ iterations to converge to the final policy.}
	\label{fig:constant_demand}
\end{figure}

\begin{figure}[htbp]
	\subfigure[$(n_0^{\mathrm{\gls{FD}}},n_0^{\mathrm{\gls{GW}}},n_0^{\mathrm{\gls{OD}}}) = (0,6,0)$.]{
		\includegraphics[width=0.3\textwidth]{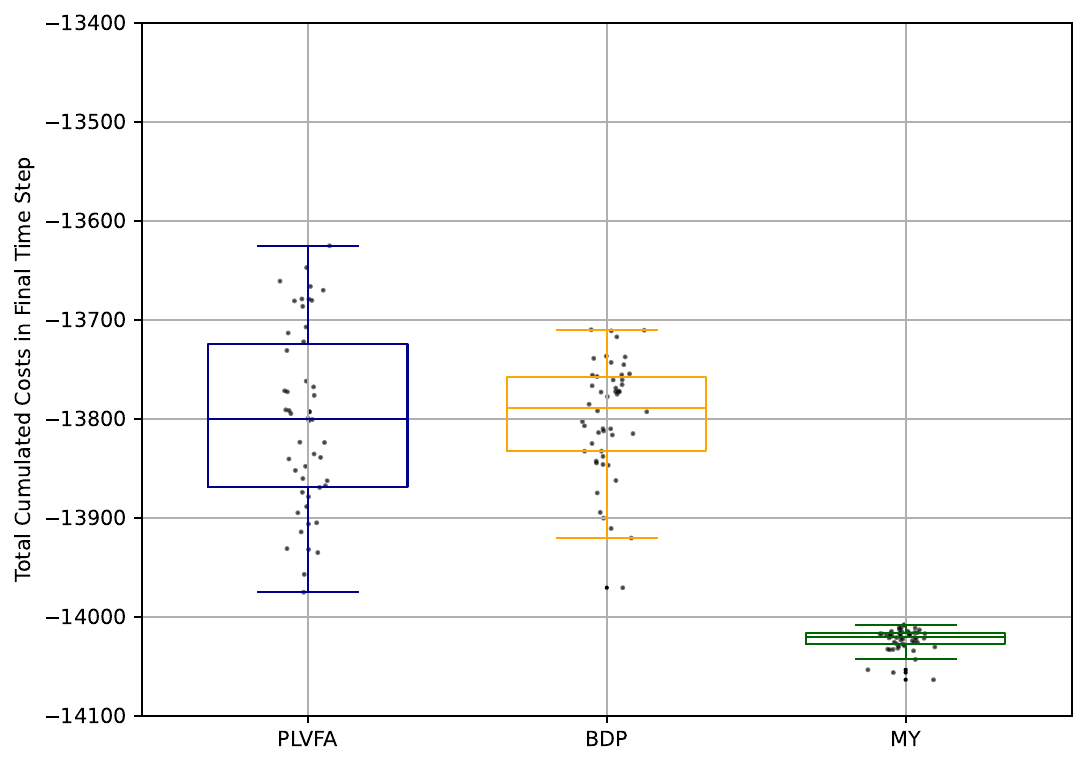}}
	\subfigure[$(n_0^{\mathrm{\gls{FD}}},n_0^{\mathrm{\gls{GW}}},n_0^{\mathrm{\gls{OD}}}) = (0,9,0)$.]{
		\includegraphics[width=0.3\textwidth]{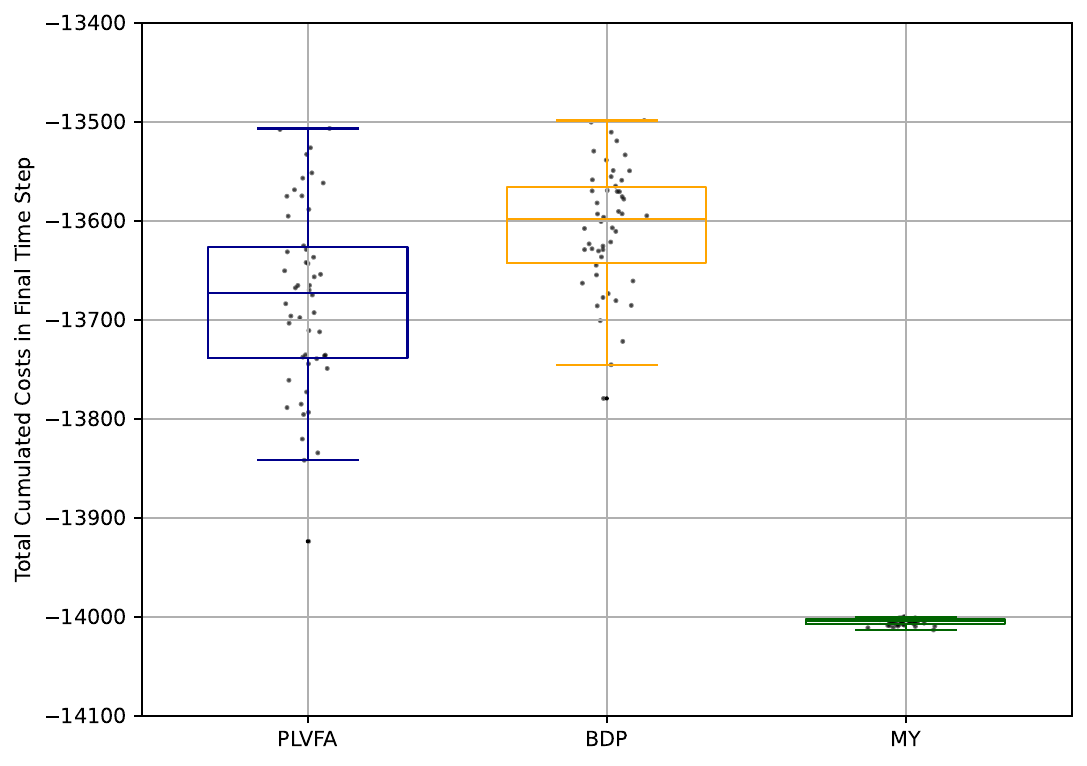}}
	\subfigure[$(n_0^{\mathrm{\gls{FD}}},n_0^{\mathrm{\gls{GW}}},n_0^{\mathrm{\gls{OD}}}) = (0,12,0)$.]{
		\includegraphics[width=0.3\textwidth]{figures/Results/inp009_growth_fleet30percentlower_validation.pdf}}
	\caption{Growth demand scenario: Total cumulated costs in~$t=T$ obtained by \gls{PL-VFA}, \gls{bdp}, and \gls{MY}. The \gls{PL-VFA} algorithm required~$10\,\mathrm{k}$ iterations to converge to the final policy.}
	\label{fig:growth_demand}
\end{figure}

\begin{figure}[htbp]
	\subfigure[$(n_0^{\mathrm{\gls{FD}}},n_0^{\mathrm{\gls{GW}}},n_0^{\mathrm{\gls{OD}}}) = (0,6,0)$.]{
		\includegraphics[width=0.3\textwidth]{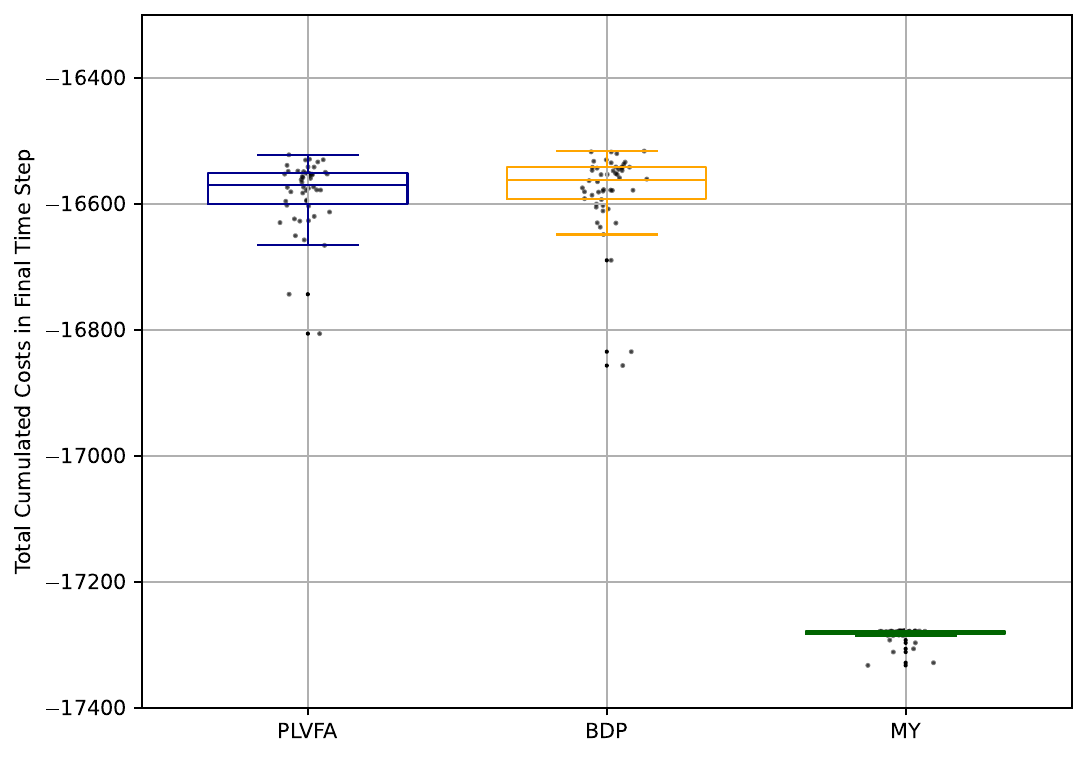}}
	\subfigure[$(n_0^{\mathrm{\gls{FD}}},n_0^{\mathrm{\gls{GW}}},n_0^{\mathrm{\gls{OD}}}) = (0,9,0)$.]{
		\includegraphics[width=0.3\textwidth]{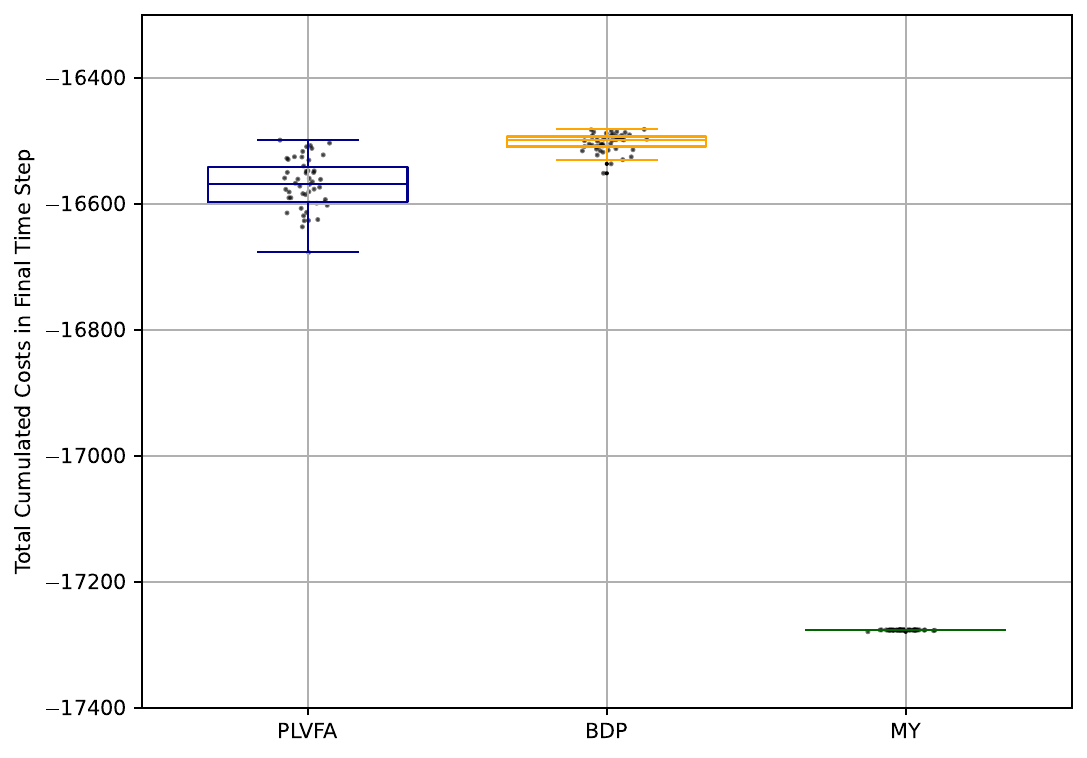}}
	\subfigure[$(n_0^{\mathrm{\gls{FD}}},n_0^{\mathrm{\gls{GW}}},n_0^{\mathrm{\gls{OD}}}) = (0,12,0)$.]{
		\includegraphics[width=0.3\textwidth]{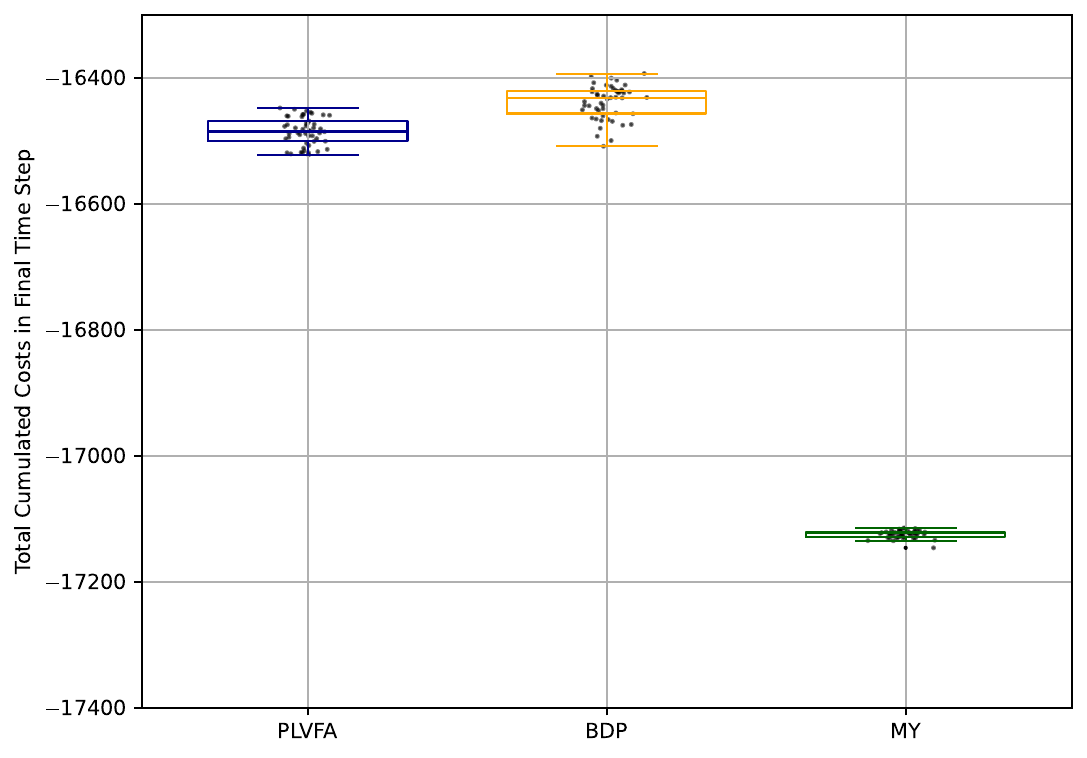}}
	\caption{Peak demand scenario: Total cumulated costs in~$t=T$ obtained by \gls{PL-VFA}, \gls{bdp}, and \gls{MY}. The \gls{PL-VFA} algorithm required~$5\,\mathrm{k}$ iterations to converge to the final policy.}
	\label{fig:peak_demand}
\end{figure}

In this section, we provide detailed box-and-whisker plots of the results in Section \ref{sec:validation_plvfa}. To this end, we report the total cumulated costs of the constant demand scenario in Figure \ref{fig:constant_demand}, of the growth demand scenario in Figure \ref{fig:growth_demand}, and of the peak demand scenario in Figure~\ref{fig:peak_demand}. To account for the smaller instance sizes, i.e. the fewer drivers, we altered the instance settings described in Section \ref{sec:design_of_experiments}. We reduced the total demand from~3,000 to 150 orders. In the constant demand scenario the total demand therefore is~${\sum_{i\in\mathcal{M}}\lambda_{it}^{\mathrm{R}} = 150,\,\forall t\in\mathcal{T}}$. In the growth demand scenario, we let the total demand grow with a rate of~$0.3\,\%$ per time step~$t$ with a total demand in time step~$t=T$ of 150 orders. In the peak demand scenario, we let the total demand grow from 150 orders in~$t=0$ up to 225 orders in time step~$t=13$ and decrease again to 150 orders in~$t=26$. To obtain this type of behavior, we added~$\lambda_{it}^{\mathrm{R}}\,0.5\,\exp(-0.1\,(t-13)^2)$ to each~$\lambda_{it}^{\mathrm{R}}$. Finally, we increased the joining rate of \glspl{GW} to~$q^{\mathrm{\gls{GW}}} = 0.3$. 

\FloatBarrier
\section{Hyperparameter tuning} \label{app:parameter_tuning}
To select optimal hyperparameters for $\alpha$, $k^{\mathrm{\gls{GW}}}$, and $k^{\mathrm{\gls{OD}}}$, we conducted a grid-search over~${\alpha\in\left\{10^{-1},10^{-2},10^{-3},10^{-4}\right\}}$ and $k^{\mathrm{\gls{GW},\gls{OD}}} \in\left\{1,50,100,150,200\right\}$. We set $k^{\mathrm{\gls{GW}}} = k^{\mathrm{\gls{OD}}}$ to limit the computational burden. To measure the performance of each parameter setting, we evaluate the relative gap (in \%) of the average total cumulated costs of a policy $\pi^{\mathrm{\gls{PL-VFA}}}$, retrieved after $100\,\mathrm{k}$ iterations of Algorithm~\ref{alg:piecewise_linear_pseudocode}, to the costs obtained with a $\pi^{\mathrm{\gls{MY}}}$, given by
\begin{equation}
\delta^{\mathrm{\gls{MY}}} = 100\,\frac{\bar{C}_T^{\mathrm{\gls{PL-VFA}}}-\bar{C}_T^{\mathrm{\gls{MY}}}}{\bar{C}_T^{\mathrm{\gls{MY}}}}.
\end{equation}
Herein, we obtain $\bar{C}_T^{\mathrm{\gls{MY}},\mathrm{\gls{PL-VFA}}}$ via Equation \eqref{eq:total_cumulated_costs}. We execute both policies 50 times in the environment with the base case parameters to obtain the total cumulated costs. Moreover, we report the standard deviation~$\sigma$ of this gap in percentage points (pp.). Table \ref{tab:grid_search} summarizes the grid search results.
\begin{table}[htbp]
	\centering
	\setlength\heavyrulewidth{1pt}
	\caption{Average gap, $\delta^{\mathrm{\gls{MY}}}$, and standard deviation, $\sigma$, of \gls{PL-VFA} to \gls{MY} after $100\,\mathrm{k}$ iterations of Algorithm \ref{alg:piecewise_linear_pseudocode} for varying $\alpha$, $k^{\mathrm{\gls{GW}}}$, and $k^{\mathrm{\gls{OD}}}$. Best solutions in \textbf{bold} and chosen hyperparameter setting denoted by $^{\textbf{*}}$.}\label{tab:grid_search}
	\begin{tiny}
	\begin{tabular}{lllllllllll}
		\toprule
		& \multicolumn{10}{c}{$k^{\mathrm{\gls{GW}}},\,k^{\mathrm{\gls{OD}}}$} \\
		\cmidrule[1pt]{2-11}
		& \multicolumn{2}{c}{1} & \multicolumn{2}{c}{50} &  \multicolumn{2}{c}{100} &  \multicolumn{2}{c}{150} &  \multicolumn{2}{c}{200} \\
		\cmidrule(r){2-3}  \cmidrule(r){4-5}  \cmidrule(r){6-7}  \cmidrule(r){8-9}  \cmidrule(r){10-11}
		$\alpha$  & $\delta^{\mathrm{MY}}\,(\%)$ & $\sigma\,(\mathrm{pp.})$ & $\delta^{\mathrm{MY}}\,(\%)$ & $\sigma\,(\mathrm{pp.})$ & $\delta^{\mathrm{MY}}\,(\%)$ & $\sigma\,(\mathrm{pp.})$ & $\delta^{\mathrm{MY}}\,(\%)$ & $\sigma\,(\mathrm{pp.})$ & $\delta^{\mathrm{MY}}\,(\%)$ & $\sigma\,(\mathrm{pp.})$ \\
		\cmidrule[1pt](r){1-11}
		&&&&&&&&&&\\[-1.25ex]
		$10^{-1}$ & 1.08                   & 1.81     & 0.50                   & 0.64     & 4.81                   & 0.47     & 72.93 & 22.76    & 241.50                 & 5.62     \\ [0.5ex]
		\hline
		& & & & & & & & & & \\[-0.5ex]
		$10^{-2}$ & -0.87                  & 1.17     & 6.72                   & 0.53     & -4.13                  & 0.42     & \textbf{-12.47}                 & 0.61     & \textbf{-17.14}                 & 0.76     \\[0.5ex]
		\hline
		& & & & & & & & & & \\[-0.5ex]
		$10^{-3}$ & -0.92                  & 0.50     & -7.17                  & 1.06     & \textbf{-9.69}$^{\textbf{*}}$                  & 0.58     & \textbf{-9.74}                  & 0.57     & \textbf{-10.38}                 & 0.59     \\[0.5ex]
		\hline
		& & & & & & & & & & \\[-0.5ex]
		$10^{-4}$ & -0.73                  & 0.42     & -2.82                  & 0.40     & -2.89                  & 0.42     & -2.89                  & 0.42     & -2.89                  & 0.42 \\
		\bottomrule    
	\end{tabular}
	\end{tiny}
\end{table}

\begin{figure}[htbp]
	\centering
	\subfigure[$(\alpha=10^{-2},k^{\mathrm{\gls{GW}},\mathrm{\gls{OD}}} = 150)$.]{
		\includegraphics[width=0.3\textwidth]{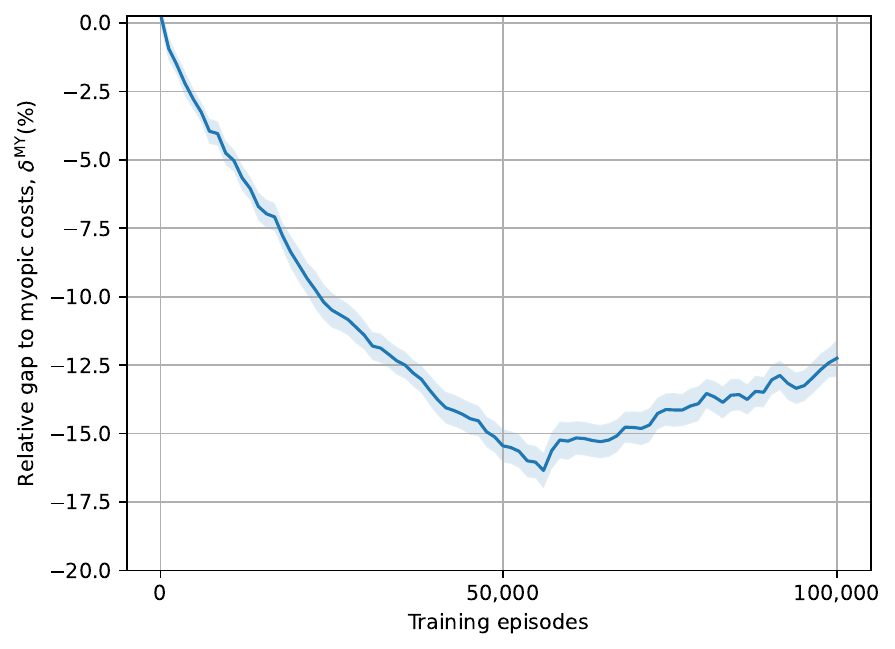}\label{fig:inp001000_15000_tuning}}
	\subfigure[$(\alpha=10^{-2},k^{\mathrm{\gls{GW}},\mathrm{\gls{OD}}} = 200)$.]{
		\includegraphics[width=0.3\textwidth]{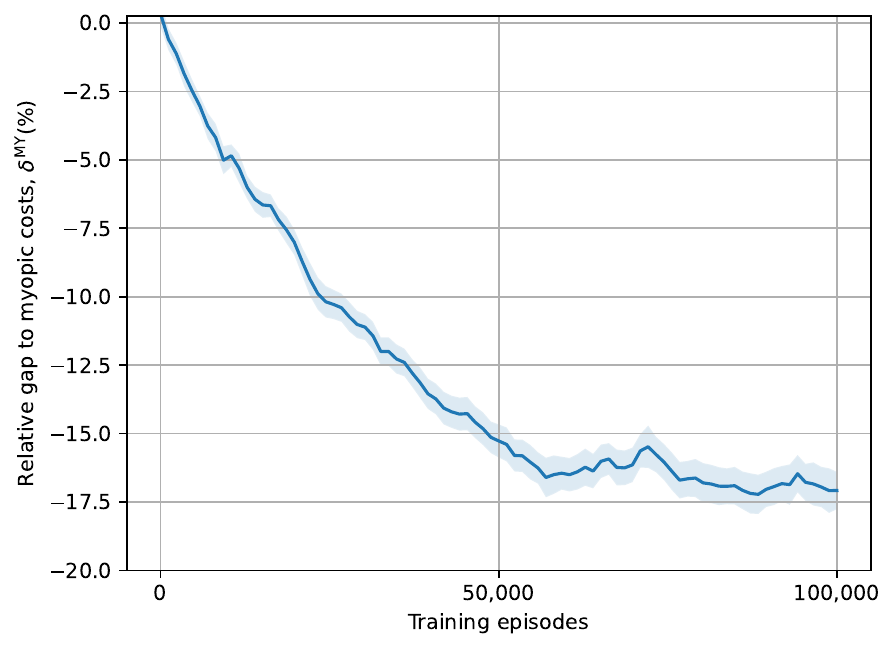}\label{fig:inp001000_20000_tuning}}\\
	\subfigure[$(\alpha=10^{-3},k^{\mathrm{\gls{GW}},\mathrm{\gls{OD}}} = 100)$.]{
		\includegraphics[width=0.3\textwidth]{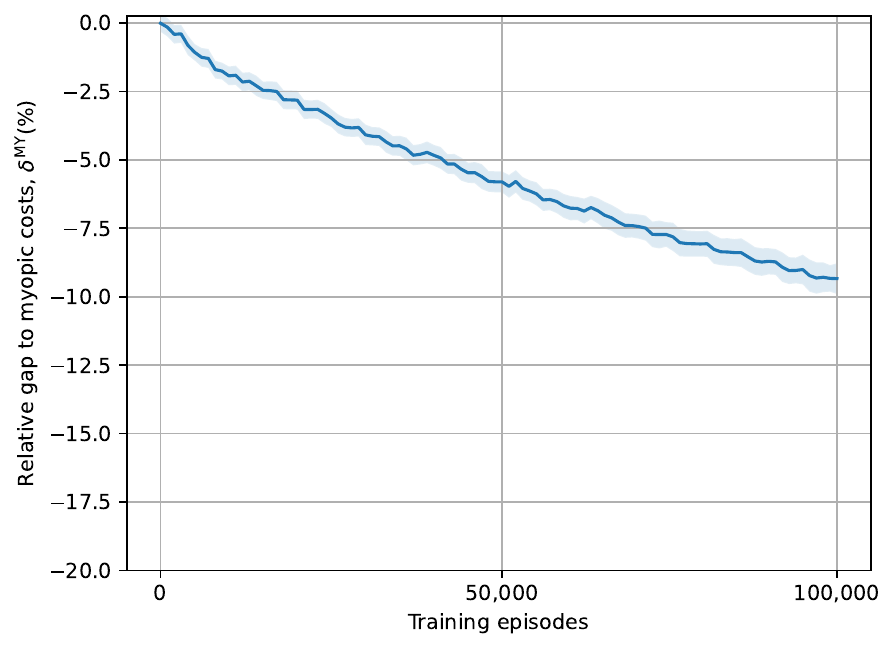}\label{fig:inp000100_10000_tuning}}
	\subfigure[$(\alpha=10^{-3},k^{\mathrm{\gls{GW}},\mathrm{\gls{OD}}} = 150)$.]{
		\includegraphics[width=0.3\textwidth]{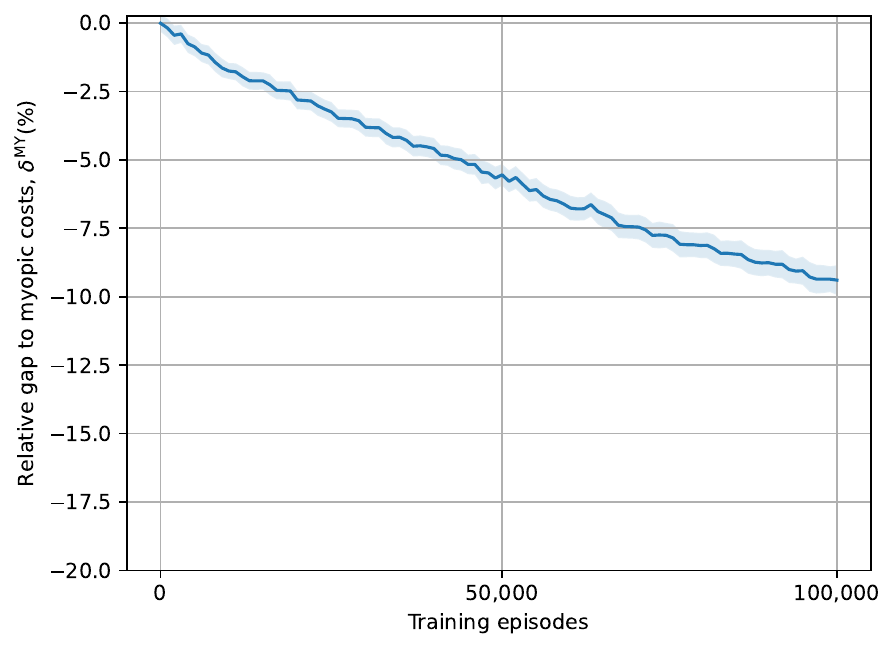}\label{fig:inp000100_15000_tuning}}
	\subfigure[$(\alpha=10^{-3},k^{\mathrm{\gls{GW}},\mathrm{\gls{OD}}} = 200)$.]{
		\includegraphics[width=0.3\textwidth]{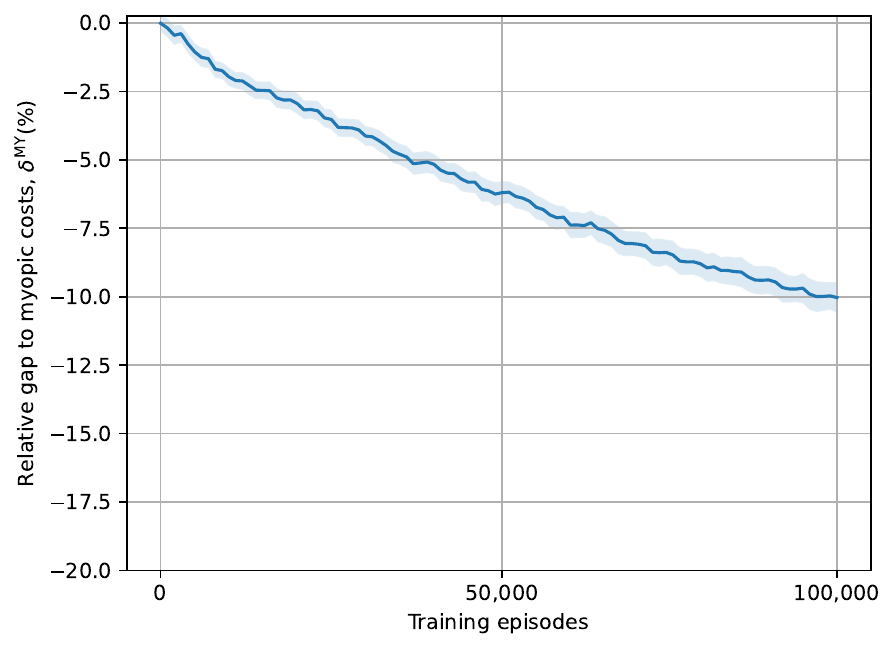}\label{fig:inp000100_20000_tuning}}
	\caption{Average $\delta^{\mathrm{\gls{MY}}}$ during training for different $\alpha$ and $k^{\mathrm{\gls{GW},\gls{OD}}}$. The shaded area corresponds to the variance corridor of one standard
		deviation over 50 policy executions.}
	\label{fig:tuning_alpha}
\end{figure}

We observe that the best results are achieved when $\alpha=10^{-2}$ and $k^{\mathrm{\gls{GW}},\mathrm{\gls{OD}}} = 200$ with $\delta^{\mathrm{\gls{MY}}} = -17.14$ and the second best results are achieved when $\alpha=10^{-2}$ and $k^{\mathrm{\gls{GW}},\mathrm{\gls{OD}}} = 150$ with $\delta^{\mathrm{\gls{MY}}} = -12.47$. In Figure \ref{fig:tuning_alpha}, we report the performance, in terms of $\delta^{\mathrm{\gls{MY}}}$, of \gls{PL-VFA} over the number of training iterations. We observe that the learning process for $(\alpha=10^{-2},k^{\mathrm{\gls{GW}},\mathrm{\gls{OD}}} = 150)$ exhibits a sudden performance collapse after approximately $50\,\mathrm{k}$ iterations (cf. Figure \ref{fig:inp001000_15000_tuning}). For $(\alpha=10^{-2},k^{\mathrm{\gls{GW}},\mathrm{\gls{OD}}} = 200)$ the performance does not collapse as significantly as for $(\alpha=10^{-2},k^{\mathrm{\gls{GW}},\mathrm{\gls{OD}}} = 150)$ but is also unstable after approximately $50\,\mathrm{k}$ iterations (cf. Figure \ref{fig:inp001000_20000_tuning}). For~${(\alpha=10^{-3},k^{\mathrm{\gls{GW}},\mathrm{\gls{OD}}} = 100,150,200)}$ the convergence curves show a more stable learning behaviour, although they did not converge after $100\,\mathrm{k}$ iterations (cf. Figures \ref{fig:inp000100_10000_tuning}, \ref{fig:inp000100_15000_tuning}, and \ref{fig:inp000100_20000_tuning}). For~${(\alpha=10^{-3},k^{\mathrm{\gls{GW}},\mathrm{\gls{OD}}} = 100,150,200)}$ the results do not differ significantly among each other. Hence, to balance learning performance and stability and to prevent too coarse aggregation factors, we set~${(\alpha=10^{-3},k^{\mathrm{\gls{GW}},\mathrm{\gls{OD}}} = 100)}$. 
\FloatBarrier
\section{Impact of underhiring on \gls{LSP}'s service level.}\label{app:sevice_level_analysis}

\begin{figure}[htbp]
	\subfigure[$q^{\mathrm{\gls{GW}}} = 0.05$]{
		\includegraphics[width=0.3\textwidth]{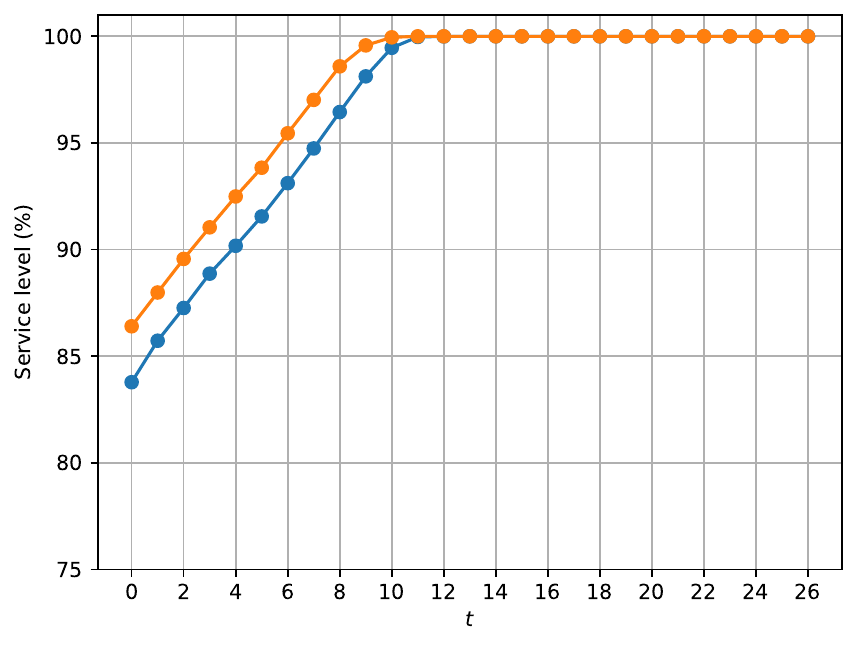}
		\label{fig:drivers_FD_GW_add_gw_inp003}}
	\subfigure[$q^{\mathrm{\gls{GW}}} = 0.09$]{
		\includegraphics[width=0.3\textwidth]{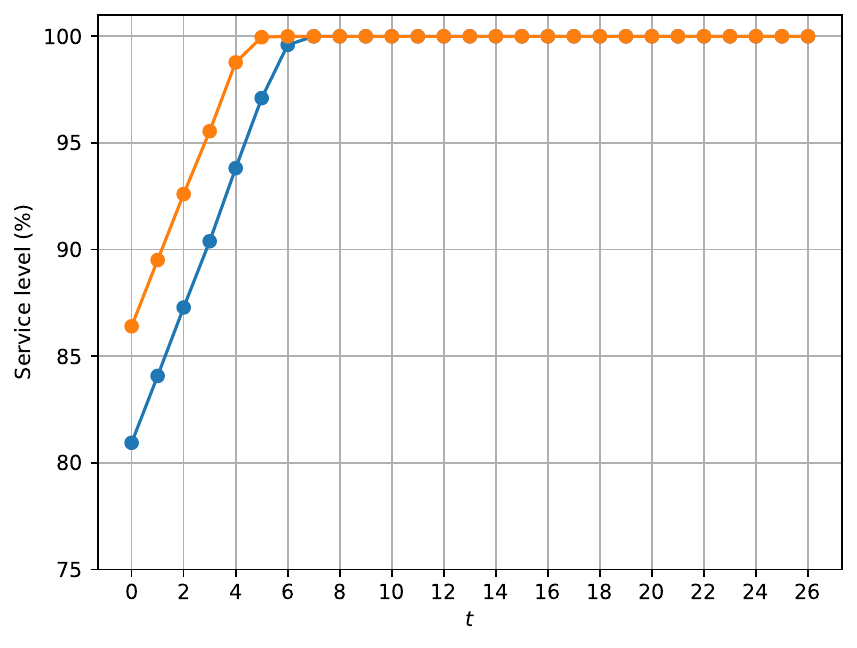}
		\label{fig:drivers_FD_GW_add_gw_inp009}}
	\subfigure[$q^{\mathrm{\gls{GW}}} = 0.17$]{
		\includegraphics[width=0.3\textwidth]{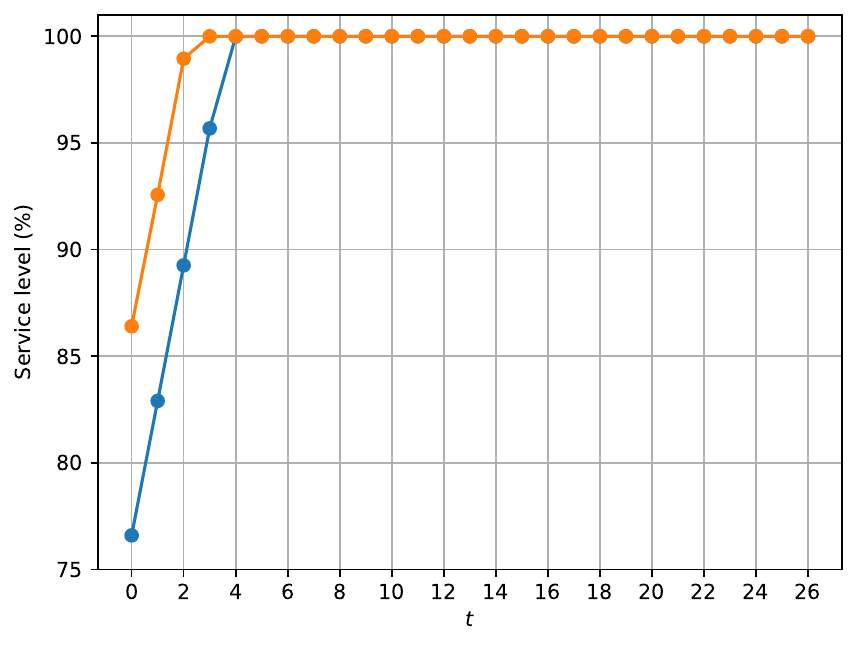}
		\label{fig:drivers_FD_GW_add_gw_inp015}}
	\caption{Average service level in each time step~$t$ when varying~$q^{\mathrm{\gls{GW}}}$.}\label{fig:variation_joining_rates_sla}
\end{figure}

To understand the impact of~$\pi^{\mathrm{\gls{PL-VFA}}}$'s underhiring on the \gls{LSP}'s service level, we visualize service levels for~$q^{\mathrm{\gls{GW}}} = 0.05$,~$q^{\mathrm{\gls{GW}}} = 0.09$, and~$q^{\mathrm{\gls{GW}}} = 0.17$ in Figure \ref{fig:variation_joining_rates_sla}. For all three joining rates, both policies accept lower service levels at the beginning of the time horizon and achieve a service level of~100\% after four (cf. Figure \ref{fig:drivers_FD_GW_add_gw_inp015}) to ten (cf. Figure \ref{fig:drivers_FD_GW_add_gw_inp003}) time steps respectively. We observe that the \gls{LSP} accepts lower service levels with increasing joining rate in early time steps of up to 77\%. However, they reach a service level of~100\% faster when the joining rate is high since lower service levels imply higher penalty costs,~$\pi^{\mathrm{\gls{PL-VFA}}}$ trades-off the cost of hiring \glspl{FD} with the penalties it has to pay for requests that were not served.

%\end{APPENDIX}